\documentclass[longauth]{aa}

\usepackage{natbib}  
\bibpunct{(}{)}{;}{a}{}{,} 

\usepackage[colorlinks = true,
            linkcolor = blue,
            urlcolor  = blue,
            citecolor = blue]{hyperref} 

\usepackage{graphicx}
\usepackage{txfonts}
\begin{document} 

   \title{TOI-5005 b: A super-Neptune in the savanna near the ridge}

    \author{A.~Castro-Gonz\'{a}lez\inst{ \ref{CAB_villafranca}}
    \and
    J.~Lillo-Box\inst{\ref{CAB_villafranca}}
    \and
    D. J. Armstrong\inst{\ref{warwik},\ref{warwik_2}}
    \and
    L.~Acuña\inst{\ref{max_planck}}
    \and
    A.~Aguichine\inst{\ref{UCSC}}
    \and
    V. Bourrier\inst{\ref{obs_geneva}}
    \and
    S. Gandhi\inst{\ref{warwik}}
    \and
    S.~G.~Sousa\inst{\ref{CAUP}, \ref{dep_astro_porto}}
    \and
    E.~Delgado-Mena\inst{\ref{CAUP}}
    \and
    A.~Moya\inst{\ref{uni_valencia}}
    \and
    V.~Adibekyan\inst{\ref{CAUP}, \ref{dep_astro_porto}}
    \and
    A.~C.~M.~Correia\inst{\ref{cfisuc-coimbra},\ref{obs_paris}}
    \and
    D.~Barrado\inst{ \ref{CAB_villafranca}}
    \and 
    M.~Damasso\inst{\ref{INAF_torin}}
    \and \\
    J.~N.~Winn\inst{\ref{princeton}}
    \and
    N.~C.~Santos\inst{\ref{CAUP},\ref{dep_astro_porto}}
    \and
    K. Barkaoui\inst{\ref{astro_liege},\ref{Depart_Earth_MIT},\ref{IAC}} 
    \and
    S.~C.~C.~Barros\inst{\ref{CAUP},\ref{dep_astro_porto}} 
    \and 
    Z.~Benkhaldoun\inst{\ref{ouka}}
    \and
    F.~Bouchy\inst{\ref{geneva_departement}}
    \and
    C.~Brice\~{n}o\inst{\ref{cerro_tololo}}
    \and
    D.~A.~Caldwell\inst{\ref{SETI_NASA_ames}}
    \and
    K. A.\ Collins\inst{\ref{Cambridge,MA}}
    \and
    Z.~Essack\inst{\ref{uni_new_mexico}}
    \and
    M. Ghachoui\inst{\ref{astro_liege},\ref{ouka}} 
    \and
    M. Gillon\inst{\ref{astro_liege}} 
    \and
    R.~Hounsell\inst{\ref{uni_maryland},\ref{nasa_goddard}} 
    \and
    E.~Jehin\inst{\ref{STAR_Uliege}} 
    \and
    J.~M.~Jenkins\inst{\ref{NASA_ames}}
    \and
    M.~A.~F.~Keniger\inst{\ref{warwik},\ref{warwik_2}}
    \and
    N.~Law\inst{\ref{north_carolina_chapel_hill}}
    \and
    A.~W.~Mann\inst{\ref{north_carolina_chapel_hill}}
    \and
    L.~D.~Nielsen\inst{\ref{munich}}
    \and
    F.~J.~Pozuelos\inst{\ref{iaa}} 
    \and
    N.~Schanche\inst{\ref{nasa_goddard},\ref{uni_maryland_2}}
    \and
    S.~Seager\inst{\ref{MIT_physics}, \ref{MIT_earth}, \ref{MIT_aeronautics}} 
    \and
    T.-G. Tan\inst{\ref{perth}}
    \and
    M. Timmermans\inst{\ref{astro_liege}} 
    \and
    J.~Villase{\~ n}or\inst{\ref{MIT_physics}}
    \and
    C. N.\ Watkins\inst{\ref{Cambridge,MA}}
    \and
    C.~Ziegler\inst{\ref{austin_uni}}
    }
    
    \institute{Centro de Astrobiolog\'{i}a, CSIC-INTA, Camino Bajo del Castillo s/n, 28692 Villanueva de la Ca\~{n}ada, Madrid, Spain\label{CAB_villafranca} \\\email{acastro@cab.inta-csic.es}
    \and
    Centre for Exoplanets and Habitability, University of Warwick, Gibbet Hill Road, Coventry, CV4 7AL, UK\label{warwik}
    \and
    Department of Physics, University of Warwick, Gibbet Hill Road, Coventry, CV4 7AL, UK\label{warwik_2}
    \and
    Max Planck Institute for Astronomy, Königstuhl 17, D-69117 Heidelberg, Germany\label{max_planck}
    \and
    Department of Astronomy and Astrophysics, University of California, Santa Cruz, CA, USA\label{UCSC}
    \and
    Observatoire Astronomique de l’Université de Genève, Chemin Pegasi 51b, CH-1290 Versoix, Switzerland\label{obs_geneva}
    \and
    Instituto de Astrof\'isica e Ci\^encias do Espa\c{c}o, Universidade do Porto, CAUP, Rua das Estrelas, 4150-762 Porto, Portugal\label{CAUP}
    \and
    Departamento de F\'isica e Astronomia, Faculdade de Ci\^encias, Universidade do Porto, Rua do Campo Alegre, 4169-007 Porto, Portugal\label{dep_astro_porto}
    \and
    Departament d’Astronomia i Astrofísica, Universitat de València, C. Dr. Moliner 50, 46100 Burjassot, Spain\label{uni_valencia}
    \and
    CFisUC, Departamento de F\'isica, Universidade de Coimbra, 3004-516 Coimbra, Portugal\label{cfisuc-coimbra}
    \and
    IMCCE, UMR8028 CNRS, Observatoire de Paris, PSL Universit\'{e}, 77 Av. Denfert-Rochereau, 75014 Paris, France\label{obs_paris}
    \and
    INAF - Osservatorio Astrofisico di Torin, Via Osservatorio 20, I-10025 Pino Torinese, Italy\label{INAF_torin}
    \and
     Department of Astrophysical Sciences, Princeton University, Princeton, NJ 08544, USA\label{princeton}
     \and
    Astrobiology Research Unit, Université de Liège, 19C Allée du 6 Août, 4000 Liège, Belgium \label{astro_liege}
    \and
    Department of Earth, Atmospheric and Planetary Science, Massachusetts Institute of Technology, 77 Massachusetts Avenue, Cambridge, MA 02139, USA \label{Depart_Earth_MIT}
    \and
    Instituto de Astrofísica de Canarias (IAC), 38205 La Laguna, Tenerife, Spain \label{IAC}
    \and
     Oukaimeden Observatory, High Energy Physics and Astrophysics Laboratory, Faculty of sciences Semlalia, Cadi Ayyad University, Marrakech, Morocco\label{ouka}
    \and
    Département d’Astronomie, Université de Genève, Chemin Pegasi 51, 1290 Versoix, Switzerland\label{geneva_departement}
    \and
    Cerro Tololo Inter-American Observatory, Casilla 603, La Serena, Chile\label{cerro_tololo}
    \and
    SETI Institute, Mountain View, CA 94043 USA/NASA Ames Research Center, Moffett Field, CA 94035 USA\label{SETI_NASA_ames}
     \and
      Center for Astrophysics \textbar \ Harvard \& Smithsonian, 60 Garden Street, Cambridge, MA 02138, USA\label{Cambridge,MA}
      \and
      Department of Physics and Astronomy, The University of New Mexico, 210 Yale Blvd NE, Albuquerque, NM 87106, USA\label{uni_new_mexico}
    \and
    University of Maryland, Baltimore County, 1000 Hilltop Cir., Baltimore, MD 21250, USA\label{uni_maryland}
    \and
     NASA Goddard Space Flight Center, 8800 Greenbelt Rd., Greenbelt, MD 20771, USA\label{nasa_goddard}
    \and
     Space Sciences, Technologies and Astrophysics Research (STAR) Institute, Université de Liège, Allée du 6 Août 19C, 4000 Liège, Belgium \label{STAR_Uliege}
    \and
     NASA Ames Research Center, Moffett Field, CA 94035, USA\label{NASA_ames}
     \and
     Department of Physics and Astronomy, The University of North Carolina at Chapel Hill, Chapel Hill, NC 27599-3255, USA\label{north_carolina_chapel_hill}
    \and
     University Observatory, Faculty of Physics, Ludwig-Maximilians-Universit{\"a}t M{\"u}nchen, Scheinerstr. 1, 81679 Munich, Germany\label{munich}
     \and
     Instituto de Astrof\'isica de Andaluc\'ia (IAA-CSIC), Glorieta de la Astronom\'ia s/n, 18008 Granada, Spain\label{iaa}
     \and
     Department of Astronomy, University of Maryland, College Park, MD  20742, USA\label{uni_maryland_2}
     \and
     Department of Physics and Kavli Institute for Astrophysics and Space Research, Massachusetts Institute of Technology, Cambridge, MA 02139, USA\label{MIT_physics}
    \and
    Department of Earth, Atmospheric and Planetary Sciences, Massachusetts Institute of Technology, Cambridge, MA 02139, USA\label{MIT_earth}
    \and
    Department of Aeronautics and Astronautics, MIT, 77 Massachusetts Avenue, Cambridge, MA 02139, USA\label{MIT_aeronautics}
    \and 
    Perth Exoplanet Survey Telescope, Perth, Western Australia\label{perth}
    \and
    Department of Physics, Engineering and Astronomy, Stephen F. Austin State University, 1936 North St, Nacogdoches, TX 75962, USA\label{austin_uni}
    }

\date{Received 25 July 2024 / Accepted 26 September 2024}

 
  \abstract
   {The Neptunian desert and savanna have  recently been found to be separated by a ridge, an overdensity of planets in the period range of  $\simeq$3--5 days. These features are thought to be shaped by dynamical and atmospheric processes. However, their roles are not yet well understood.}
   {Our aim was to confirm and characterize the super-Neptune TESS candidate TOI-5005.01, which orbits a moderately bright (V = 11.8) solar-type star (G2 V) with an orbital period of 6.3 days. With these properties, TOI-5005.01 is located in the Neptunian savanna near the ridge.}
   {We used Bayesian inference to analyse 38 HARPS radial velocity measurements, three sectors of TESS photometry, and two PEST and TRAPPIST-South transits. We tested a set of models involving eccentric and circular orbits, long-term drifts, and Gaussian processes to account for correlated stellar and instrumental noise. We computed the Bayesian evidence to find the model that best represents our dataset and infer the orbital and physical properties of the system.}
   {We confirm TOI-5005~b to be a transiting super-Neptune with a radius of $R_{\rm p}$ = $6.25\pm 0.24$~$\rm R_{\rm \oplus}$ ($R_{\rm p}$ = $0.558\pm 0.021$ $\rm R_{\rm J}$) and a mass of $M_{\rm p}$~= $32.7\pm 5.9$ $\rm M_{\oplus}$ ($M_{\rm p}$ = $0.103\pm 0.018$ $\rm M_{\rm J}$), which corresponds to a mean density of $\rho_{\rm p}$ = $0.74 \pm 0.16$ $\rm g \, cm^{-3}$.  Our internal structure modelling indicates that the core mass fraction (CMF = $0.74^{+0.05}_{-0.45}$) and envelope metal mass fraction ($Z_{\rm env}$ = $0.08^{+0.41}_{-0.06}$) of TOI-5005~b are degenerate, but the overall metal mass fraction is well constrained to a value slightly lower than that of Neptune and Uranus ($Z_{\rm planet}$ = $0.76^{+0.04}_{-0.11}$). The $Z_{\rm planet}$/$Z_{\rm star}$ ratio is consistent with the well-known mass-metallicity relation, which suggests that TOI-5005~b was formed via core accretion. We also estimated the present-day atmospheric mass-loss rate of TOI-5005~b, but found contrasting predictions depending on the choice of photoevaporation model ($0.013\pm 0.008$~$\rm M_{\oplus}$~Gyr$^{-1}$ vs $0.17\pm 0.12$ ~$\rm M_{\oplus}$~Gyr$^{-1}$). At a population level, we find statistical evidence ($p$-value = $0.0092^{+0.0184}_{-0.0066}$) that planets in the savanna such as TOI-5005~b tend to show lower densities than planets in the ridge, with a dividing line around 1 $\rm g \, cm^{-3}$, which supports the hypothesis of different evolutionary pathways populating the two regimes.}
   {TOI-5005~b is located in a region of the period-radius space that is key to studying the transition between the Neptunian ridge and the savanna. It orbits the brightest star of all such planets known today, which makes it a target of interest for atmospheric and orbital architecture observations that will bring a clearer picture of its overall evolution.}


   \maketitle

%

\section{Introduction}

The discovery of 51 Peg b \citep{1995Natur.378..355M} and subsequent detections of Jupiter-like planets in close-in orbits \citep[e.g.][]{1997ApJ...474L.115B,1998PASP..110.1389B,2000A&A...356..599S,2000ApJ...529L..45C,2000ApJ...529L..41H} revealed that the Solar System is not an archetypal planetary system in our Galaxy. The paradigm shift was strengthened with the detection of giant planets with sizes and masses between those of Neptune and Saturn \citep[e.g.][]{2004ApJ...617..580B,2004ApJ...614L..81M,2010A&A...520A..66B,2010Sci...330...51H,2011ApJ...728..138H}. These planets are commonly known as transitional, intermediate, or Neptunian planets, and populate an extensive region of the parameter space with no representation in the Solar System. 

The core accretion theory for planet formation \citep{1996Icar..124...62P} predicts that giant planets can only form at large orbital distances, beyond the ice line, where the solid material in the protoplanetary disk can build planetary cores massive enough to trigger runaway gas accretion \citep[$\sim$10$\rm M_{\oplus}$;][]{2006ApJ...648..666R,2015ApJ...811...41L,2019ApJ...878...36L}. The cores of Neptunian planets are expected to be massive enough to initiate gas accretion, so this process must have been interrupted for those failed giants. This could be caused by a late formation of the planetary core or by an early dissipation of the disk gas \citep[e.g.][]{2011A&A...526A.111M,2015IJAsB..14..201M,2016ApJ...829..114B}. Interestingly, planet occurrence studies based on transit \citep[e.g.][]{2012ApJS..201...15H}, radial velocity \citep[e.g.][]{2021AJ....162..243B}, and microlensing \citep[e.g.][]{2016ApJ...833..145S} surveys show that the occurrence rates of Neptunes and Jupiters are comparable in a wide range of orbital distances. Therefore, the interruption of the core accretion process seems to be a frequent phenomenon during the formation of giant planets.

The physical properties of Neptunian planets at large orbital distances are thought to be primarily forged during their formation. However, close-in Neptunes ($P_{\rm orb}$ < 30 days) are known to be affected by complex atmospheric and dynamical processes that can modify both their orbital and physical properties. Close-in giants can migrate inwards soon after their formation, before the protoplanetary disk has dissipated, in a process called disk-driven migration \citep{1979ApJ...233..857G,1996Natur.380..606L,2016SSRv..205...77B}.  They can also undergo high-eccentricity tidal migration  \citep[HEM;][]{2003ApJ...589..605W,2008ApJ...686..621F,2008ApJ...686..580C,2011CeMDA.111..105C,2012ApJ...751..119B}, which can occur at any time of a planet's lifetime due to an outer massive perturber. Once they reach a close-in configuration, Neptunian planets can undergo significant physical changes due to evaporation, as evidenced by the observed high mass-loss rates in GJ 436~b \citep{2015Natur.522..459E}, GJ 3470 b \citep{2018A&A...620A.147B},  HAT-P-11 b \citep{2022NatAs...6..141B}, and HAT-P-26~b \citep{2022AJ....164..234V}. 

Both migration and evaporation processes are thought to be the main agents that shape the distribution of close-in Neptunes. In a recent work, \citet{2024A&A...689A.250C} study planet occurrences in the Neptunian domain and find evidence of an overdensity of planets in the orbital period range of $\simeq$3-5 days, which they called the Neptunian ridge. The ridge thus appears as a true physical feature separating the Neptunian desert \citep[i.e. a dearth of Neptunes on the shortest-period orbits;][]{2011A&A...528A...2B,2011ApJ...727L..44S,2011ApJ...742...38Y,2013ApJ...763...12B,2016MNRAS.455L..96H,2016NatCo...711201L,2016A&A...589A..75M} and the Neptunian savanna \citep[i.e. a moderately populated region
at larger orbital distances;][]{2023A&A...669A..63B}. Interestingly, an important fraction of Neptunes at the edge of the desert (i.e. in the newly identified ridge) have been observed in eccentric and polar orbits \citep[e.g.][]{2020A&A...635A..37C,2023A&A...669A..63B}, which favours HEM processes as the main migration mechanism populating the ridge. In contrast, planets in the savanna tend to show small orbital misalignments and circular orbits (see Fig. 1 of Bourrier et al., in prep.). These emerging trends can be explained in two different ways. On the one hand, the ridge may be primarily populated by HEM processes, while the planets in the savanna could reach these locations through disk-driven migration. On the other hand, HEM processes could dominate planet migration throughout the entire Neptunian period range, being the spin-axis angle and eccentricity trends a consequence of photoevaporation and tidal forces exerted by the star. In both hypotheses, we would only detect those Neptunes in the ridge that migrated recently, and hence did not have time to circularize their orbits, align their spin-axis angles, and completely evaporate their atmospheres \cite[see][for an extended discussion]{2018Natur.553..477B,2020A&A...635A..37C,2021A&A...647A..40A,2024A&A...689A.250C}.

Our current understanding of the Neptunian desert, ridge, and savanna is limited by the scarcity of observational constraints. In a first step, obtaining a large sample of confirmed planets with precise radii, masses, and orbital eccentricities is critical to infer possible differences in the formation and evolution mechanisms that gave rise to those Neptunian features. In a second step, coupling the aforementioned constraints with follow-up observations of the spin-axis angle and atmospheric escape rates will provide a clearer picture of the origins and evolution of close-in Neptunes at a population level. In this regard, the Transiting Exoplanets Survey Satellite \citep[TESS;][]{2014SPIE.9143E..20R} is playing a key role. Its photometric precision is high enough to enable the detection of Neptunian planets, which typically show large transit depths of a few parts per thousand. In addition, its focus on bright stars and the monitoring of practically the entire sky is boosting the number of close-in Neptunian candidates around stars amenable for detailed follow-up studies.

The HARPS-NOMADS collaboration (PI Armstrong, programmes 108.21YY.001 and 108.21YY.002) is an observational effort to confirm, characterize, and eventually perform statistical studies of close-in Neptunes detected by TESS \citep[e.g.][]{2023MNRAS.524.5804A,2023MNRAS.524.3877H,2023A&A...669A.109L,2023MNRAS.526..548O,2024MNRAS.532.1612H}. This work is part of such a collaboration, where we confirm and characterize the close-in ($P_{\rm orb}$ = 6.3 days) super-Neptune ($R_{\rm p}$ = 6.3 $\rm R_{\rm \oplus}$) TOI-5005~b, which orbits a moderately bright ($\rm V$ = 11.8) solar-type star (G2 V, $T_{\rm eff}$ = 5750 K).  With these properties, TOI-5005~b is located in the Neptunian savanna near the ridge, a poorly populated but key region of the parameter space for understanding the transition between those regimes. In Sect.~\ref{sec:observations} we describe the TESS, HARPS, and additional PEST and TRAPPIST-South photometric observations. In Sect.~\ref{sec:stellar_charact} we present our stellar characterization based on the HARPS spectra. In Sect.~\ref{sec:analysis_results} we describe our analysis of photometric and
spectroscopic data and present the derived system parameters. In Sect.~\ref{sec:discussion} we discuss the results, and we conclude in Sect.~\ref{conclusions}.

\section{Observations}
\label{sec:observations}

\subsection{TESS high-precision photometry}
\label{sec:obs_tess}

\begin{table*}[]
\caption{TESS observations of TOI-5005.}
\renewcommand{\arraystretch}{1.4}
\setlength{\tabcolsep}{5.5pt}
\begin{tabular}{cccccccccc}
\hline \hline
Sector & Cycle & Start date  & End date     & Camera & CCD & FFIs   & TPFs    & Cadence      & Photometry pipelines \\ \hline
12     & 1     & 21 May 2019 & 18 June 2019 & 1      & 1   & 1234    & 0       & 30 min       & QLP                  \\
39     & 3     & 27 May 2021 & 24 June 2021 & 1      & 1   & 3865 & 3865       & 10 min       & TESS-SPOC, QLP       \\
65     & 5     & 4 May 2023  & 2 June 2023  & 1      & 2   & 11663 & 19515 & 200 s, 120 s & SPOC, TESS-SPOC, QLP \\
\hline
\end{tabular}
\label{tab:TESS_summary}
\end{table*}

The star TOI-5005 (TIC 282485660) has been observed by TESS in sectors 12, 39, and 65 (hereafter S12, S39, and S65).  In Table \ref{tab:TESS_summary}, we summarize the details of the observations. The full-frame images (FFIs) of the three sectors were processed through the Quick Look Pipeline \citep[QLP;][]{2020RNAAS...4..204H}, which computed simple aperture photometry (SAP) for all sources in the TESS Input Catalog \citep[TIC;][]{2018AJ....156..102S,2019AJ....158..138S} with magnitudes up to T = 13.5 mag. The S39 and S65 FFIs were also processed by the TESS-SPOC pipeline \citep{2020RNAAS...4..201C}, and the S65 TPFs by the SPOC pipeline \citep{2016SPIE.9913E..3EJ}. SPOC and TESS-SPOC operate at the Science Processing Operations Center (NASA Ames Research Center) under the same codebase and provide SAP \citep{twicken:PA2010SPIE,morris:PA2020KDPH} and Presearch Data Conditioned Simple Aperture Photometry (PDCSAP). The PDCSAP is the SAP processed by the PDC algorithm, which corrects
the photometry of instrumental systematics that are common to all stars in the same CCD \citep{2012PASP..124.1000S,2012PASP..124..985S}. The complete QLP, TESS-SPOC, and SPOC datasets are available at the Mikulski Archive for Space Telescopes (MAST).\footnote{\url{https://mast.stsci.edu/portal/Mashup/Clients/Mast/Portal.html}}

In January 2022, the QLP-based faint-star search pipeline \citep{2022ApJS..259...33K} detected in S39 a periodic transit-like flux decrease that was alerted by the TESS Science Office as a TESS Object of Interest (TOI-5005.01) that would benefit from follow-up observations \citep{2021ApJS..254...39G}. The detection yielded a period of $6.30847 \pm 0.00003$ days, a transit duration of $2.9 \pm 0.3$ hours, and a transit depth of $3690 \pm 4$ ppm (parts per million).

\subsubsection{\texttt{TLS} and \texttt{GLS} periodograms}

In Fig.~\ref{fig:tls_to_TESS}, we show the sector-by-sector transit least squares periodograms \citep[\texttt{TLS;}][]{2019A&A...623A..39H} of the QLP/SAP (S12), TESS-SPOC/PDCSAP (S39), and SPOC/PDCSAP (S65) photometry flattened; that is, de-trended from low-frequency trends of stellar or instrumental origin. In the three sectors, the maximum power peak corresponds to a 6.3-day periodicity, which coincides with that of TOI-5005.01. This peak has a signal detection efficiency (SDE) of 10.3 in S12, 16.0 in S39, and 16.5 in S65. Those SDE are above the commonly used empirical thresholds for transit detection, namely, SDE $>$ 6.0 \citep{2015ApJ...807...45D}, SDE $>$ 6.5 \citep{2018AJ....156..277L}, SDE $>$ 7 \citep{2012ApJ...761..123S}, and SDE $>$ 10 \citep{2018MNRAS.473L.131W}. Therefore, although TOI-5005.01 was not announced until S39 was observed, we find that its transit-like signature can be detected in the three sectors, with two transit events in S12, four in S39, and three and a half in S65. We also computed the \texttt{TLS} periodogram of the complete dataset and recovered the 6.3-day signal with an SDE of 51.8. We repeated the process with the TOI-5005.01 transits masked to search for additional transit-like signatures and found no further significant periodicities. We note that the SPOC pipeline also detected the 6.3-day transit signal with a multiple event detection statistic (MES) of 21.7 in S39 and 13.3 in S65 \citep{2002ApJ...575..493J,2020TPSkdph}, which correspond to model-fitted signal-to-noise ratios (S/N) of 24.5 and 18.3, respectively \citep{Twicken:DVdiagnostics2018,Li:DVmodelFit2019}.

In Fig.~\ref{fig:gls_to_TESS}, we show the sector-by-sector generalized Lomb-Scargle periodograms \citep[\texttt{GLS;}][]{2009A&A...496..577Z} of the TESS time series after masking the TOI-5005.01 transit-like features. The periodograms of the systematics-corrected time series (i.e. PDCSAP of S39 and S65) show maximum power periods of 6.4 and 6.3 days with False Alarm Probabilities (FAPs) below $10^{-40}\,\%$. These periodicities coincide with the orbital period of TOI-5005.01. In the right panels of Fig.~\ref{fig:gls_to_TESS}, we show the TESS photometry folded in phase to the orbital period of TOI-5005.01, which illustrates the existence of a sinusoidal-like modulation of 0.9 ppt (parts per thousand) in S39 and 0.7 ppt in S65 synchronized with the orbit of the planet candidate. We note, however, that while the signal periodicity persists, the signal phase is modified from one sector to another. The periodogram of the systematics-uncorrected QLP/SAP photometry shows a maximum power period of 6.9 days with a FAP below $10^{-30}\,\%$. This periodicity differs from that of the TOI-5005.01 orbit by 8$\%$, but the photometry folded to its orbital period shows a tentative sinusoidal behaviour. With an orbital period of 6.3 days, TOI-5005.01 is located within the sub-Alfvénic radius of its parent star. Therefore, this potentially synchronized signal suggests a planet-induced origin most likely due to magnetic star-planet interactions \cite[MSPIs; e.g.][]{2000ApJ...533L.151C,2003A&A...406..373S,2003ApJ...597.1092S,2009EM&P..105..373P,2019NatAs...3.1128C,2024A&A...684A.160C}. This signal also seems to be present within the systematics-uncorrected S39 and S65 SAP fluxes. In the S39 SAP fluxes, the 6.3-day periodicity appears as the second-highest peak. In the S65 SAP fluxes, the signal does not appear, but after performing a simple linear de-trending the maximum power period also indicates a 6.0-day periodicity. Consistently, the S39 and S65 SAP fluxes folded in phase to the planetary orbit show a sinusoidal modulation. We illustrate the \texttt{GLS} periodograms of the S39 and S65 SAP fluxes in Fig.~E.1 (available on Zenodo). In this work, we delve into the PDC correction to unveil whether the prominent 6.3-day signal has a true stellar origin, or if it could have been generated by residual uncorrected systematics. Before doing such an analysis, it is important to assess the possibility of flux contamination within the photometric aperture. Given the large TESS pixel sizes (21 $\times$ 21 arcsec), it is very common that the measured fluxes do not  come from the target star exclusively, but also from nearby stars, which could eventually be the sources of the signals found. Indeed, the TESS photometry of TOI-5005 receives flux contributions from several nearby stars. 

\subsubsection{The \texttt{TESS-cont} algorithm}

\begin{figure*}
    \includegraphics[width=0.48\textwidth]{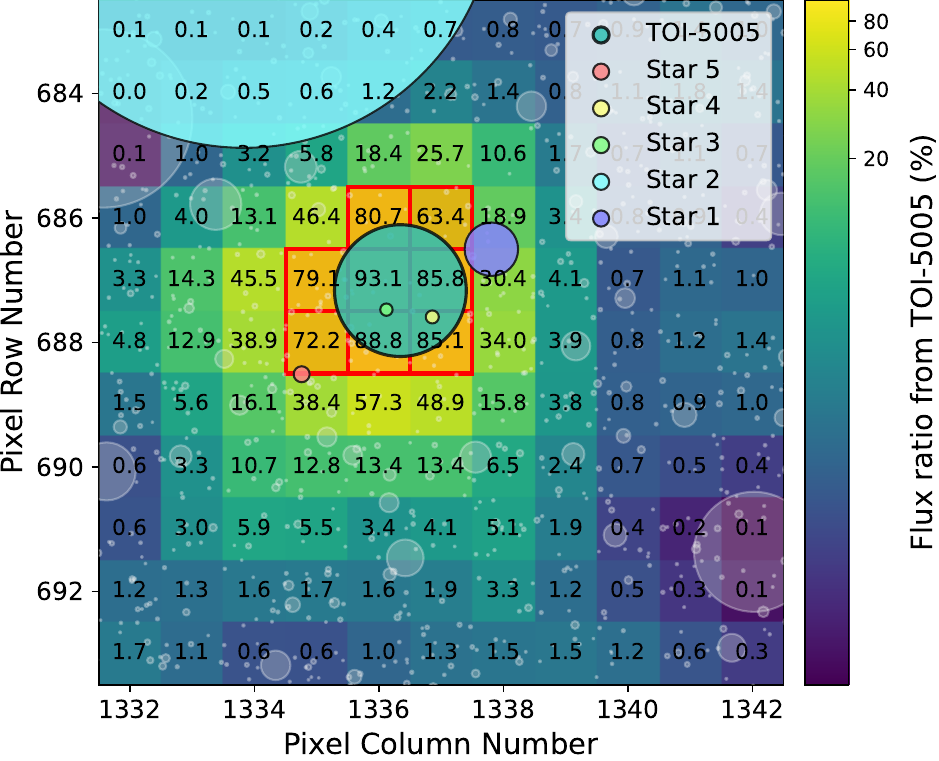}
    \includegraphics[width=0.48\textwidth]{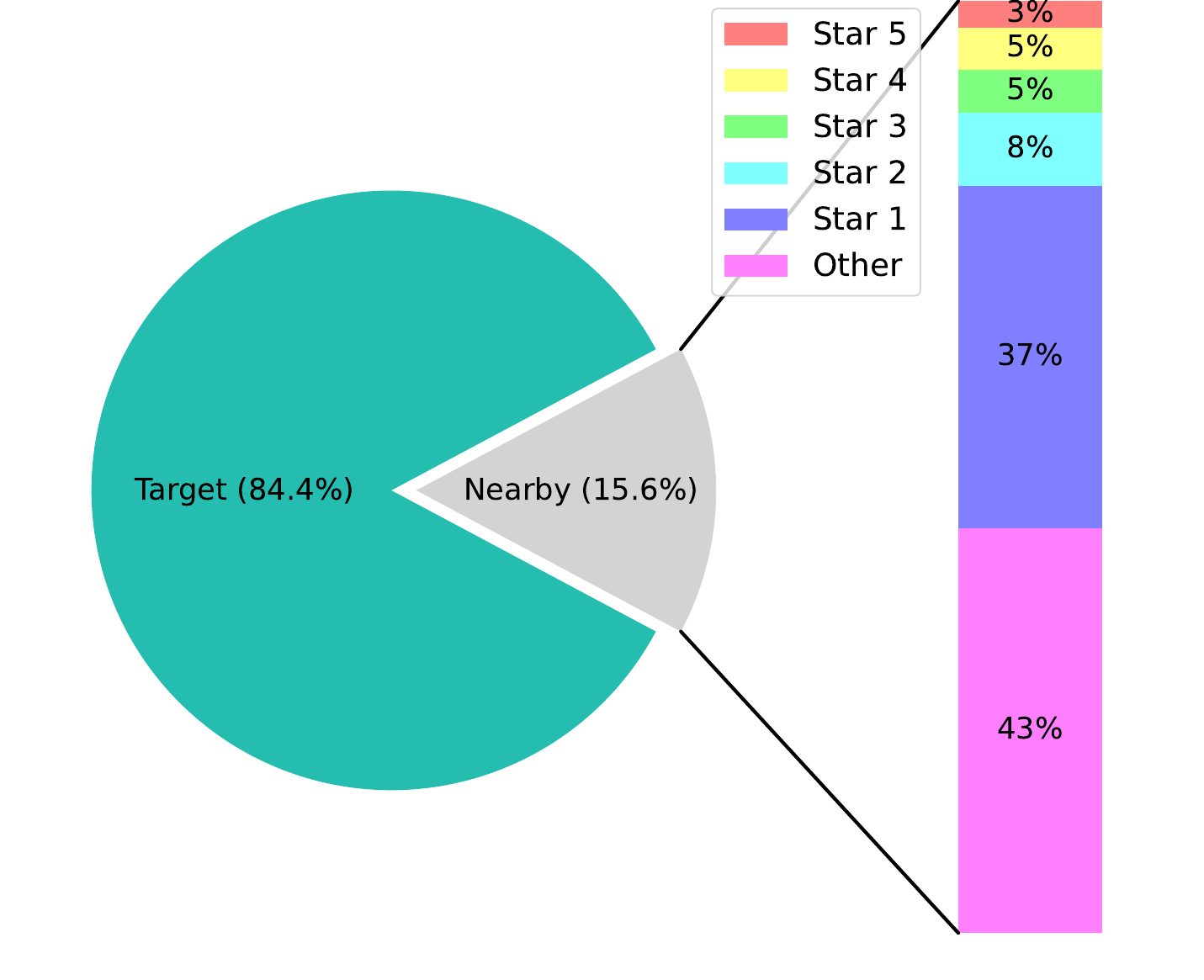}
    \caption{Nearby sources contaminating the  TOI-5005 photometry. Left: TPF-shaped heatmap with the pixel-by-pixel flux fraction from TOI-5005 in S65. The red grid is the SPOC aperture. The pixel scale is 21 arcsec $\rm pixel^{-1}$. The white disks represent all the \textit{Gaia} sources, and the five sources that most contribute to the aperture flux are highlighted in different colours. The disk areas scale with the emitted fluxes. Right: Flux contributions to the SPOC aperture from the target and most contaminant stars. This plot was created through \texttt{TESS-cont} (\url{https://github.com/castro-gzlz/TESS-cont}).}
    \label{fig:TESS-cont}
\end{figure*}

We developed a Python package to quantify the flux contribution from nearby sources in the TESS photometry: \texttt{TESS-cont}.\footnote{Available at \url{https://github.com/castro-gzlz/TESS-cont}.} In this section we describe the main aspects of its operation.

The \texttt{TESS-cont} algorithm (1) identifies the main contaminant sources, (2) quantifies their individual and total contributions to the selected aperture (i.e. SPOC or custom), and (3) determines whether any of these sources could be the origin of the observed transit or variability signals. The package first searches for all the nearby \textit{Gaia} DR2 \citep{2018A&A...616A...1G} or DR3 \citep{2023A&A...674A...1G} sources using the \texttt{get\_gaia\_data} function of \texttt{tpfplotter} \citep{2020A&A...635A.128A} and then constructs their Point Spread Functions (PSFs) to estimate the flux distribution across the TPF or FFI of the target star. The TESS PSFs are not Gaussian and vary across the focal plane mainly due to the optics. Therefore, instead of PSFs, TESS has Pixel Response Functions (PRFs) that better represent the flux distribution of the point sources. TESS PRFs were created by the SPOC pipeline based on micro-dithered data taken during the commissioning phase. These PRFs were built for a discrete number of CCD locations, while real sources can appear at any location. Therefore, to better represent the flux distribution of any source, \texttt{TESS-cont} uses the \texttt{TESS\_PRF} module \citep{2022ascl.soft07008B} to perform a bilinear interpolation between the four nearest SPOC PRFs. The obtained PRFs are then scaled to the stellar relative fluxes and placed in a TPF-shaped array. For each TPF pixel, \texttt{TESS-cont} computes the flux contribution from each source, and this information is used to compute the total flux contributions within the photometric aperture.

\subsubsection{Contamination analysis}

In Fig.~\ref{fig:TESS-cont}, we illustrate the \texttt{TESS-cont} output for TOI-5005 in S65. The left panel consists of a TPF-shaped heatmap with the pixel-by-pixel flux fraction from the target star, and the right panel shows the aperture flux contributions from the target star and main contaminant sources. The TOI-5005 flux falling inside the SPOC photometric aperture is 86.1$\%$ and 84.4$\%$ in S39 and S65, respectively, which is in agreement with the CROWDSAP metric estimated by SPOC. There is a relatively bright source ($G$ = 13.6 mag) surrounding the aperture (TIC 282476646, labelled as Star~1) that contributes 37$\%$ of the contaminant flux (i.e. 6$\%$ of the total flux). The following most contaminant sources are TIC 282485676, TIC 282476644, TIC 282478138, and TIC 282480455, which are labelled as Star~2, Star~3, Star~4, and Star~5 in Fig.~\ref{fig:TESS-cont}, and contribute 8$\%$, 5$\%$, 5$\%$, and 3$\%$ to the total contaminant flux, respectively. The remaining 43$\%$ of the contaminant flux is contributed by other nearby sources. We studied whether the dilution-corrected SPOC-derived $3690 \pm 4$ ppm transit and the 0.9 ppt sinusoidal modulation could have originated in any of the contaminant sources. To do so, we used the \texttt{TESS-cont} \texttt{DILUTION} feature to dilute the dilution-corrected transit depth assumed to come from TOI-5005, and de-blend it by considering that it comes from the contaminant sources  \citep[][]{2018AJ....156..277L,2020MNRAS.499.5416C,2021MNRAS.508..195D}. We obtain dilution-corrected transit depths of 5$\%$, 26$\%$, 43$\%$, 44$\%$, and 70$\%$ for Star~1, Star~2, Star~3, Star~4, and Star~5, respectively. For the 6.3-day sinusoidal-like signal we obtain dilution-corrected amplitudes between 1$\%$ (Star~1) and 17$\%$ (Star~5). Therefore, being those values lower than 100$\%$, this analysis cannot discard that the transit and sinusoidal signals could have originated in any of the nearby contaminant sources. We note that the ground-based photometry (Sect.~\ref{sec:ground_based}) and high-resolution spectroscopy (Sect.~\ref{sec:obs_harps}) allowed us to discard these sources as the origin of the planetary signal. However, having no counterpart in independent observations, the origin of the 6.3-day photometric modulation remains uncertain. Among the five most contaminant sources, Star~1 and Star~2 have available QLP photometry at MAST. Star~1 shows no signs of stellar variability, but Star~2 exhibits a strong variability with a periodicity of 0.7~days. In addition, it shows a hint of a $\simeq$6.7-day modulation in the second half of S39. Since this modulation is not detected in S12, S65 or the first half of S39, and its amplitude is comparable to that of the 0.7-day variability, it cannot correspond to the 6.3-day persistent signal. However, it makes us suspect that the pixels near TOI-5005 could be affected by uncorrected systematics (see Sect.~\ref{subsubsec:photometric_variability} and Appendix~\ref{sec:cbv_correction}).

\subsubsection{Dataset selection}

In Sects.~\ref{subsec:TESS_analysis}, \ref{subsec:joint_analysis}, and \ref{subsec:stellar_signals_analysis}, we analyse the photometric signals based on SAP fluxes corrected for crowding only. We chose SAP instead of PDCSAP because of two main reasons. First, we aim to investigate whether the detected sinusoidal modulation has a stellar origin or if it could have been artificially originated by residual uncorrected systematics. Second, there are no SPOC/PDCSAP fluxes for the S12 FFIs, which prevents us from building a homogeneous dataset based on PDCSAP photometry. In fact, in terms of homogeneity, QLP/SAP photometry was also extracted differently than SPOC/SAP. While SPOC/SAP is simply the sum of the calibrated TPF fluxes within a pixel-based grid aperture, QLP/SAP uses several circular apertures and it has a higher level of processing \citep[e.g. see][]{2019ApJ...871L..24V,2020RNAAS...4..251F}. Therefore, we decided to analyse a homogeneous dataset composed of SPOC/SAP photometry (S39 and S65) and SAP photometry that we extracted similarly to SPOC (S12). The S12 photometric aperture was automatically selected by \texttt{TESS-cont} to minimize contamination; that is, we only considered pixels with a target flux contribution larger than 60$\%$, similarly to the S39 and S65 SPOC apertures. In Fig.~E.2 (available on Zenodo), we show our selected S12 aperture over an 11 $\times$ 11 pixel FFI cutout and a similarly shaped heatmap containing the pixel-by-pixel target flux fractions. In Table.~\ref{tab:TESS_SAP}, we present the complete TESS SAP dataset together with their associated quality flags (QFs) determined by SPOC. We discarded those observations with SPOC QFs different from zero, which were mainly flagged because of stray light coming from the Earth or Moon. For completeness, we repeated the analysis in Sect.~\ref{subsec:joint_analysis} by considering the highest level dataset available (QLP/SAP for S12, and SPOC/PDCSAP for S39 and S65), and found a consistent solution for the planetary and orbital parameters within 1$\sigma$. We also present this dataset in Table.~\ref{tab:TESS_QLP_PDCSAP}.

\subsection{Ground-based photometry}
\label{sec:ground_based}

We observed two transits of TOI-5005.01 from ground-based facilities to attempt to determine the true source of the TESS detection and to obtain precise ephemeris to facilitate the scheduling of follow-up observations. 

\subsubsection{PEST}
\label{sec:obs_pest}

We observed a full transit window of TOI-5005.01 in the Sloan $r'$ filter on 5 May 2022 from the Perth Exoplanet Survey Telescope (PEST) located near Perth, Australia. These observations are part of the TESS Follow-up Observing Program \citep[TFOP;][]{collins:2019}.\footnote{\url{https://tess.mit.edu/followup}} We scheduled the transit observations through the {\tt TESS Transit Finder}, which is a customized version of the {\tt Tapir} package \citep{Jensen:2013}. The 0.3-m telescope is equipped with a $5544\times3694$ QHY183M camera.  Images are binned $2\times2$ in software giving an image scale of 0.7$\arcsec$ pixel$^{-1}$, resulting in a $32\arcmin\times21\arcmin$ field of view. A custom pipeline based on {\tt C-Munipack}\footnote{Available at \url{http://c-munipack.sourceforge.net}} was used to calibrate the images and extract the differential photometry. We used circular photometric apertures with a radius of 6.4$\arcsec$. The target star aperture excluded most of the flux from the nearest known neighbours in the \textit{Gaia} DR3 catalogue, TIC 282476644 (Star~3) and TIC 282478138 (Star~4), which are 7.1$\arcsec$ and 15.5$\arcsec$ from TOI-5005, respectively. The light curve data are available in Table~\ref{tab:pest_data} and {\tt EXOFOP-TESS}.\footnote{\url{https://exofop.ipac.caltech.edu/tess/}}

\subsubsection{TRAPPIST-South}
\label{sec:obs_TRAPPIST_South}
We observed one transit of TOI-5005.01 with the TRAPPIST-South telescope \citep{TS_Gillon,TS_Jehin} located at ESO's La Silla Observatory in Chile on 22 March 2024. As the planet was already confirmed on target by the PEST transit and the HARPS radial velocity measurements (see Sect.~\ref{sec:obs_harps}), we obtained this transit observation to keep the ephemeris up to date. TRAPPIST-South is a 0.6-m telescope equipped with an FLI ProLine camera and a back-illuminated CCD which has a pixel size of 0.64$\arcsec$ and provides a field of view of $22\arcmin \times 22\arcmin$. It is a robotic Ritchey-Chrétien telescope with F/8 and it is equipped with a German equatorial mount. The transit was obtained in the Sloan $z'$ filter with an exposure time of 20 s. We reduced the data and performed aperture and differential photometry using a custom pipeline built with the \texttt{prose} package\footnote{Available at \url{https://github.com/lgrcia/prose/}} \citep{prosesoft,2022_prose}. To minimize red and white noise in the transit light curve, we selected four comparison stars and an uncontaminated circular aperture of 4.1$\arcsec$ for a full width at half-maximum (FWHM) of 2.6$\arcsec$. A meridian flip occurred at 2460392.8576 BJD which caused an offset in the normalized flux. The light curve data are available in Table~\ref{tab:trappist_south_data}.

\subsection{SOAR high-resolution imaging}

\begin{figure}
    \centering
    \includegraphics[width=0.48\textwidth]{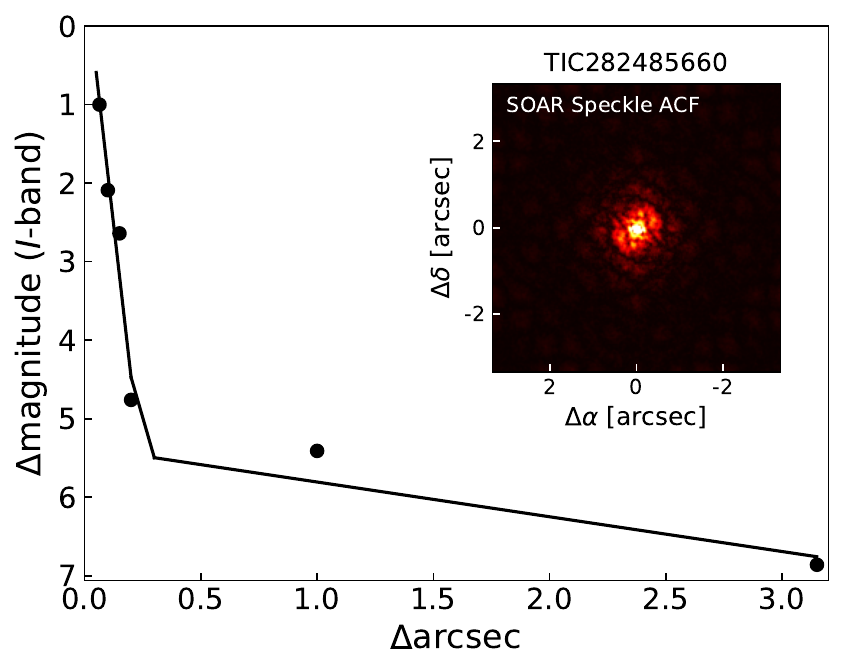}
    \caption{Detection sensitivity (5$\sigma$)  to nearby companions for the SOAR speckle observation of TOI-5005 as a function of separation from the target star and magnitude difference with respect to the target. The inset shows the speckle auto-correlation function.}
    \label{fig:SOAR_speckle}
\end{figure}

High-angular resolution imaging is needed to search for nearby sources that can contaminate the TESS photometry, resulting in an underestimated planetary radius, or be the source of astrophysical false positives, such as background eclipsing binaries. We searched for stellar companions to TOI-5005 with speckle imaging on the 4.1-m Southern Astrophysical Research (SOAR) telescope \citep{2018PASP..130c5002T} on 15 April 2022, observing in Cousins $I$-band, a similar visible bandpass as TESS. This observation was sensitive with 5$\sigma$ detection to a 5.5-magnitude fainter star at an angular distance of 1$\arcsec$ from the target. More details of the observations within the SOAR TESS survey are available in \citet{2020AJ....159...19Z}. The 5$\sigma$ detection sensitivity and speckle auto-correlation functions from the observations are shown in Fig.~\ref{fig:SOAR_speckle}. No nearby stars were detected within 3$\arcsec$ of TOI-5005 in the SOAR observations.

\subsection{HARPS high-resolution spectroscopy}

\label{sec:obs_harps}

We observed TOI-5005 with the High Accuracy Radial velocity Planet Searcher spectrograph \citep[HARPS;][]{2003Msngr.114...20M}, which is mounted on the ESOS's 3.6m telescope located at La Silla Observatory in Chile. HARPS is a fibre-fed cross-dispersed echelle spectrograph stabilized in a vacuum vessel. It covers a wavelength range between 378 and 691 nm and has a spectral resolution power of 115\,000.

We acquired a total of 38 HARPS spectra between 15 March 2023 and 18 August 2023 under the HARPS-NOMADS programmes 108.21YY.001 and 108.21YY.002 (PI: Armstrong). The nightly seeing conditions ranged from 0.76 to 2.67 arcsec, with a median value of 1.24 arcsec. We performed the acquisitions with a typical exposure time of 1800 s, which resulted in S/N per pixel between 14.3 and 36.7, with a median value of 24.1. We used the High Accuracy Mode (HAM) with a 1 arcsec science fibre centred on the star and a second fibre in the sky to monitor the sky background.

We used the HARPS Data Reduction Software to reduce the raw spectra and extract the RVs through the cross-correlation technique based on a G2 template \citep{1996A&AS..119..373B,2002A&A...388..632P}. We also used the HARPS DRS to extract several activity indicators such as the full width at half maximum (FWHM) of the
cross-correlation function (CCF), the contrast of the CCF, the $\rm H_{\alpha}$ and sodium doublet (NaD) line depths, and the S-index. We present the complete dataset in Table \ref{tab:harps_rvs}. The HARPS RV uncertainties range from 2.4 to 7.6 $\rm m\,s^{-1}$, with a median value of 3.8 $\rm m\,s^{-1}$ and a root-mean-square (rms) of 4.3 $\rm m\,s^{-1}$. These values contrast with the dispersion of the RV measurements, which have a standard deviation of 13.1 $\rm m\,s^{-1}$. This factor three between the RV dispersion and uncertainties indicates that our data are not white-noise-dominated but could contain planetary or stellar signals. 

In Fig.~\ref{fig:gls_to_HARPS}, we show the HARPS RVs and activity indicators together with their \texttt{GLS} periodograms. Several time series follow upward or downward trends (e.g. RV, FWHM, and Contrast), which could be due to physical (e.g. outer long-period massive companion, stellar magnetic cycle) or instrumental (e.g. night-to-night drifts) effects. The RV periodogram shows a maximum power period of 8.5 days. This periodicity has a FAP of 6$\%$, so we do not consider it significant.\footnote{The most commonly adopted criterion to consider a \texttt{GLS} peak significant is to have a FAP $<$ 0.1 $\%$.}  Interestingly, the second-highest peak coincides with the 6.3-day periodicity of TOI-5005.01, although it is not significant either.  We recomputed the periodogram after subtracting a simple linear trend that we previously fit to the data. The periodogram of the de-trended RVs shows a significant maximum power period of 6.3 days (FAP = 0.05 $\%$), which matches the TOI-5005.01 periodicity and ephemeris. The periodograms of the FWHM, S-index, Contrast, and Ca activity indicators show maximum power periods of 21.3, 20.8, 21.0, and 20.5 days, with FAPs of 10, 1.1, 0.040, and 10 $\%$, respectively. This recurring $\simeq$21-day periodicity indicates the existence of an activity-related signal that most likely corresponds to the rotation period of the star (see Sect.~\ref{subsec:prot}).  The periodogram of the $\rm H_{\alpha}$ indicator shows a maximum power period of 6.2 days with a FAP of 91 $\%$. This periodicity coincides with that of the TOI-5005.01 orbit. As mentioned in Sect.~\ref{sec:obs_tess}, given the closeness of this Neptune-sized planet to its host star, this periodicity could be interpreted as a planet-induced enhancement of chromospheric activity due to magnetic star-planet interactions. However, in this case, the signal significance is extremely small, and the phase-folded data do not show a clear modulation such as that seen within the TESS photometry. Therefore, the weakness of the signal, together with the absence of additional spectroscopic activity signals matching the TOI-5005.01 periodicity, suggests that the possible magnetic star-planet interactions detected within the TESS data cannot be measured in our HARPS data. In Fig~\ref{fig:gls_to_HARPS}, we show the window function of the observations and its periodogram, which shows a maximum power period of 129.8 days. This periodicity does not coincide with any of the previously mentioned planet and activity signals.

\section{Stellar characterization}
\label{sec:stellar_charact}


\begin{table}
\renewcommand{\arraystretch}{1.3}
\setlength{\tabcolsep}{3pt}
\caption{General properties of TOI-5005.}
\label{tab:stellar_general_prop}
\begin{tabular}{llc}
\hline \hline
Parameter                               & Value                     & Reference \\ \hline
\multicolumn{3}{l}{Identifiers}                                                 \\ \hline
TOI                                     & 5005                      & (1)       \\
TIC                                     & 282485660                 & (2)       \\
2MASS                                   & J15522597-4808419         & (3)       \\
Gaia DR3                                & 5984530842395365248       & (4)       \\ \hline
\multicolumn{3}{l}{Astrometric properties}                                      \\ \hline
RA, Dec                                 & 15:52:25.97, -48:08:42.37 & (4)       \\
$\rm \mu_{\alpha}$ ($\rm mas\,yr^{-1}$) & -7.830 $\pm$ 0.018        & (4)       \\
$\rm \mu_{\delta}$ ($\rm mas\,yr^{-1}$) & -24.642 $\pm$ 0.015       & (4)       \\
Parallax (mas)                          & 4.760 $\pm$ 0.017         & (4)       \\
Distance (pc)                           & 210.08 $\pm$ 0.75         & (4)       \\
RV ($\rm km\,s^{-1}$)                   & 16.58 $\pm$ 0.77          & (4)       \\ \hline
Photometric properties                  &                           &           \\ \hline
TESS (mag)                              & 11.1300 $\pm$ 0.0061      & (2)       \\
G (mag)                                 & 11.63820 $\pm$ 0.00045    & (2)       \\
J (mag)                                 & 10.393 $\pm$ 0.023        & (3)       \\
H (mag)                                 & 10.058 $\pm$ 0.022        & (3)       \\
$\rm K_{s}$ (mag)                       & 10.004 $\pm$ 0.021        & (3)       \\
B (mag)                                 & 12.572 $\pm$ 0.013        & (5)       \\
V (mag)                                 & 11.822 $\pm$ 0.050        & (5)       \\
g' (mag)                                & 12.150 $\pm$ 0.026        & (5)       \\
r' (mag)                                & 11.576 $\pm$ 0.079        & (5)       \\
i' (mag)                                & 11.362 $\pm$ 0.142        & (5)       \\ \hline
\end{tabular}
\tablefoot{(1) \citealp{2021ApJS..254...39G}; (2) \citealp{2019AJ....158..138S}; (3) \citealp{skrutskie2006}; (4) \citealp{2023A&A...674A...1G}; (5) \citealp{2018AAS...23222306H}.}
\end{table}

\subsection{General description}
\label{subsec:general_description}

TOI-5005 is a moderately bright \citep[V = 11.822 $\pm$ 0.050 mag;][]{2018AAS...23222306H} early-type G-dwarf star visible from the southern sky. According to the measured \textit{Gaia} DR3 parallax ($\pi$ = 4.760 $\pm$ 0.017 mas), TOI-5005 is located 210.08 $\pm$ 0.75 pc away from the Sun. In Table \ref{tab:stellar_general_prop}, we show the main astrometric and photometric properties compiled from the literature. The TESS Input Catalog \citep[TIC v8.2;][]{2019AJ....158..138S} used publicly available photometry to estimate an effective temperature of $T_{\rm eff}$ = 5840 $\pm$ 125 K, surface gravity of log $g$ = 4.484 $\pm$ 0.075 dex, stellar radius of $R_{\star}$ = 0.972 $\pm$ 0.045 $\rm R_{\odot}$, and stellar mass of $\rm M_{\star}$ = 1.05 $\pm$ 0.13 $\rm M_{\odot}$. In the next sections, we describe our stellar characterization based on a high-resolution, high S/N spectrum obtained from the combination of the 38 individual HARPS spectra.

\subsection{Stellar physical parameters}
\label{subsec:atmospheric_parameters}

We derived the stellar atmospheric parameters ($T_{\mathrm{eff}}$, $\log g$, microturbulence, and [Fe/H]) using ARES+MOOG following the methodology described in \citet[][]{Sousa-21, Sousa-14, Santos-13}. We used the ARES code\footnote{The last version of ARES code (ARES v2) can be downloaded at \url{https://github.com/sousasag/ARES}} \citep{Sousa-07, Sousa-15} to consistently measure the equivalent widths (EW) of selected iron lines based on the line list presented in \citet[][]{Sousa-08}. This was done on a 
combined HARPS spectrum of TOI-5005. We then used a minimization process to find the ionization and excitation equilibrium and converge to the best set of spectroscopic parameters. This process makes use of a grid of Kurucz model atmospheres \citep{Kurucz-93} and the radiative transfer code MOOG \citep{Sneden-73}. A trigonometric surface gravity was also derived using \textit{Gaia} DR3 data following the same methodology as described in \citet[][]{Sousa-21}. In this process, we derived the stellar mass using the calibration presented in \citet[][]{Torres-2010}: $M_{\star}$ = 0.97 $\pm$ 0.02~$\rm M_{\odot}$. Following a similar calibration, presented in the same work, we obtained the stellar radius: $R_{\star}$ = 0.93 $\pm$ 0.03~$\rm R_{\odot}$. These values are consistent with the photometry-based estimations of the TIC catalogue within 1$\sigma$. We list the stellar atmospheric parameters, mass, and radius of TOI-5005 in Table \ref{tab:stellar_parameters}.

\subsection{Chemical abundances}
\label{subsec:chemical_abundances}

We derived stellar abundances of different elements using the classical curve-of-growth analysis method assuming local thermodynamic equilibrium. We used the same radiative transfer code (MOOG) and the same model atmospheres that we previously used for the stellar parameters determinations. We followed the methods described in \citet[]{Adibekyan-12, Adibekyan-15, Delgado-17} to derive chemical abundances of refractory elements, and followed the method of \cite{Delgado-21, Bertrandelis-15} to derive abundances of volatile elements such as carbon and oxygen. The oxygen abundances are based on two weak atomic lines which present large uncertainties, especially when the S/N of the spectra are not very high. One of the lines was contaminated by an Earth airglow and hence we only used 14 spectra to determine its abundance. We obtained all the [X/H] ratios by doing a differential analysis with respect to a high S/N solar (Vesta) spectrum from HARPS. 

In addition, we obtained the abundance of lithium by performing spectral synthesis with MOOG, following the same procedure as in \citet{Delgado-14}. We first fixed the macroturbulence velocity to 3.2 km\,s$^{-1}$ \citep[based on the empirical calibration by][]{doyle-14} to estimate the $v \, \textrm{sin} \, i_{\star}$ from two Fe lines in the region, leading to a value of 1.35 $\pm$ 0.10 km\,s$^{-1}$. We obtained an abundance A(Li)\,=\,1.65 $\pm$ 0.10\,dex, which is a relatively high value for a star of this temperature, suggesting that TOI-5005 is younger than the Sun. We present the abundances of all the elements in Table \ref{tab:stellar_parameters}.

\subsection{Rotation period}
\label{subsec:prot}
The HARPS activity indicators FWHM, S-index, Contrast, and Ca, show periodic signals at $\simeq$21 days, which most likely reflect the rotation period ($P_{\rm rot}$) of TOI-5005 (Sect.~\ref{sec:obs_harps}). This signal is significantly detected in the S-index and Contrast indicators (FAPs of 1.1$\%$ and 0.040$\%$, respectively), and is detected with a weaker significance in the FWHM and Ca indicators (FAPs of 10$\%$). Interestingly, it appears to not have a significant effect on the RVs (see Sect.~\ref{subsec:harps_rv_analysis} for a detailed analysis), and it is not detected in the TESS photometry either (probably because of insufficient photometric precision or an incomplete phase coverage). In this section we study additional observables to compare with the periodicity detected in the HARPS activity indicators. 

The $v \, \textrm{sin} \, i_{\star}$ obtained in Sect.~\ref{subsec:chemical_abundances} (1.35 $\pm$ 0.10 km\,s$^{-1}$) together with the stellar radius derived in Sect.~\ref{subsec:atmospheric_parameters} corresponds to a $P_{\rm rot}  \, \textrm{sin} \, i_{\star}$ of 35 $\pm$ 3 days. Hence, a certain inclination angle relative to our line of sight seems to be required to match the observed $\simeq$21 day activity signal. We note, however, that measuring accurate rotation velocities for slow rotators ($v \, \textrm{sin} \, i_{\star}$ $<$ 3$\rm \, km \, s^{-1}$) is a difficult task and hence our estimate should be taken with care. In particular, the reported $v \, \textrm{sin} \, i_{\star}$ uncertainty might be underestimated. We also used the S-index provided by the HARPS pipeline, which is calibrated to the Mount Wilson scale \citep{1978PASP...90..267V}, to obtain the $\textrm{log} \, R'_{HK}$ of each spectrum. To do so, we considered the B-V colour from the APASS catalogue \citep[][B-V = 0.75 $\pm$ 0.05, see Table \ref{tab:stellar_general_prop}]{2018AAS...23222306H} and used the \texttt{pyrhk} code\footnote{Available at \url{https://github.com/gomesdasilva/pyrhk}.} to obtain an average $\textrm{log} \, R'_{HK}$ of -4.819 $\pm$ 0.052 dex with bolometric corrections from \citet{1982A&A...107...31M}. We used the \citet{2016A&A...595A..12S} activity-rotation empirical calibrations and obtained a rotation period of $P_{\rm rot}$ = 24.9 $\pm$ 5.0 days, which is consistent with the $\simeq$21-day activity signal within 1$\sigma$. An accurate determination of the stellar rotation period requires a more thorough analysis of the activity time series. In Sect~\ref{subsubsec:HARPS_activity_indicators}, we use an activity model to jointly analyse different HARPS activity indicators. From this analysis, we obtain a precise rotation period of $P_{\rm rot}$ = $21.01^{+0.46}_{-0.60}$ days, which we adopt as our final estimate.

\subsection{Age}
\label{subsec:age}
Chemical abundances can be used for estimating stellar ages. Some chemical elements are related to different astrophysical origin channels and can be used as tracers of the time a star was born. Those elements are called chemical clocks (CCs) and their use for dating stars has been proposed in many works \citep[e.g. see][and references therein]{2024MNRAS.528.3464R}. Almost all those works use linear regressions to describe the relation between the CCs and the stellar age.  \citet{2019A&A...624A..78D} presented a complete set of multidimensional linear regressions using all the chemical clocks that could be made with that dataset. One of the main problems of this technique is that by using different CCs we can obtain slightly different stellar age estimations. To solve this, one option is to combine these estimations to obtain a more robust age estimator, but a simple mean, for example, cannot be used here since all the chemical clocks are very correlated. In a recent work, \citet{2022A&A...660A..15M} constructed a Hierarchical Bayesian model (HBM) for estimating stellar ages combining the results from different chemical clocks and their multidimensional linear regressions also properly propagating uncertainties all along the procedure. We used this HBM for estimating the age of TOI-5005. Using [Y/Mg], [Sr/Mg], [Y/Si], [Y/Ti], and [Y/Zn], the stellar $T_{\rm eff}$, $\textrm{log}\,g$, [Fe/H], and the corresponding uncertainties, we obtained the posterior probability distribution for its age. This distribution is compatible with zero and indicates a 3$\sigma$ upper limit of 3.6 Gyr.

The stellar rotation period has been also used to date stars through gyrochronology \citep[e.g.][]{2003ApJ...586..464B,2007ApJ...669.1167B,2008ApJ...687.1264M,2015MNRAS.450.1787A}. We used the empirical relations by \citet{2019AJ....158..173A} implemented in \texttt{stardate}\footnote{Available at \url{https://github.com/RuthAngus/stardate}.} to estimate the age of TOI-5005. Based on the \textit{Gaia} parallax (Sect.~\ref{subsec:general_description}), stellar atmospheric parameters (Sect.~\ref{subsec:atmospheric_parameters}), and measured rotation period ($P_{\rm rot}$ = $21.01^{+0.46}_{-0.60}$ days; see Sects.~\ref{subsec:prot} and \ref{subsubsec:HARPS_activity_indicators}), we obtain a stellar age of $2.20^{+0.61}_{-0.28}$ Gyrs, which is compatible with the CC estimate. We include both the CC 3$\sigma$ upper limit ($\rm Age_{CC}$) and the gyrochronology age ($\rm Age_{gyro}$) in Table~\ref{tab:stellar_parameters}. We note that given its better precision, for subsequent analysis we adopt the gyrochronology age.

\subsection{Galactic membership}
\label{subsec:galactic_membership}

We computed the Galactic space velocity (\textit{UVW}) of TOI-5005 based on its radial velocity, parallax, RA/Dec coordinates, and proper motion as measured by \textit{Gaia} (Table~\ref{tab:stellar_general_prop}). We adopted the solar peculiar motion from \citealp{2003A&A...409..523R} ($U_{\odot} = -10.3 \rm \, km\,s^{-1}$, $V_{\odot} = 6.3 \, \rm km\,s^{-1}$, and $W_{\odot} = 5.9 \, \rm km\,s^{-1}$), and derived $U = -15.40 \pm 0.67 \, \rm km\,s^{-1}$, $V = -21.10 \pm 0.37 \, \rm km\,s^{-1}$, and $W = -6.79 \pm 0.08 \, \rm km\,s^{-1}$ with respect to the local standard of rest (LSR). Similarly to \citet{2003A&A...410..527B}, we assumed that the Galactic space velocities of the stellar populations follow a multi-dimensional Gaussian distribution

\begin{equation}
    f(U,V,W) = k \cdot \textnormal{exp}  \left(     
    -\frac{U^{2}}{2\sigma_{U}^{2}}
    -\frac{(V-V_{\rm asym})^{2}}{2\sigma_{V}^{2}}
    -\frac{W^{2}}{2\sigma_{W}^{2}}
    \right)
,\end{equation}

\noindent where $k = \left( 2\pi \right)^{-3/2} \sigma_{U}^{-1} \sigma_{V}^{-1} \sigma_{W}^{-1}$, being $\sigma_{U}$, $\sigma_{V}$, and $\sigma_{W}$ the characteristic velocity disperions, and $V_{\rm asym}$ the asymmetric drift. We estimated the probabilities that TOI-5005 belongs to the thin disk (TD), the thick disk (D), and the halo (H) by multiplying $f(U,V,W)$ by the probability that a star in our neighbourhood belongs to those three Galactic populations. In order to make our estimation self-consistent, we adopted the relative likelihoods of belonging to each population as well as the characteristics for stellar components
in the solar neighbourhood from \citet{2003A&A...409..523R}. As a result, we obtain TD = 98.865 $\%$, D = 1.133 $\%$, and H = 0.002 $\%$. Therefore, it is very likely that TOI-5005 is a
member of the Galactic thin disk population. This result is consistent with the stellar age estimation (Sect.~\ref{subsec:age}), since most thin disk stars have ages less than 8 Gyr \citep[e.g.][]{1998A&A...338..161F,2001ASPC..245..207B,2003A&A...410..527B}.


\begin{table}[]
\renewcommand{\arraystretch}{1.3}
\setlength{\tabcolsep}{15.7pt}
\caption{Stellar properties of TOI-5005 derived in this work.}
\label{tab:stellar_parameters}
\begin{tabular}{lll}
\hline \hline
Parameter               & Value  & \multicolumn{1}{c}{Section}      \\ \hline
\multicolumn{3}{l}{Atmospheric parameters and spectral type}                                      \\ \hline
$T_{\rm eff}$ (K)                & $5749 \pm 61$      & Sect. \ref{subsec:atmospheric_parameters} \\
log $g$ (dex)                    & $4.43 \pm 0.10$    & Sect. \ref{subsec:atmospheric_parameters} \\
log $g_{\rm GAIA}$ (dex)             & $4.56 \pm 0.03$    & Sect. \ref{subsec:atmospheric_parameters} \\
{[}Fe/H{]} (dex)                 & $0.15 \pm 0.04$    & Sect. \ref{subsec:atmospheric_parameters} \\
$\xi_{\mathrm{t}}$ ($\rm km\,s^{-1}$)        & $0.97 \pm 0.02$    & Sect. \ref{subsec:atmospheric_parameters} \\
SpT                              & G2 V               & Sect. \ref{subsec:atmospheric_parameters} \\ \hline
\multicolumn{3}{l}{Physical parameters}                                                           \\ \hline
$R_{\star}$ $(\rm R_{\odot})$        & $0.93 \pm 0.03$    & Sect. \ref{subsec:atmospheric_parameters} \\
$M_{\star}$ $(\rm M_{\odot})$        & $0.97 \pm 0.02$    & Sect. \ref{subsec:atmospheric_parameters} \\
$v \, \textrm{sin} \, i_{\star}$ ($\rm km\,s^{-1}$) & $1.35 \pm 0.10$                 & Sect. \ref{subsec:chemical_abundances}    \\
$\textrm{log} \, R'_{HK}$ (dex) & -4.819 $\pm$ 0.052 & Sect.~\ref{subsec:prot}  \\
$P_{\rm rot}$ (days)             & $21.01^{+0.46}_{-0.60}$                &  Sect.~\ref{subsec:prot}                                       \\
$\rm Age_{CC}$ (Gyr)            & < 3.6                & Sect. \ref{subsec:age}                    \\
$\rm Age_{gyro}$ (Gyr)          & $2.20^{+0.61}_{-0.28}$                & Sect. \ref{subsec:age}                                         \\ \hline
\multicolumn{3}{l}{Chemical abundances}                                                           \\ \hline
{[}Mg/H{]} (dex)                 & $0.100 \pm 0.030$  & Sect. \ref{subsec:chemical_abundances}    \\
{[}Si/H{]} (dex)                 & $0.100 \pm 0.040$  & Sect. \ref{subsec:chemical_abundances}    \\
{[}Ni/H{]} (dex)                 & $0.100 \pm 0.020$  & Sect. \ref{subsec:chemical_abundances}    \\
{[}Ti/H{]} (dex)                 & $0.150 \pm 0.030$  & Sect. \ref{subsec:chemical_abundances}    \\
{[}O/H{]} (dex)                  & $0.141 \pm 0.187$  & Sect. \ref{subsec:chemical_abundances}    \\
{[}C/H{]} (dex)                  & $-0.035 \pm 0.034$ & Sect. \ref{subsec:chemical_abundances}    \\
{[}Cu/H{]} (dex)                 & $0.015 \pm 0.051$  & Sect. \ref{subsec:chemical_abundances}    \\
{[}Zn/H{]} (dex)                 & $0.011 \pm 0.025$  & Sect. \ref{subsec:chemical_abundances}    \\
{[}Sr/H{]} (dex)                 & $0.280 \pm 0.077$  & Sect. \ref{subsec:chemical_abundances}    \\
{[}Y/H{]} (dex)                  & $0.237 \pm 0.053$  & Sect. \ref{subsec:chemical_abundances}    \\
{[}Zr/H{]} (dex)                 & $0.194 \pm 0.061$  & Sect. \ref{subsec:chemical_abundances}    \\
{[}Ba/H{]} (dex)                 & $0.267 \pm 0.027$  & Sect. \ref{subsec:chemical_abundances}    \\
{[}Ce/H{]} (dex)                 & $0.246 \pm 0.035$  & Sect. \ref{subsec:chemical_abundances}    \\
{[}Nd/H{]} (dex)                 & $0.192 \pm 0.036$  & Sect. \ref{subsec:chemical_abundances}    \\ \hline
\multicolumn{3}{l}{Galactic space velocities and membership}                                      \\ \hline
U ($\rm km \, s^{-1}$)           & $-15.40 \pm 0.67$  & Sect. \ref{subsec:galactic_membership}    \\
V ($\rm km \, s^{-1}$)           & $-21.10 \pm 0.37$  & Sect. \ref{subsec:galactic_membership}    \\
W ($\rm km \, s^{-1}$)           & $-6.79 \pm 0.08$   & Sect. \ref{subsec:galactic_membership}    \\
Gal. population                  & Thin disk          & Sect. \ref{subsec:galactic_membership}    \\ \hline
\end{tabular}
\end{table}

\section{Analysis and results}
\label{sec:analysis_results}

\subsection{Model inference and parameter determination}
\label{sec:model_sel_param_det}

We analysed the TESS, HARPS, PEST, and TRAPPIST-South datasets through Bayesian inference. Our main goal is to find the model that best represents our data, and use it to derive accurate physical parameters. In Appendix~\ref{bayesian}, we describe the mathematical framework behind our Bayesian analysis, where we include details on the considered likelihood functions, prior distributions, and their implementation. For each tested model, we used a Markov chain Monte Carlo (MCMC) affine-invariant ensemble sampler \citep{2010CAMCS...5...65G} as implemented in \texttt{emcee}\footnote{Available at \url{https://github.com/dfm/emcee}.} \citep{2013PASP..125..306F} to sample the parameter space and generate marginal posterior distributions associated with each parameter. We used eight times as many walkers as the number of parameters and performed two consecutive runs. The first run consisted of 200\,000 iterations, and the second run consisted of 100\,000 iterations. Between both runs, we reset the sampler and considered the initial values from the last iteration of the first run. Following \citet{2013PASP..125..306F}, we examined the convergence by estimating the autocorrelation times and checking that they are all at least 50 times smaller than the chain length.

To infer the model that best represents our data we estimated the Bayesian evidence through \texttt{bayev}\footnote{Available at \url{https://github.com/exord/bayev}.} \citep{2016A&A...585A.134D}, which is based on the formalism described in \citet{PERRAKIS201454}. The package uses the marginalized posterior distributions obtained by \texttt{emcee}, the likelihood functions (Eqs. \ref{eq:log-like_uncorrelated} and \ref{eq:log-like_correlated}), and the prior distributions (Eqs. \ref{eq:unifor_priors} and \ref{eq:gaussian_priors}) to obtain the model evidence $\mathcal{Z}$.\footnote{When describing the mathematical framework in Appendix.~\ref{bayesian}, we refer to the model evidence as $P(D)$. However, for simplicity, in the main text, we adopt the widely used notation $\mathcal{Z}$.} For the model selection, we considered a criterion based on Occam's razor principle. That is, we always selected the simplest model, unless there is a more complex model with significantly more evidence. Following the Jeffreys' scale \citep{jeffreys1961theory}, we consider that a complex model has strong evidence against a simple model if the logarithmic difference $\mathcal{B}$ = ln$(\mathcal{Z}_{\rm complex})$ - ln$(\mathcal{Z}_{\rm simple})$ is larger than five.

\subsection{HARPS radial velocity analysis}
\label{subsec:harps_rv_analysis}

We analysed the HARPS dataset described in Sect.~\ref{sec:obs_harps} through the procedure described in Sect.~\ref{sec:model_sel_param_det}. The RV periodogram showed a significant 6.3-day signal that matches the periodicity and ephemeris of TOI-5005~b, and the periodograms of four activity indicators showed maximum power periods at $\simeq$21 days, which indicated the existence of an activity-related signal that corresponds to the rotation period of the star. This signal, however, did not appear within the RV periodograms, which suggests that the stellar activity did not significantly influence the measured RVs. Under this situation, the main objectives of this section are the following. First, we aim to study whether our RV model should incorporate a stellar noise component. Secondly, we aim to figure out whether there are additional planetary signals in the system. Finally, we aim to select the simplest model that best represents our data. 

We built 21 models with different components aimed at describing the phenomena that might be affecting our dataset. Those components are an instrumental model, a linear drift, a Keplerian, a stochastic process composed of a mean function and an autocorrelation function for describing unknown correlated noise (i.e. a Gaussian process; see Appendix \ref{bayesian}), and a white noise model (i.e. a jitter term, see Appendix \ref{bayesian}).

The instrumental model consists of an offset that describes the systemic RV of the star as measured by HARPS ($v_{\rm HARPS}$). The linear drift adds a slope ($\gamma_{\rm HARPS}$) to the instrumental model to account for possible long-term trends that can be approximated by a straight line. The Keplerian model describes the planetary orbit. We implemented it through \texttt{radvel}\footnote{Available at \url{https://radvel.readthedocs.io}.} \citep{2018PASP..130d4504F} by considering the parametrization $\left\lbrace P_{\rm orb}, T_{0}, K, \sqrt{e} \cos(\omega), \sqrt{e} \sin(\omega)\right\rbrace$, where $P_{\rm orb}$ is the planetary orbital period, $T_{0}$ is the time of inferior conjuntion, $K$ is the semi-amplitude, $e$ is the orbital eccentricity, and $\omega$ is the argument of the periastron. To model the unknown correlated noise we defined a Gaussian process (GP) with a quasi-periodic covariance function that can be written as \citep{2015ITPAM..38..252A,2016A&A...588A..31F}

\begin{equation}
    \label{eq:QP}
    K_{QP} (\tau) = \eta_{1}^{2} \rm{exp} \left[ - \frac{\tau^{2}}{2\eta_{2}^{2}} - \frac{2sin^{2} \left(  \frac{\pi \tau}{\eta_{3}} \right)}{\eta_{4}^{2}}  \right],
\end{equation}where $\tau = x_{i} - x_{j}$ is the separation between two time stamps, and $\eta_{1}$, $\eta_{2}$, $\eta_{3}$, and $\eta_{4}$ are hyperparameters defining the GP. Our GP covariance choice is motivated by its physical interpretation. The quasi-periodic covariance (hereinafter QP) has been designed so that $\eta_{3}$ can be interpreted as the stellar rotation period, $\eta_{2}$ is a measure of the timescale of appearance and disappearance of the active regions, and $\eta_{4}$ indicates the complexity of the harmonic content of the activity signal (see \citealt{2014MNRAS.443.2517H} and \citealt{2018MNRAS.474.2094A} for further details on the physical interpretation of the QP hyperparameters). Finally, to account for the existence of uncorrelated noise not taken into account in the error bars, we included a jitter term that we added quadratically to the estimated HARPS uncertainties.

We built different models involving circular and eccentric orbits, one and two planets, linear trends, and correlated noise components. We refer to each model following a similar notation as that in \citet{2021A&A...654A..60L}; that is, $\rm Xp \left[ P_{i}c \right] \left[ Y \right]$, where X is the total number of Keplerians, $\rm P_{i}$ indicates which planets are considered to have a circular orbit, and Y indicates whether the model has a linear drift (L) or a correlated noise component (QP). We tested the following models: 0p, 1p1c, 1p, 2p1c2c, 2p, 2p1c, 2p2c, 0pL, 1p1cL, 1pL, 2p1c2cL, 2pL, 2p1cL, 2p2cL, 0pQP, 1p1cQP, 1pQP, 2p1c2cQP, 2pQP, 2p1cQP, and 2p2cQP. In this work, planet `1' corresponds to TOI-5005.01. To ensure this in all cases, we set uninformative but restrictive priors on the orbital period of planet 1; that is, $P_{\rm orb, 1}\,\in$ $\mathcal{U}(5, 7)$ days. To avoid degeneracies in the time of mid-transit due to the infinite possibilities of $T_{0}+n \times P$, being $n$ $\in$ $\mathbb{N}$,  we sampled $T_{0}$ between 0 and $P$; that is, we sampled $T_{0}$ in the orbital phase space. For planet 2, given that we do not have any hint of its existence based on TESS data, we considered a wide uninformative prior for its orbital period $P_{\rm orb, 2}\,\in$ $\mathcal{U}(0, 1000)$ days. For the remaining parameters, we used wide uninformative priors large enough to ensure that they do not bias the analysis. 

\begin{figure*}
\centering
\includegraphics[width=0.98\textwidth]{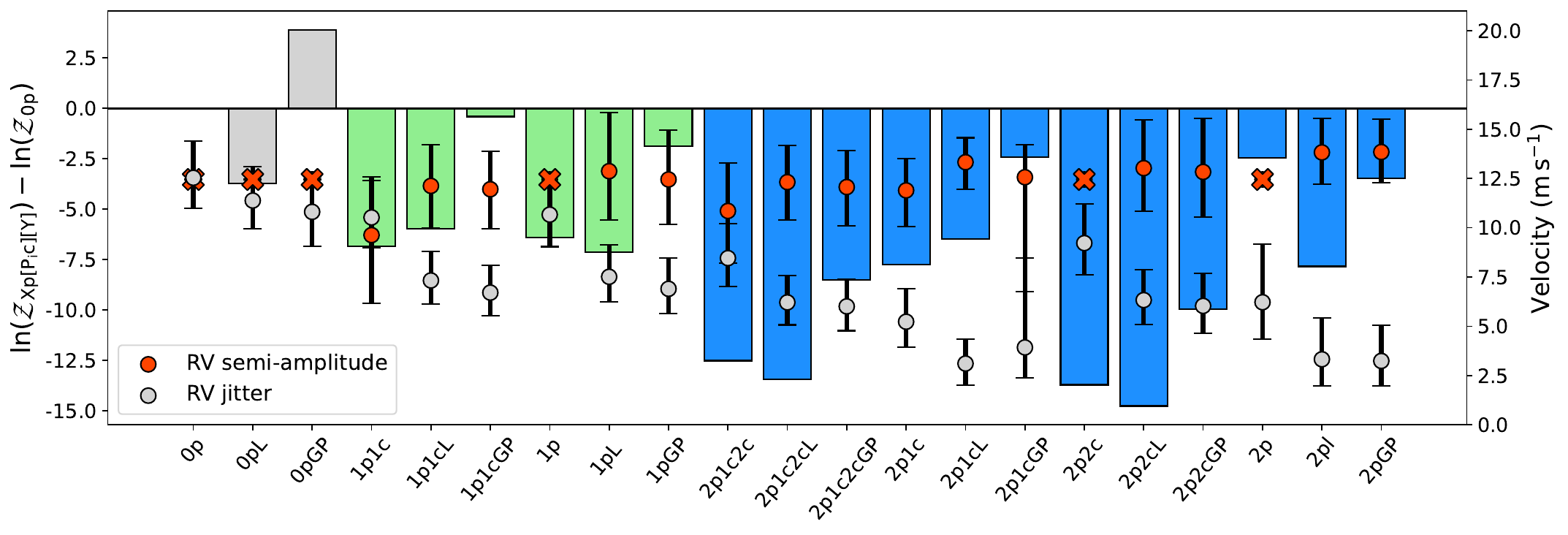}
    \caption{Bar chart showing the differences of the log-evidence of the 21 tested models (labelled on the $X$-axis) and the 0p model. The grey, green, and blue bars correspond to models with zero, one, and two planets, respectively. The red and grey circles indicate the posterior semi-amplitudes and RV jitters of the tested models. The red crosses indicate null semi-amplitude values, either because they correspond to a model without planets, or because the MCMC fit did not converge.}
    \label{fig:evidences_harps}
\end{figure*}

In Fig.~\ref{fig:evidences_harps}, we show the log-evidence of each model subtracted from the evidence of the 0p model; that is, $\rm ln(\mathcal{Z}_{Xp[P_{i}c][Y]}) - ln(\mathcal{Z}_{0p})$. In the same figure, we also illustrate the posterior Keplerian semi-amplitudes of TOI-5005 b and the jitter components of each tested model. The main conclusion of this analysis is that there is no model including planets that has a larger Bayesian evidence than the 0p model. Therefore, even though the planetary models converge towards a 6.3-day orbit for planet 1 with a $\simeq$5$\sigma$ detection of the RV semi-amplitude, this signal cannot be confirmed via Bayesian inference and model comparison based on the measured HARPS RVs alone when considering wide uninformative priors. This result illustrates the well-known difficulties of RV blind searches \citep[e.g.][]{2016SPIE.9908E..12Q,2022A&A...667A.102L}, which typically require large amounts of RV data points to consider planetary candidate signals as confirmed planets \citep[e.g.][]{2020A&A...639A..77S,2022A&A...658A.115F,2024A&A...690A..79G}. To select the model that best represents our dataset we need to include additional information that constrains the presence of the planetary signal. That is, we need to incorporate the TESS transits in the tested models. In Sect.~\ref{subsec:TESS_analysis}, we describe our TESS transit modelling, and in Sect.~\ref{subsec:joint_analysis} we incorporate it into the RV model comparison analysis.

\subsection{TESS photometric transits analysis}
\label{subsec:TESS_analysis}

We analysed the SAP photometry corrected for contamination described in Sect.~\ref{sec:obs_tess} to infer several orbital and physical parameters of TOI-5005~b. We built a model composed of two components: a transit model and a GP to account for the correlated photometric noise.

We considered the \citet{2002ApJ...580L.171M} quadratic limb darkened model as implemented in \texttt{batman}\footnote{Available at \url{https://github.com/lkreidberg/batman}.} \citep{2015PASP..127.1161K}. The model is described by the orbital period of the planet ($P_{\rm orb}$), the time of inferior conjunction ($T_{0}$), the orbital inclination ($i$), the quadratic limb darkening (LD) coefficients $u_{1}$ and $u_{2}$, the planet-to-star radius ratio ($R_{p} / R_{\star}$), the semi-major axis scaled to the stellar radius ($a/R_{\star}$), the orbital eccentricity ($e$), and the argument of the periastron ($\omega$). We followed the prescription by \citet{2013MNRAS.435.2152K} and parametrized the LD coefficients as $q_{1} =(u_{1} + u_{2})^{2}$ and $q_{2} = 0.5 \, u_{1} (u_{1} + u_{2})^{-1}$. Given that $a/R_{\star}$ is typically poorly constrained based on transit data alone, we parametrized it through the Kepler's third law and the measured stellar mass ($M_{\star}$) and radius ($R_{\star}$) similarly to previous works \citep[e.g.][]{2007ApJ...664.1190S,2020A&A...642A.121L,2022MNRAS.509.1075C}. We also parametrized $e$ and $\omega$ as in Sect.~\ref{subsec:harps_rv_analysis}. We considered uninformative priors for all the parameters except for those for which we have prior independent information, which we constrained through Gaussian priors. Those parameters are the stellar radius and mass, which we derived in our spectroscopic analysis (Sect.~\ref{sec:stellar_charact}), and the LD coefficients, which we computed based on the \texttt{ldtk} package \citep{2015MNRAS.453.3821P}. The package relies upon the TESS transmission curve, $T_{\rm eff}$, log $g$, and [Fe/H], and uses the synthetic spectra library from \citet{2013A&A...553A...6H} to infer the LD coefficients of a given law. To account for possible systematics in the estimated LD coefficients \citep[e.g.][]{2022AJ....163..228P}, we considered conservative uncertainties of 0.2 in $u_{1}$ and $u_{2}$.

The TESS SAP photometry of TOI-5005 is considerably affected by correlated noise. In Sect.~\ref{sec:obs_tess}, we saw that such noise is not dominated by stellar rotation (see also Sect.~\ref{sec:obs_harps}), so it most likely has an instrumental origin. Based on such noise properties, we chose a simple autocorrelation function $k(x_{i}, x_{j})$ (also called a GP kernel; see Appendix~\ref{bayesian} for further details) described as

\begin{equation}
    k(x_{i}, x_{j}) = \eta_{\sigma}^{2} \left[ \left(1 + \frac{1}{\epsilon} \right) e^{-(1-\epsilon) \sqrt{3} \tau / \eta_{\rho}} \cdot \left(1 - \frac{1}{\epsilon} \right) e^{-(1+\epsilon) \sqrt{3} \tau / \eta_{\rho}} \right],
\end{equation}where $\tau = x_{i} - x_{j}$ is the temporal separation between two time stamps, and $\eta_{\sigma}$ and $\eta_{\rho}$ are two hyperparameters that represent the characteristic amplitude and timescale of the correlated variations, respectively. This kernel is called approximate Matérn-3/2, and it has an additional parameter $\epsilon$ that controls the approximation to the exact kernel, which we fixed to its default value of $10^{-2}$ (see \citealt{2017AJ....154..220F} for further details). Given its simplicity and flexibility, this kernel has been extensively used to model TESS photometry where the correlated noise is dominated by unknown instrumental systematics \citep[e.g.][]{2023A&A...675A..52C,2023A&A...679A..33D,2023A&A...677A.182M}, as is the case of TOI-5005. Similarly to the transit component, we considered wide uninformative priors for both hyperparameters since we do not have a priori information about them. Also, given that the amplitudes and timescales of the TESS systematics typically vary from one sector to another, we fit those parameters independently. Hence, we denoted them as $\eta_{\sigma_{i}}$ and $\eta_{\rho_{i}}$, where $i$ refers to the corresponding sector.

\begin{figure*}
    \centering
    \includegraphics[width=17cm]{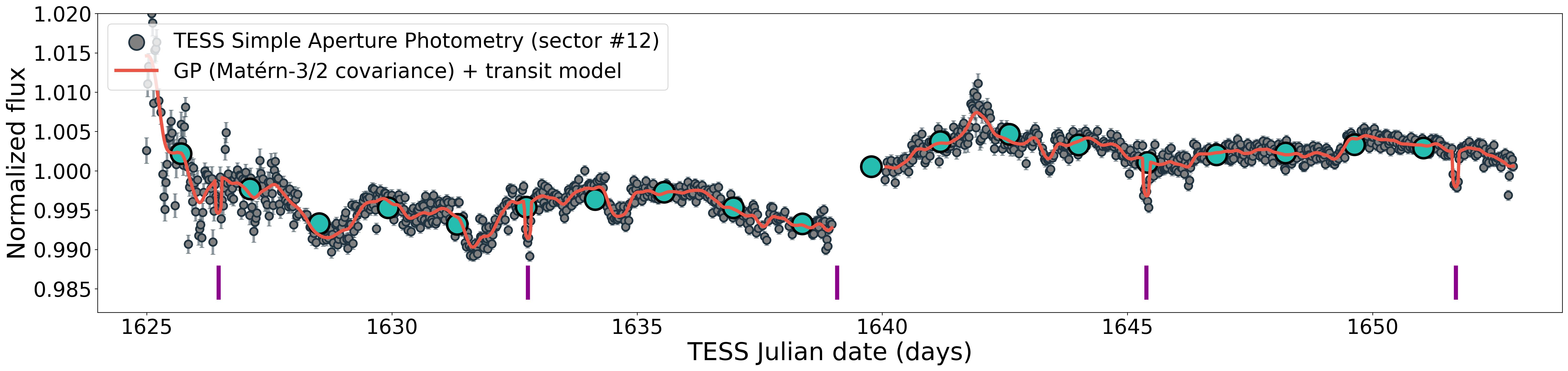}
    \includegraphics[width=17cm]{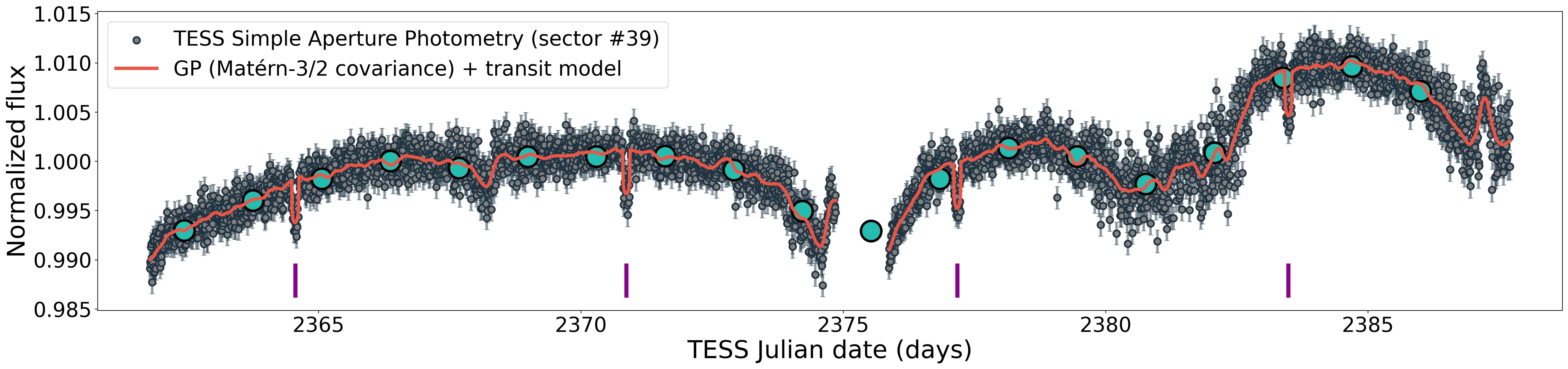}
    \includegraphics[width=17cm]{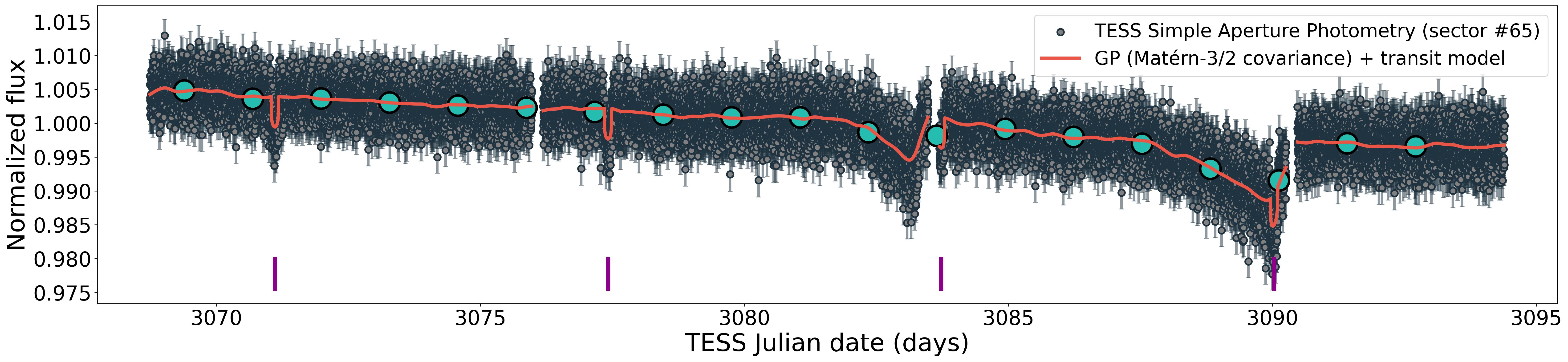}
    \caption{TESS simple aperture photometry of TOI-5005 with the median posterior model (transit + GP) superimposed. The green data points correspond to 1.5-day binned data. The vertical magenta line indicates the locations of the planetary transits. The TESS Julian date corresponds to Julian date $-$ 2457000 days.}
    \label{fig:transit+GP}
\end{figure*}

We inferred the parameters that best represent the TESS dataset through Bayesian inference as described in Sec.~\ref{sec:model_sel_param_det}. As a sanity check, we repeated the analysis by modelling the highest level public dataset (i.e. QLP/SAP in S12 and SPOC/PDCSAP in S39 and S65), and also by setting wide uninformative priors in the LD coefficients $\mathcal{U}(0,1)$. In both cases, we retrieved consistent planetary parameters at 1$\sigma$. In Fig.~\ref{fig:transit+GP}, we show the complete TESS SAP dataset together with the inferred transit+GP model evaluated on the fit parameters. In Fig.~E.3 (available on Zenodo), we show the transit+GP model over each individual observed transit event. We note that the obtained orbital and physical parameters based on the TESS data alone are consistent within 1$\sigma$ with the parameters derived in the joint analysis described in the following section. Therefore, for clarity, we only present the final characterization based on the complete dataset.

\subsection{Joint analysis and system characterization}
\label{subsec:joint_analysis}

We jointly analysed the HARPS RVs, TESS photometry, and ground-based photometry with the ultimate goal of inferring the orbital and physical parameters of the system. In Sect.~\ref{sec:HARPS_TESS}, we describe the joint analysis of the HARPS and TESS datasets, and in Sect.~\ref{sec:HARPS_TESS_PEST_TRAPPIST} we test the inclusion of the PEST and TRAPPIST-South single transits.

\subsubsection{HARPS and TESS}
\label{sec:HARPS_TESS}
\begin{figure}
    \centering\includegraphics[width=0.5\textwidth]{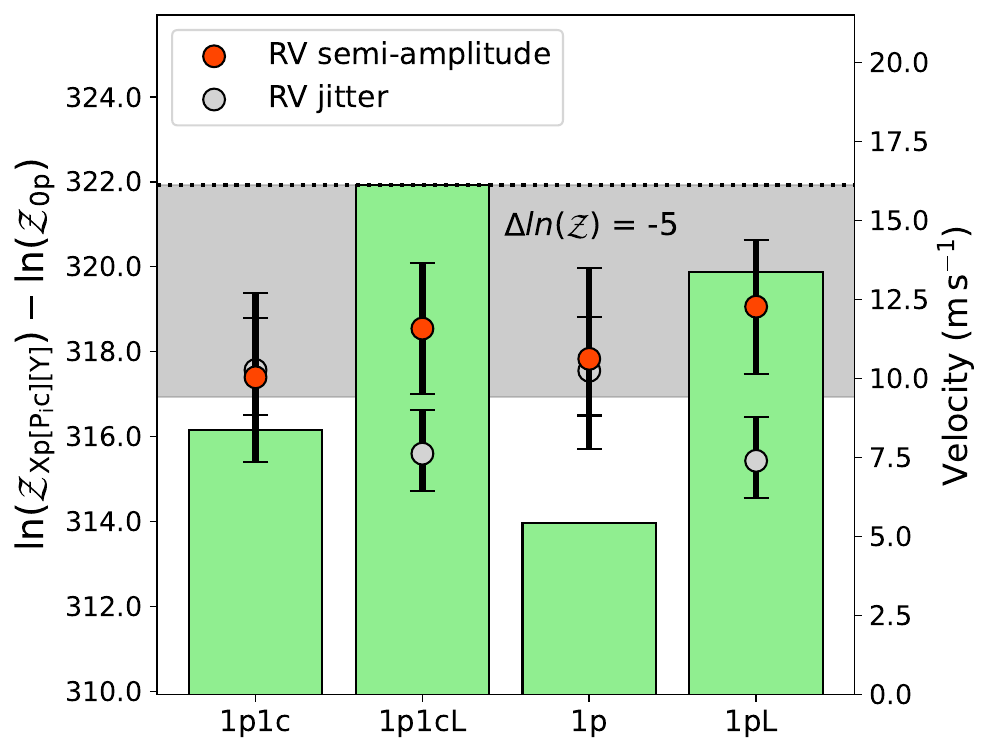}
    \caption{Bar chart showing the log-evidence of the four preferred models with respect to the 0p model in the TESS+HARPS joint analysis (Sect.~\ref{subsec:joint_analysis}). The grey shaded region indicates the $\rm 0 \geq \Delta ln \mathcal{Z} \geq$ -5 region from the largest evidence model (1p1cL).}
    \label{fig:evidences_tess_harps}
\end{figure}

\begin{figure*}
    \centering
    \includegraphics[width=0.98\textwidth]{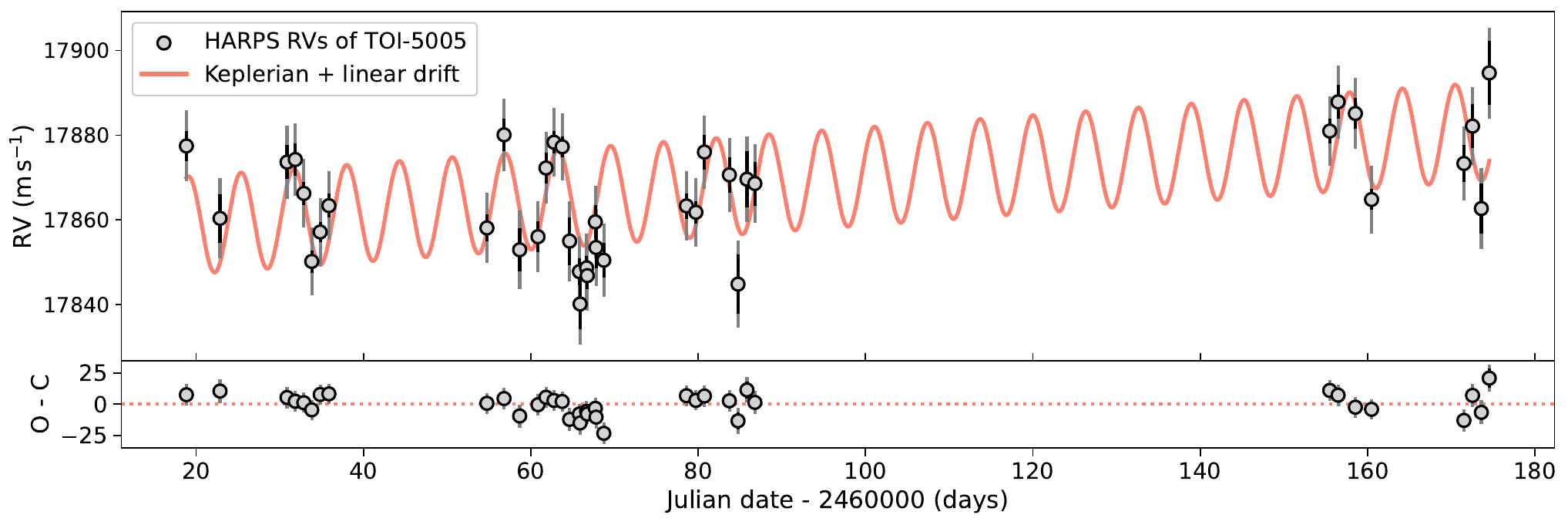}
    \caption{HARPS radial velocities of TOI-5005. The red solid line indicates the simplest model that best represents our data (i.e. a Keplerian model plus a linear drift; 1p1cL), selected based on our joint analysis of the HARPS and TESS datasets (Sect.~\ref{subsec:joint_analysis}). The black error bars represent the uncertainties as estimated by the HARPS DRS ($\sigma_{i,\rm HARPS}$; Sect.~\ref{sec:obs_harps}), and the grey error bars represent the total uncertainties computed as $(\sigma_{i,\rm HARPS}^{2} +  \sigma_{jit,\rm HARPS}^{2})^{0.5}$, where $\sigma_{jit,\rm HARPS}$ is the posterior median jitter term. }
    \label{fig:rv_model_complete}. 
\end{figure*}

\begin{figure*}
    \centering
    \includegraphics[width=0.445\textwidth]{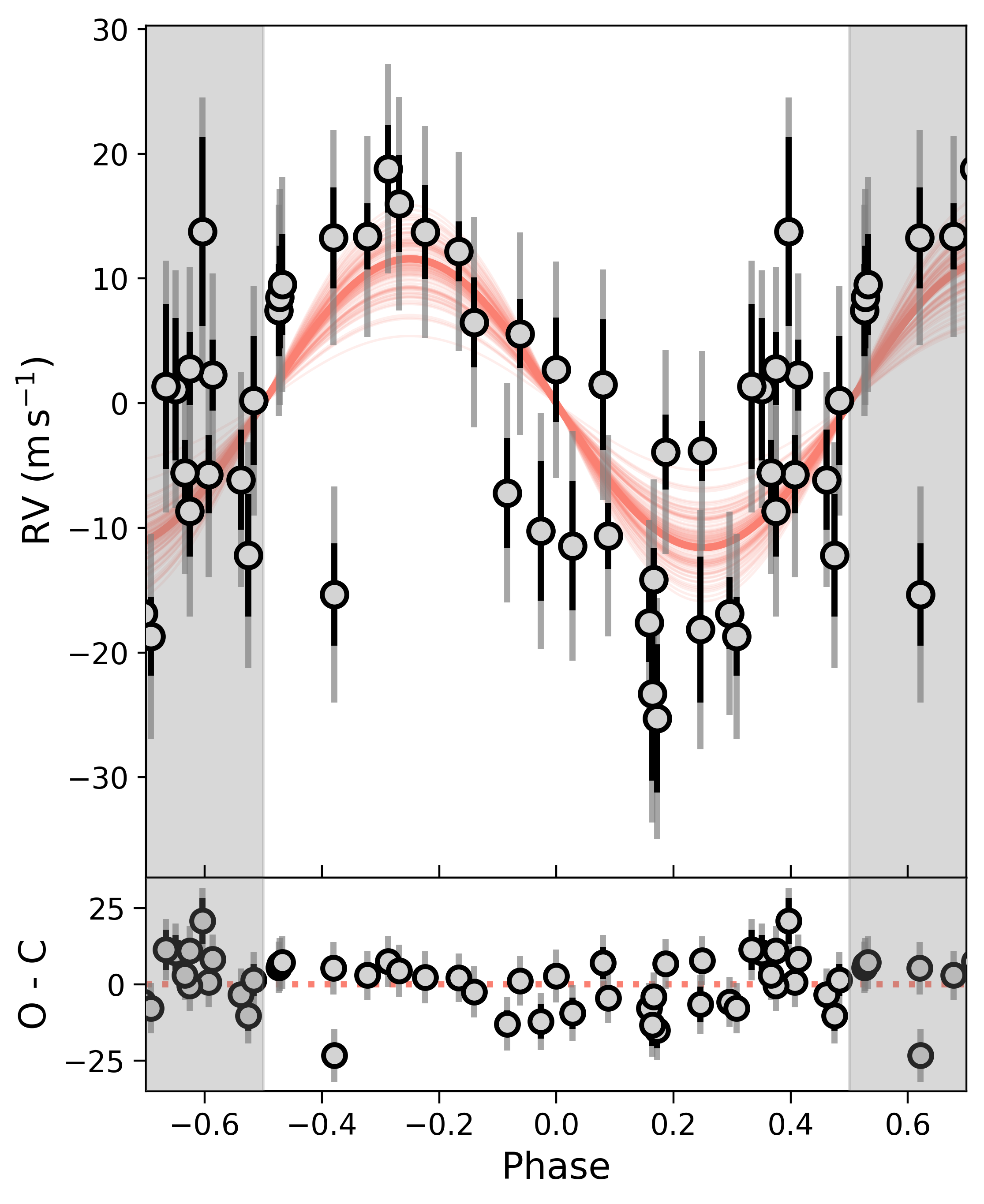}
    \includegraphics[width=0.47\textwidth]{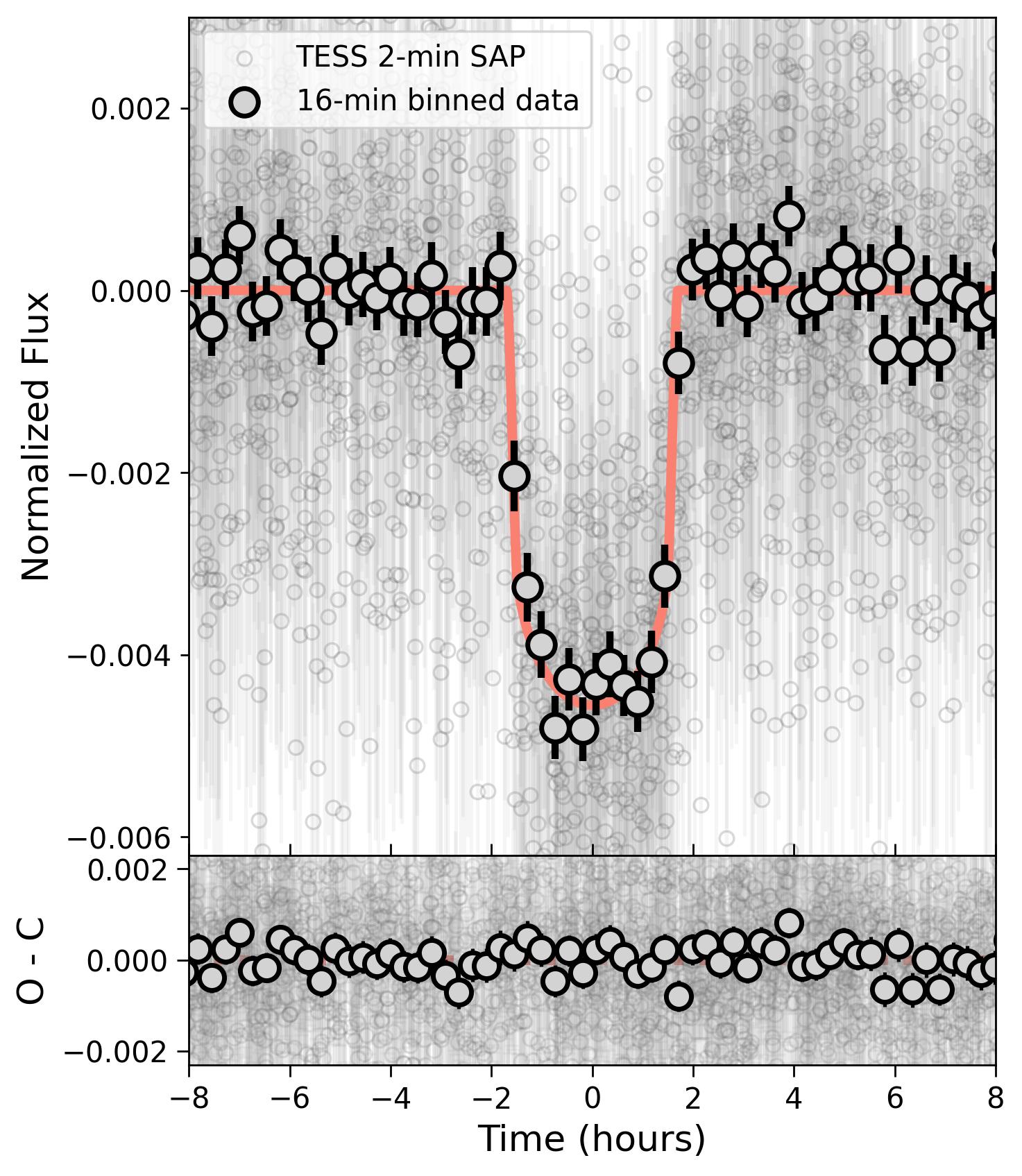}
    \caption{HARPS RVs (left) and TESS photometry (right) subtracted from the linear drift and GP component, and folded to the orbital period of TOI-5005~b. The black error bars represent the instrument uncertainties and the grey error bars represent the total uncertainties obtained by quadratically adding the corresponding jitter term. The red solid lines indicate the median posterior models. In the HARPS RVs plot we also include 100 random posterior models that illustrate the level of modelling uncertainty.}
    \label{fig:phase_folded_plots}
\end{figure*}

As shown in Sect.~\ref{subsec:harps_rv_analysis}, the RV-only analysis is not able to decide which model best represents our data. Therefore, we repeated the model comparison analysis by including the TESS transit analysis described in Sect.~\ref{subsec:TESS_analysis}. Among the 21 tested models, there are four that significantly stand out, showing Bayesian evidence much larger than the remaining 17 models: 1p1c, 1p1cL, 1p, and 1pL. This result discards the detection of additional planets in the system and also discards the need to use a GP component to describe our dataset. In Fig.~\ref{fig:evidences_tess_harps}, we show the log-evidence of those four models. They all well meet the condition $\mathcal{B} > 5$ when compared to the 0p model, so we can now confirm the planetary nature of TOI-5005~b via Bayesian inference and model comparison with uninformative priors. Among those four models, the 1p1cL model stands out with the largest evidence and meets the condition $\mathcal{B} > 5$ ($\mathcal{B}$ = 5.8) when compared to the simpler 1p1c model. This result indicates that our dataset is better described when incorporating a linear drift. Also, the models involving eccentric orbits show lower Bayesian evidence than the simpler circular models. Overall, the 1p1cL model is the simplest model that best represents our data when jointly considering the TESS and HARPS observations, so we chose it to infer the orbital and physical parameters of the planetary system following the procedure in Sect.~\ref{sec:model_sel_param_det}.  In Fig.~\ref{fig:rv_model_complete}, we show the HARPS dataset together with the fitted 1p1cL model. In Fig.~\ref{fig:phase_folded_plots}, we show the HARPS RVs and TESS photometry subtracted from the linear drift and GP component and folded to the period of TOI-5005~b. 

\subsubsection{HARPS, TESS, PEST, and TRAPPIST-South}
\label{sec:HARPS_TESS_PEST_TRAPPIST}

We tested the inclusion of the PEST and TRAPPIST-South single transits (Sects.~\ref{sec:obs_pest} and \ref{sec:obs_TRAPPIST_South}) in the joint analysis. For both datasets, we considered a transit model as described in Sect.~\ref{subsec:TESS_analysis} together with a systematics model to account for possible correlations with different parameters. These are the full width at half maximum of the target point spread function (\textit{fwhm}), airmass ($\chi$), position (\textit{x},\textit{y}), displacement (\textit{dx}, \textit{dy}), and distance to the detector centre (\textit{dist}) of the target star, and background flux (\textit{sky}). We assume linear dependencies, so we can write the systematics model as $c_{\rm inst, 0}$ + $\sum_{i = 1}^{N_{\rm inst}} c_{\rm inst, i} \times p_{\rm inst, i}$, where $p_{\rm inst, i}$ are the detrend parameters described above, $N_{\rm inst}$ is the number of such parameters, and $c_{\rm inst, i}$ are the linear combination coefficients. Similarly to the TESS analysis, we considered a jitter term per instrument.  

\begin{figure}
    \centering
\includegraphics [width=0.48\textwidth]{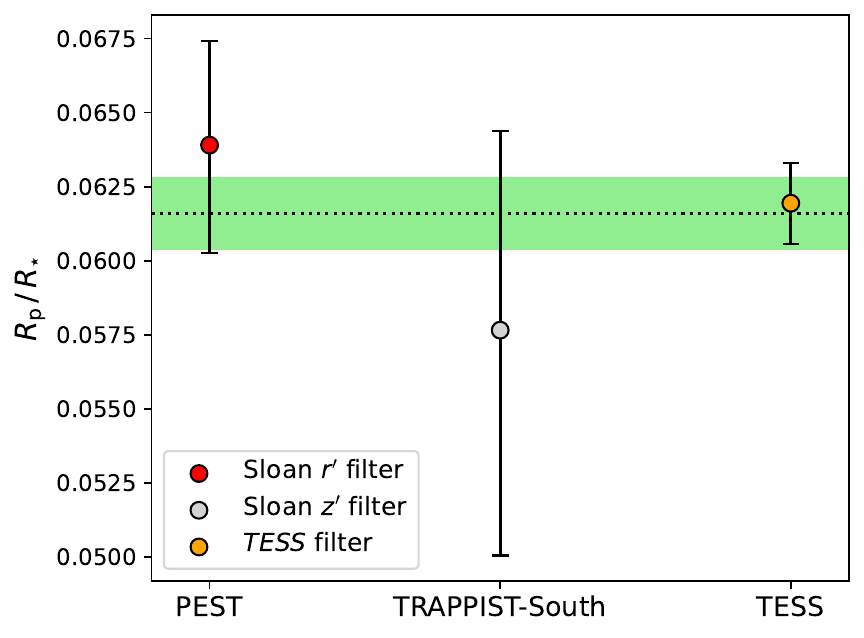}
    \caption{Planet-to-star radius ratios measured by TESS, PEST, and TRAPPIST-South. The horizontal dotted black line and green shadow indicate the median and $\pm$1$\sigma$ interval of the planet-to-star radius ratio measured jointly by the three instruments. }
    \label{fig:rprs_comparison}
\end{figure}

\begin{figure*}
    \centering
    \includegraphics[width=0.48\textwidth]{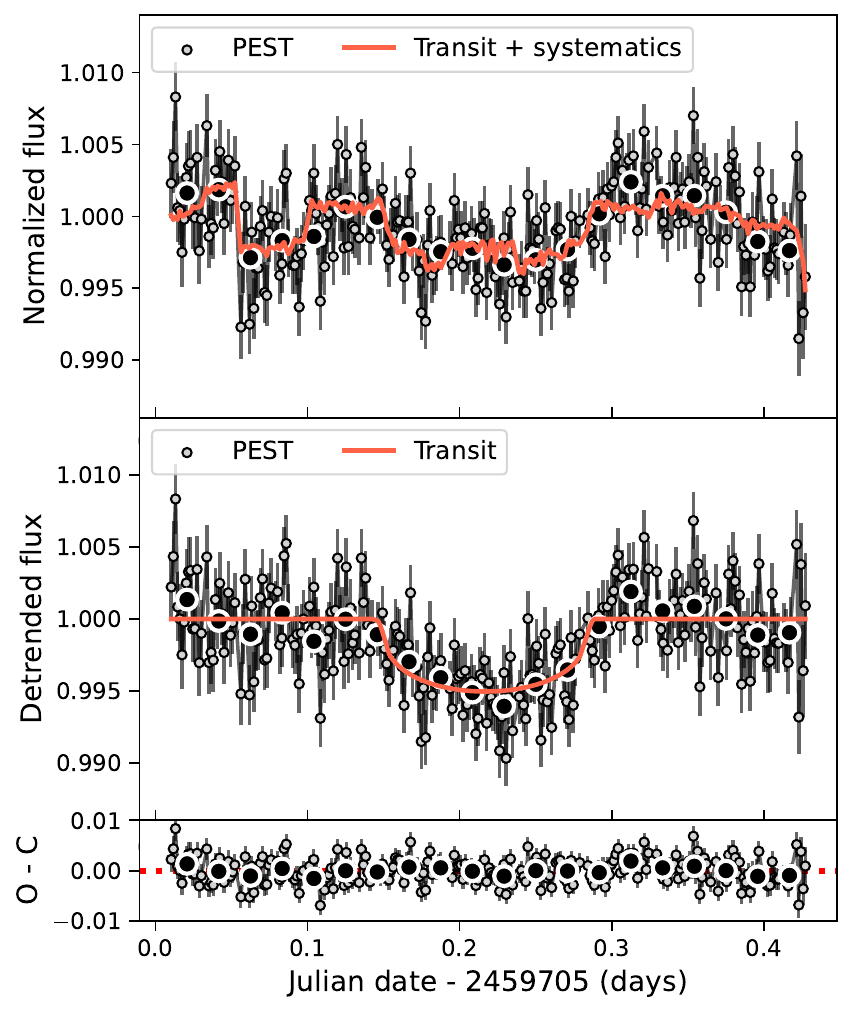}
    \includegraphics[width=0.48\textwidth]{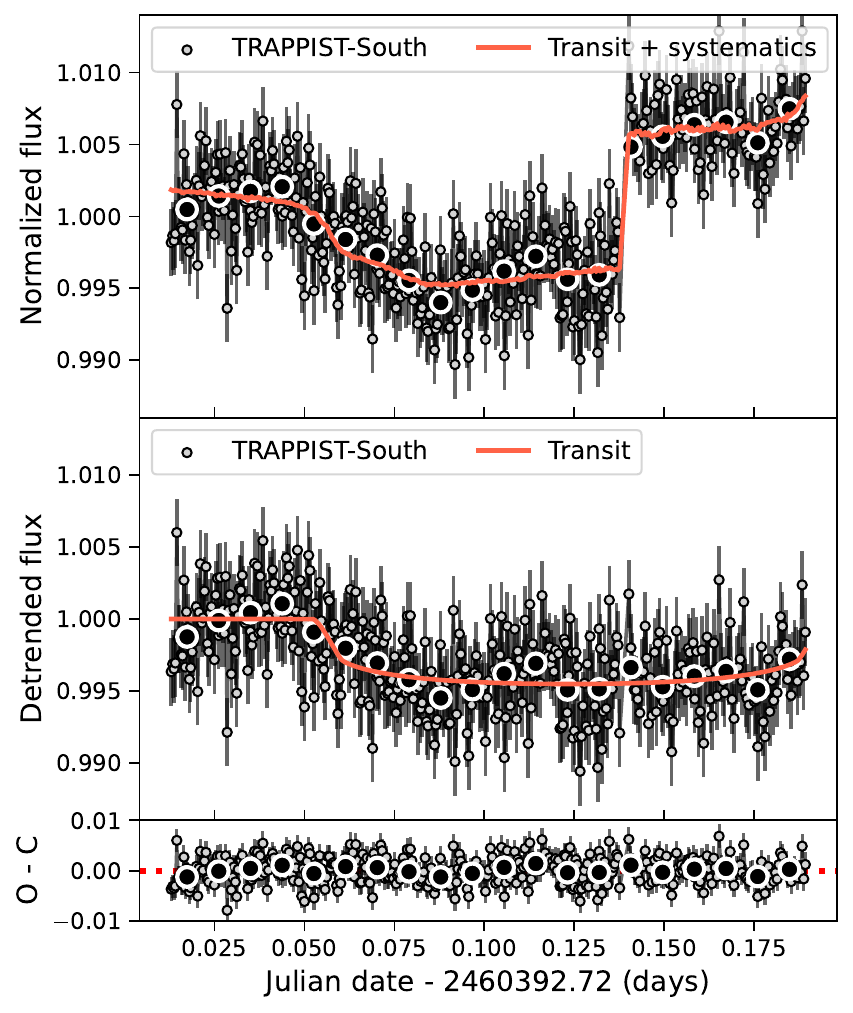}
    \caption{PEST and TRAPPIST-South photometry of TOI-5005~b acquired on 5 May 2022 and 22 March 2024, respectively. The upper panels show the raw photometry extracted from the instrument pipelines (see Sects.~\ref{sec:obs_pest} and \ref{sec:obs_TRAPPIST_South}), the lower panels show the detrended photometry, i.e. the raw photometry subtracted from the systematics model described in Sect.~\ref{sec:HARPS_TESS_PEST_TRAPPIST}, and the lower panels show the residuals of the joint fit analysis. The red solid lines indicate the median posterior models.}.
    \label{fig:pest_trappist}
\end{figure*}

The TESS, PEST, and TRAPPIST-South photometric data were acquired in different bands: TESS, Sloan \textit{r'}, and Sloan \textit{z'}, respectively. Hence, we considered different $q_{1}$ and $q_{2}$ limb-darkening coefficients for each instrument and set wide Gaussian priors based on \texttt{ldtk} as described in Sect.~\ref{subsec:TESS_analysis}. We first performed a joint analysis that considers different $R_{\rm p}/R_{\rm \star}$ ratios per instrument. We compare the posterior $R_{\rm p}/R_{\rm \star}$ in Fig.~\ref{fig:rprs_comparison}. The obtained values are consistent at the 1$\sigma$ level, which indicates that our photometric dataset is not sensitive to a possible transit depth chromaticity. This allowed us to perform a joint analysis with a common $R_{\rm p}/R_{\rm \star}$ for the three instruments. The derived orbital and physical parameters are consistent at the 1$\sigma$ level with those from the HARPS+TESS analysis (Sect.~\ref{sec:HARPS_TESS}), and several transit parameters show slightly lower uncertainties. We also find that the Bayesian Evidence difference with that of a global model with no planets is larger when including the PEST and TRAPPIST-South data: $(\rm \Delta ln \mathcal{Z})_{HARPS+TESS+PEST+TRAPPIST}$ = +361 against $(\rm \Delta ln \mathcal{Z})_{HARPS+TESS}$ = +322. Therefore, we find beneficial the inclusion of the ground-based photometry in the joint analysis.  In Fig.~\ref{fig:pest_trappist}, we show the PEST and TRAPPIST-South raw and systematics-corrected photometry together with the corresponding posterior models. In Table~\ref{tab:parameters_joint}, we present the median and 1$\sigma$ intervals of the complete parameter set that describes the joint model.  In Fig.~E.4 (available on Zenodo), we represent a corner plot with the 1D and 2D posterior distributions of the main fitted orbital and physical parameters. 

\subsection{Stellar signals analysis}
\label{subsec:stellar_signals_analysis}

\subsubsection{HARPS activity indicators}
\label{subsubsec:HARPS_activity_indicators}

Four HARPS activity indicators (FWHM, S-index, Contrast, and Ca) of TOI-5005 show a sinusoidal signal with a periodicity of $\simeq$21 days (Sect.~\ref{sec:obs_harps}, Fig.~\ref{fig:gls_to_HARPS}). This signal most likely reflects the rotation period of the star and is compatible with independent $P_{\rm rot}$ estimates (Sect.~\ref{subsec:prot}). In this section we follow the procedure described in Sect.~\ref{sec:model_sel_param_det} to obtain an accurate determination of the signal periodicity and assess its significance. We built an activity model based on a GP with a quasi-periodic covariance function (Eq.~\ref{eq:QP}), and fitted it to the individual activity indicators time series starting from wide uninformative priors. The MCMC analysis shows that the active regions' timescale ($\eta_{2}$) and harmonic complexity of the signal ($\eta_{4}$) cannot be constrained by our dataset. However, the stellar rotation period ($\eta_{3}$ or $P_{\rm rot}$) converges at $\simeq$21 days. To get a robust estimate and assess how well the activity model describes our dataset, we jointly modelled the four indicators with a common $P_{\rm rot}$. We obtain $P_{\rm rot} = 21.01^{+0.46}_{-0.60}$~days and a Bayesian Evidence against the null hypothesis of $\rm \Delta ln \mathcal{Z}$ = +8.6. We include this value in Table~\ref{tab:stellar_parameters}.

\subsubsection{TESS photometric variability}
\label{subsubsec:photometric_variability}

\begin{table}[]
\caption{Main properties of the TESS photometric variations synchronized with the orbital period of TOI-5005~b (Sect.~\ref{subsubsec:photometric_variability}).}
\renewcommand{\arraystretch}{1.4}
\setlength{\tabcolsep}{8.6pt}
\begin{tabular}{cccccc}
\hline
Sector & $f_{\rm inc}$ & $(f_{\rm inc})_{\rm vis}$ & $(f_{\rm inc})_{\rm hid}$ & $\phi_{\rm max}$ & $\phi_{\rm min}$ \\ \hline
S39    & 0.56          & 0.83                      & 0.30                      & +0.33            & -0.21            \\
S65    & 0.49          & 0.26                      & 0.72                      & -0.33            & +0.21            \\ \hline
\end{tabular}
\tablefoot{$f_{\rm inc}$ is the orbital phase fraction in which the TESS photometric flux increases. $(f_{\rm inc})_{\rm vis}$ and $(f_{\rm inc})_{\rm hid}$ represent the same fraction, but limited to the phases where a hypothetical co-rotating active region would be visible and hidden from Earth, respectively. $\phi_{\rm max}$ and $\phi_{\rm min}$ are the phase offsets between the TOI-5005~b transit time and the maximum and minimum flux emissions, respectively.}
\label{tab:properties_variations}
\end{table}

\begin{figure*}
    \centering
    \includegraphics[width=0.25\textwidth]{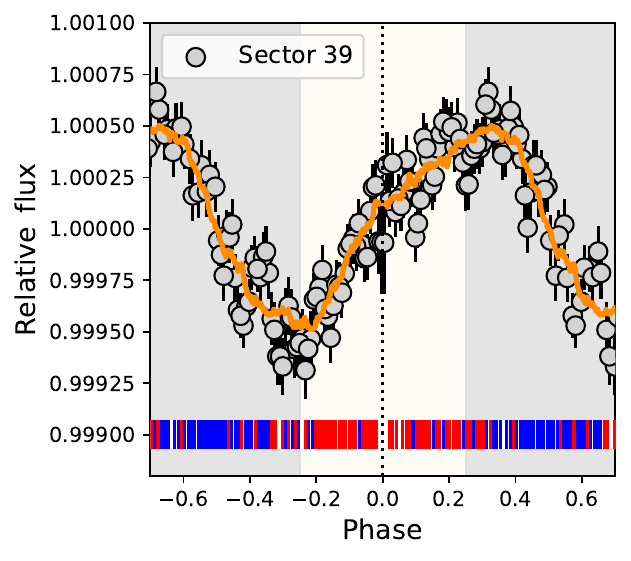}
    \includegraphics[width=0.23\textwidth]{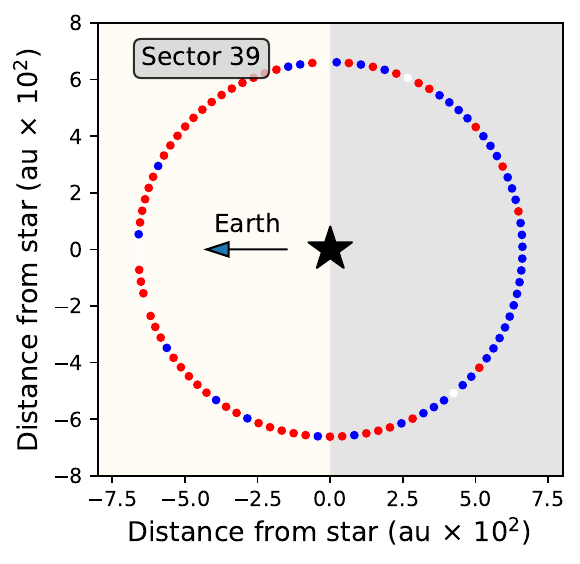}
    \includegraphics[width=0.25\textwidth]{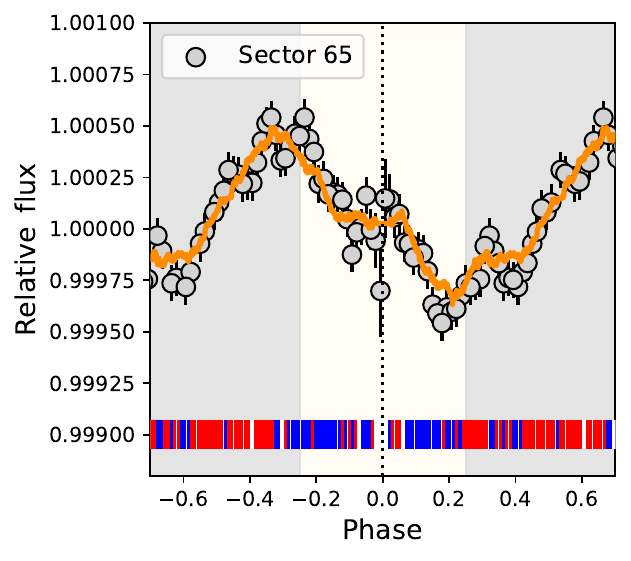}
    \includegraphics[width=0.23\textwidth]{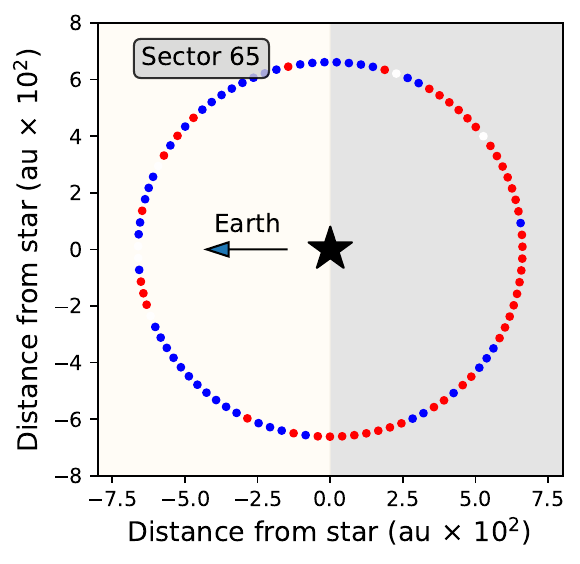}
    \caption{Links between the orbital motion of TOI-5005~b and the TESS photometric variability of TOI-5005. Left panels: TESS photometry of TOI-5005 folded in phase with the orbital period of TOI-5005~b (Sect.~\ref{subsec:joint_analysis}, Table \ref{tab:parameters_joint}) and binned with $\simeq$5$\%$ phase bins. The orange line corresponds to the filtered photometry through a median filter with a kernel size of 1401 cadences. The blue, red, and white vertical lines indicate whether the photometry increases, decreases, or remains stable in time lapses of $\simeq$1.5 h. Right panels: Orbital path of TOI-5005 b, which follows the anticlockwise direction. The circles represent the location of TOI-5005 b every $\simeq$1.5 h and are coloured similarly to the vertical lines in the left panels. In all panels the white and grey backgrounds represent the orbital regions in which a hypothetical co-rotating active region would be visible and not visible from Earth, respectively.}
    \label{fig:MSPIs}
\end{figure*}

The \texttt{GLS} periodograms of the PDCSAP photometry of TOI-5005 show a significant sinusoidal modulation with a periodicity of $\simeq 6.3$~days (Sect.~\ref{sec:obs_tess}, Fig.~\ref{fig:gls_to_TESS}). This periodicity matches the orbital period of TOI-5005~b, which suggests a planet-induced origin through MSPIs. TESS photometry, however, is frequently affected by significant instrumental systematics. If not modelled properly, they can generate instrumental signals that could be mistakenly interpreted as stellar. We studied whether the observed modulation could have an instrumental origin. To that end, we performed an independent photometric correction, which consists of fitting all the major instrumental systematics observed in the TESS CCDs without being a priori  constrained by the observed systematics in nearby stars (see Appendix~\ref{sec:cbv_correction} for further details). We find that no combination of the major TESS systematics can explain the observed $\simeq 6.3$-day modulation, which strongly favours a stellar origin. We note that the signal could still be instrumental in the case that the main pixels collecting the stellar flux are affected by particular systematics different from those observed in other regions of the CCD. However, we find this possibility very unlikely since the same signal appears in two different sectors, where TOI-5005 falls in different pixels of different CCDs. Another factor to consider is the flux contamination from nearby sources. Contrary to the planetary signal of TOI-5005~b, we do not have independent observations that confirm that the $\simeq 6.3$-day sinusoidal modulation comes from TOI-5005. However, the observed modulation would require very large photometric variations if it came from the faint contaminant sources (see Sect.~\ref{sec:obs_tess}), which makes it a very unlikely scenario. Taking all the mentioned analyses into account, we consider that the $\simeq 6.3$-day signal most likely has a stellar origin coming from TOI-5005.  

An in-depth analysis of the measured signal is beyond the scope of this work. However, we are interested in studying how the flux variations are related to the orbital motion of TOI-5005~b since it can give us important insight into the potential MSPIs. Similarly to \citet{2024A&A...684A.160C}, we divided the complete S39 and S65 phase-folded photometry into 100 bins of $\simeq$1.5 h long and studied when the flux increases and decreases. Short-term variations induced by the photometric scatter were previously filtered out through a median filter with a kernel size of 1401 cadences; that is, $\simeq$10$\%$ of the orbital phase. We find that the orbital phase fraction in which the photometric flux increases ($f_{\rm inc}$) roughly corresponds to half an orbit, which indicates a high degree of signal symmetry. We also computed the orbital fraction of increasing photometry limited to the phases where a hypothetical co-rotating and small active region would be visible $(f_{\rm inc})_{\rm vis}$ and hidden $(f_{\rm inc})_{\rm hid}$ from Earth. Interestingly, we find a strong imbalance between increasing and decreasing regions (see Table~\ref{tab:properties_variations} and Fig.~\ref{fig:MSPIs}). In S39, the TESS flux primarily increases when such a hypothetical region is visible from Earth, while in S65 it primarily increases when this region is hidden. We note that this behaviour discards the possibility that the co-rotating region is small, since, in that case, we would observe a flux increase until TOI-5005~b's transit time (phase = 0), where the flux would decrease until the spot reaches the stellar limb (phase = 0.25), where the flux would become flat until reaching the following limb (phase = 0.75). Instead, the observed photometric variations can be explained by an extensive co-rotating active region (i.e. comparable to the stellar disk) trailing or leading TOI-5005~b's orbit with a bright or dark contribution to the total stellar flux. We also computed the phase offset between TOI-5005~b's transit time and the maximum ($\phi_{\rm max}$) and minimum ($\phi_{\rm min}$) flux emission, and found a 0.5 phase difference between S39 and S65 time series (see Table~\ref{tab:properties_variations}). This phase offset shows that the signal shape changes in short time scales, and it can be either interpreted as the potential planet-induced active region changing from bright to dark contrast or alternating its location with respect to the planetary orbit from a trailing to a leading configuration and vice versa. Interestingly, very similar offsets between the planet transit time and the stellar activity extremes, as well as between different planetary orbits have been previously detected in other similar systems with signs of MSPIs \citep[e.g.][]{2005ApJ...622.1075S,2008ApJ...676..628S,2008A&A...482..691W,2019NatAs...3.1128C,2024A&A...684A.160C}.


\section{Discussion}
\label{sec:discussion}

The mass ($M_{\rm p}$ = $32.7\pm 5.9$ $\rm M_{\oplus}$; $M_{\rm p}$ = $0.103\pm 0.018$ $\rm M_{\rm J}$) and radius ($R_{\rm p}$ = $6.25\pm 0.24$~$\rm R_{\rm \oplus}$; $R_{\rm p}$ = $0.558\pm 0.021$ $\rm R_{\rm J}$) derived in the joint analysis place TOI-5005~b approximately halfway between Neptune and Saturn. We adopt the most commonly used convention \citep[e.g.][]{2011ApJ...726...52H,2014A&A...572A...2B,2015ApJ...813..111B} and refer to TOI-5005~b as a super-Neptune. 

There are very few planets with physical properties similar to those of TOI-5005~b. For example, only 119 planets (i.e. 2$\%$ of the known population) have been found within a radius range 5 $\rm R_{\oplus}$ < $R_{\rm p}$ < 7~$\rm R_{\oplus}$. This number is reduced to ten detections if we limit to a period range of 5 days < $P_{\rm orb}$ < 7 days, of which only two have a precise mass measurement (CoRoT-8 b; \citealt{2010A&A...520A..66B}, and TOI-4010 c; \citealt{2023AJ....166....7K}). Interestingly, among the ten detections, TOI-5005~b orbits the brightest host, which makes it the most amenable planet for follow-up observations in this parameter space. Other super-Neptunes similar to TOI-5005~b orbiting relatively bright hosts ($V$ < 12 mag) are K2-39~b \citep{2016AJ....152..143V}, HATS-38~b \citep{2020AJ....160..222J}, K2-334~b \citep{2021MNRAS.508..195D}, TOI-181~b \citep{2023MNRAS.521.1066M}, K2-141~c \citep{2018AJ....155..107M}, TOI-5126~b \citep{2024MNRAS.527.8768F}, and TOI-1248~b \citep{2024ApJS..272...32P}.

In Sects.~\ref{subsec:period-radius} and \ref{subsec:LDSP}, we contextualize TOI-5005~b in different parameter spaces and discuss its observed properties according to different evolutionary hypotheses and additional observational constraints. In Sects.~\ref{sec:internal_structure}, \ref{subsec:mass-loss}, and \ref{subsec:prospects_atm}, we use the system parameters to infer the internal structure of TOI-5005~b, constrain its mass-loss rate, and discuss the prospects for characterizing its atmosphere.

\subsection{TOI-5005~b in the period-radius diagram: A new super-Neptune in the savanna near the ridge}
\label{subsec:period-radius}
\begin{figure}
    \centering
    \includegraphics[width=0.48\textwidth]{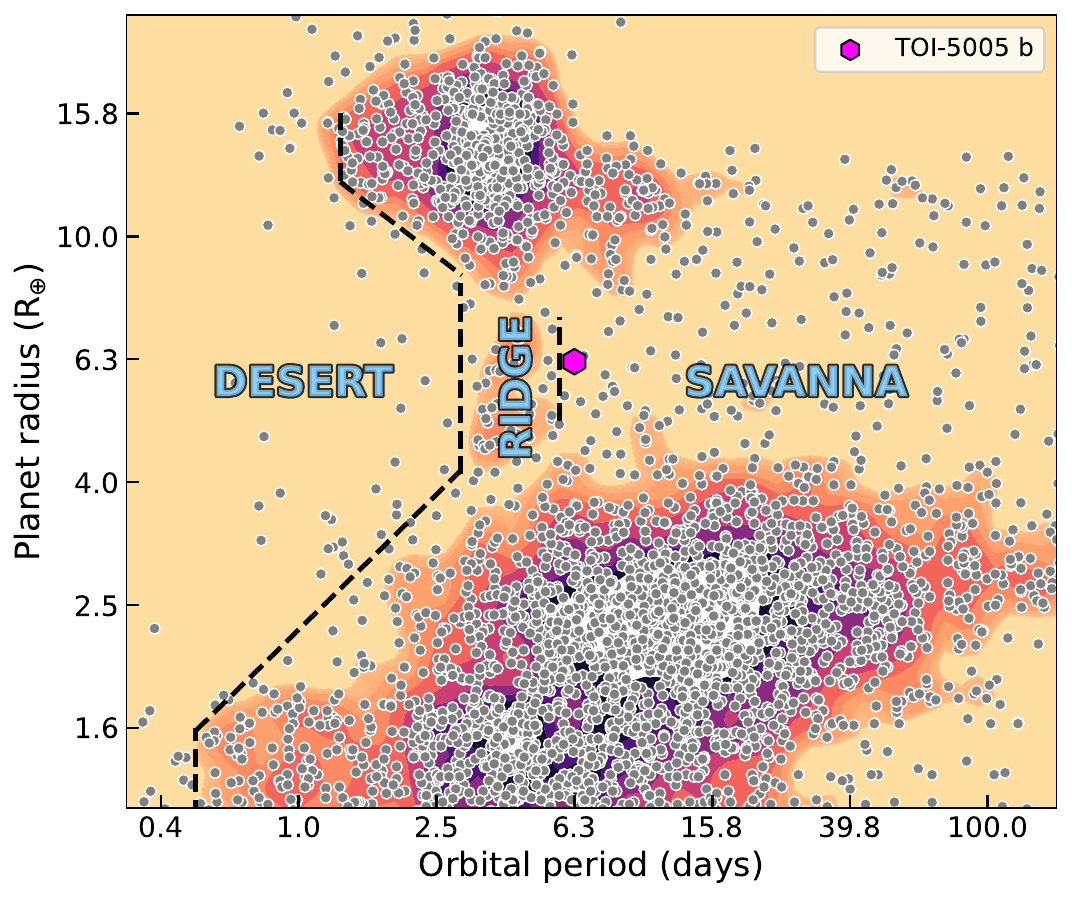}
    \caption{TOI-5005~b in the period-radius diagram of close-in exoplanets, where we highlight the population-based boundaries of the Neptunian desert, ridge, and savanna derived in \citet{2024A&A...689A.250C}. The data were collected from the NASA Exoplanet Archive \citep{2013PASP..125..989A} on 20/09/2024. This plot was generated with \texttt{nep-des} (\url{https://github.com/castro-gzlz/nep-des}).}
    \label{fig:nep_desert}
\end{figure}

In Fig.~\ref{fig:nep_desert}, we contextualize TOI-5005~b in the period-radius diagram of known close-in planets. Having an orbital period of 6.3 days, TOI-5005~b lies in the Neptunian savanna near the ridge, a recently identified overdensity of planets at $\simeq$3-5 days \citep{2024A&A...689A.250C}. The authors argue that the dynamical mechanism that brings planets preferentially to the ridge might be similar to the mechanism that brings larger planets to the $\simeq$3-5 day hot-Jupiter pileup \citep[e.g.][]{2007ARA&A..45..397U,2009ApJ...693.1084W}. Interestingly, different works are revealing a large number of eccentric and misaligned Jupiter- and Neptune-sized planets within and near the $\simeq$3-5 day overdensity \citep[e.g.][]{2012ApJ...757...18A,2020A&A...635A..37C,2023A&A...669A..63B}, which suggests that it could be primarily populated by HEM processes \citep[see][]{2012ApJ...754L..36N,2017AJ....154..106N,2018ARA&A..56..175D,2021JGRE..12606629F}. Unfortunately, our HARPS RV dataset did not allow us to constrain the orbital eccentricity of TOI-5005~b. In this hypothesis, planets within the ridge would be expected to have an outer massive companion that triggered such migration processes \citep[e.g.][]{2003ApJ...589..605W,2008ApJ...686..621F,2011CeMDA.111..105C}. Our HARPS RV dataset shows a long-term linear trend that could be caused by an outer companion (see Sects.~\ref{sec:obs_harps} and \ref{subsec:joint_analysis}). We note, however, that several activity indicators of TOI-5005 also show similar trends (Sect.~\ref{sec:obs_harps}), so we cannot discard that they could all reflect the magnetic cycle of the star. Indeed, the RV and FWHM trends are inverse to the Contrast trend, which is what we would observe if the magnetic cycle was causing them \citep{2011arXiv1107.5325L}. Additional long-term high-resolution spectroscopic measurements or high-precision astrometry are needed to resolve this dichotomy. Overall, based on our current data we cannot infer whether TOI-5005~b reached its location through disk-driven migration or HEM processes.

Atmospheric escape is also thought to shape the close-in period-radius distribution depicted in Fig.~\ref{fig:nep_desert}. While some theoretical works predict that Jupiter-sized planets could be eroded into super-Earths \citep[e.g.][]{2014ApJ...783...54K}, the majority of models and observational constraints indicate that these planets are too massive to significantly evaporate \citep[see][for a review]{2018ARA&A..56..175D,2021JGRE..12606629F}. In contrast, the atmospheres of several Neptunian planets within and near the ridge have been observed to be eroding at very high rates \citep[e.g.][]{2015Natur.522..459E,2018A&A...620A.147B}. Hence, as suggested by \citet[][]{2018Natur.553..477B,Attia2021,2024A&A...689A.250C}, Neptunes in the ridge might survive evaporation for a limited time, so that the ones that we detect today would have arrived relatively recent, presumably through HEM processes. Interestingly, this hypothesis would answer the question of why do warm Neptunes present nonzero eccentricity studied in \citet{2020A&A...635A..37C}. Unfortunately, atmospheric escape has not been extensively probed at larger orbital distances (i.e. within the savanna), and evaporation models show important discrepancies in the still poorly explored Neptunian domain (see Sect.~\ref{subsec:mass-loss} for further discussion). Hence, given its unusual location in the period-radius diagram and the brightness of the host star, TOI-5005~b represents a unique opportunity to probe how deep into the savanna atmospheric escape plays a relevant role. This, together with observations of the spin-orbit angle and long-term monitoring to detect massive companions will provide a clearer picture of the overall evolution of this system.

\subsection{TOI-5005~b in the $M_{\rm p}-R_{\rm p}$ and $\rho_{\rm p}-P_{\rm  orb}$ diagrams: A new member of the low-density savanna planets}
\label{subsec:LDSP}

\begin{figure*}
    \centering
    \includegraphics[width=0.48\textwidth]{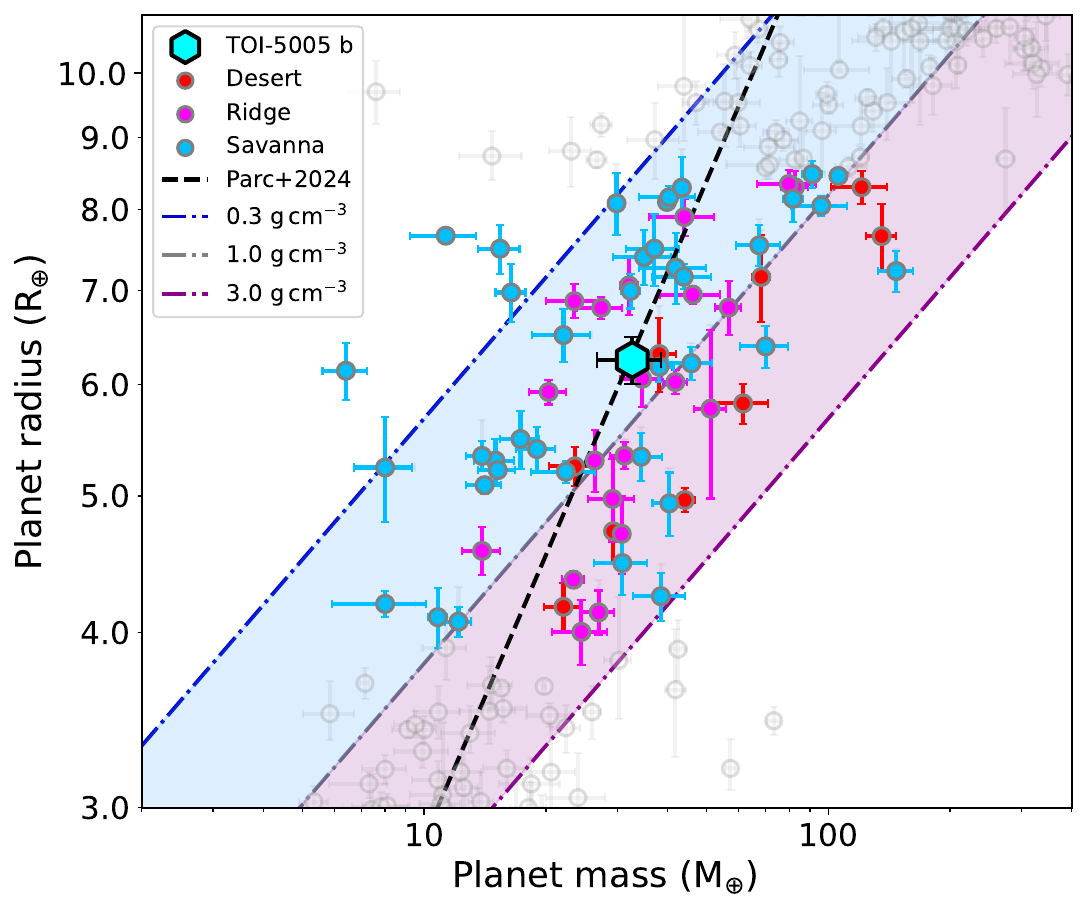}
    \includegraphics[width=0.476\textwidth]{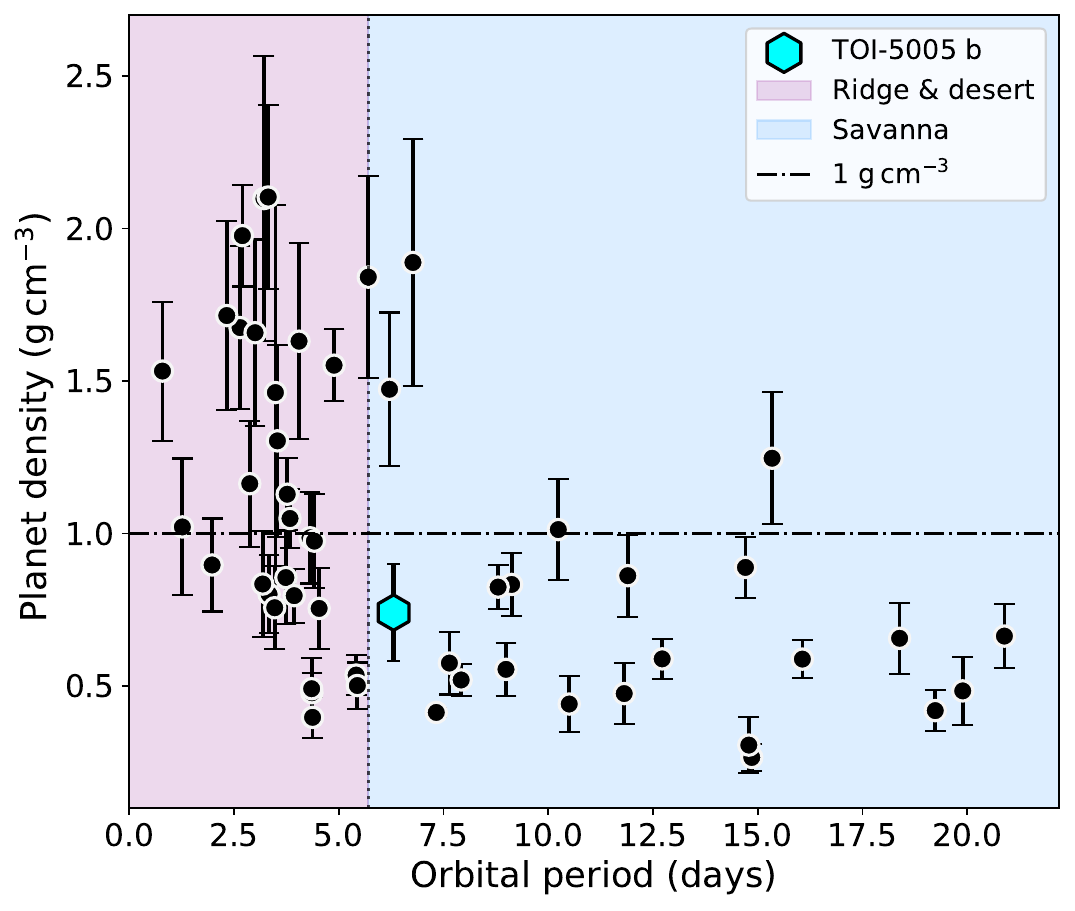}
    \caption{TOI-5005~b in the mass-radius and density-period diagrams. Left: Mass-radius diagram of all Neptunian planets with masses and radii with precisions better than 20$\%$. Right: Density-period diagram of the same sample. The data were collected from the NASA Exoplanet Archive on 20/09/2024. This plot was generated with \texttt{mr-plotter} \citep[\url{https://github.com/castro-gzlz/mr-plotter};][]{2023A&A...675A..52C}. }
    \label{fig:mass-radius-density-period}
\end{figure*}

In Fig.~\ref{fig:mass-radius-density-period}, left panel, we contextualize TOI-5005~b in the mass-radius diagram of precisely characterized planets. Its location is consistent with the mass-radius relation for volatile-rich exoplanets derived by \citet{2024A&A...688A..59P}. If we focus on the entire population, we see that there is a large dispersion in this region of the parameter space. Hence, continuing with the discussion in Sect.~\ref{subsec:period-radius}, we aim to study whether there exists any link between the observed planet densities and the new Neptunian landscape presented in \citet{2024A&A...689A.250C}. In the left panel of Fig.~\ref{fig:mass-radius-density-period}, we highlight the planets in the desert, ridge, and savanna in different colours. In the right panel of Fig.~\ref{fig:mass-radius-density-period}, we represent the density-period distribution of Neptunian planets. Interestingly, we find that planets in the ridge (and desert) tend to have densities larger than 1 $\rm g \, cm^{-3}$ (median value of 1.3~$\rm g \, cm^{-3}$), while planets in the savanna show lower densities, typically below 1~$\rm g \, cm^{-3}$ (median value of 0.6~$\rm g \, cm^{-3}$). We note that this difference is unlikely to be attributed to observational biases since denser planets are easier to detect. We quantified the significance of this trend through a Kolmogorov–Smirnov (KS) statistical test between the ridge and savanna populations, which we performed by taking into account the uncertainties of the densities through bootstrapping. We obtain a $D$-statistic of 0.39 $\pm$ 0.04 ($p$-value of $0.0092^{+0.0184}_{-0.0066}$), which allows us to confidently reject the hypothesis that both samples are drawn from the same distribution. We also searched for possible trends in the mass-period and radius-period spaces (see Fig.~E.5, available on Zenodo). We obtain $D$-statistics 0.33 $\pm$ 0.03 and 0.22 $\pm$ 0.03 ($p$-values of $0.037^{+0.030}_{-0.018}$ and $0.33^{+0.19}_{-0.13}$) for the mass and radius sample, respectively. On the one hand, the mass-period $p$-value is lower than
the most commonly assumed threshold for considering statistical significance (i.e. $p$-value < 0.05), although we note that its upper limit is larger (0.068). On the other hand, the radius-period $p$-value is well above such a threshold. We thus conclude that the radius distribution across the orbital period space is insensitive to the ridge and savanna, but Neptunes in the ridge tend to be more massive and dense than those in the savanna, being the planet density the key parameter that maximizes the significance. Having a density of $\rho_{\rm p}$ = $0.74 \pm 0.16$ $\rm g \, cm^{-3}$ and being located in the savanna with an orbital period of 6.3 days, TOI-5005~b is consistent with this newly identified trend.

The density-period trend observed in the Neptunian population might be explained through atmospheric, formation, and dynamical processes. However, determining which is the predominant agent shaping it is not trivial. On the one hand, Neptunian planets could all have been formed with low densities (e.g. < 1 $\rm g \, cm^{-3}$), presumably with extensive H/He atmospheres, and only those receiving high irradiations (i.e. those orbiting at short orbital distances) would be able to undergo strong enough evaporation to increase the bulk planet density. On the other hand, the formation and migration history of different types of Neptunes could also play an important role. While most close-in giant planets are thought to have undergone disk-driven migration soon after their formation, there is increasing evidence that many Neptunes in the ridge would have undergone late HEM processes (see  Sect.~\ref{subsec:period-radius}). In that sense, the density difference between Neptunes in the ridge and savanna could reflect the existence of two different populations of planets that formed and migrated through different channels. To date, there is not enough observational evidence to discern between the different hypotheses. Hence, confirming and characterizing planets located around the boundary between the desert and savanna such as TOI-5005~b will provide important insight into the transition between two possible different populations of Neptunian worlds.

\subsection{Internal structure} \label{sec:internal_structure}

\begin{table}[]
\centering
\renewcommand{\arraystretch}{1.35}
\setlength{\tabcolsep}{20pt}
\caption{Main physical parameters of the interior structure retrieval of TOI-5005~b (Sect.~\ref{sec:internal_structure}).}
\label{tab:interior_retrieval}
\begin{tabular}{ll}
\hline \hline
Parameter               & Value                  \\ \hline
Core mass fraction, CMF & $0.74^{+0.05}_{-0.45}$ \\
Atmospheric metallicity, log(Fe/H)            & $0.73^{+1.18}_{-1.64}$ \\
Envelope metal mass fraction, $Z_{\rm env}$           & $0.08^{+0.41}_{-0.06}$ \\
Total metal mass fraction, $Z_{\rm planet}$        & $0.76^{+0.04}_{-0.11}$ \\
Internal temperature, $T_{\rm int}$ ($K$)          & $97^{+15}_{-17}$ \\

\hline  
\end{tabular}
\end{table}

We performed an MCMC retrieval of the interior structure of TOI-5005~b. We employed GASTLI (GAS gianT modeL for Interiors, \citealp[]{Acuna21,2024A&A...688A..60A}) as a forward interior model, which is a one-dimensional, coupled interior-atmosphere model for warm gas giants. Our interior model is stratified into two layers: a core that consists of a 1:1 rock and water mixture, and an envelope. The mass of the core is computed from the planet's interior mass, $M_{\rm interior}$, and the core mass fraction (CMF), which are our input parameters, $M_{\rm core} =$ CMF $\times \ M_{\rm p}$. The composition of the envelope is determined by the input parameter $Z_{\rm env}$, which is the mass fraction of metals in the envelope, represented by water, whereas (1-$Z_{\rm env}$) is the mass fraction of H/He in the envelope. Hence, the total metal mass fraction in the planet is computed as $Z_{\rm planet}$ = CMF + (1-CMF) $\times \ Z_{\rm env}$. The interior structure models solve the equations of hydrostatic equilibrium, adiabatic temperature profile, mass conservation and Gauss’s theorem to compute the pressure, temperature and gravity profiles, respectively. The density and entropy are obtained with the respective equation of state (EOS) of each material. We adopt state-of-the-art EOS for silicates \citep{sesame,Miguel22}, H/He \citep{Chabrier21,HG23}, and water \citep{Mazevet19,Haldemann20}. Furthermore, we solve the equation of thermal cooling to determine the planetary luminosity as a function of age, $L = 4 \pi \sigma R^{2} T_{\rm int}^4$, where $T_{\rm int}$ is the internal (or intrinsic) temperature \citep{Fortney07,Thorngren16}. The boundary condition of the interior model is determined by coupling it to a grid of atmospheric models obtained with \texttt{petitCODE} \citep{Molliere15,Molliere17}. In the atmospheric models, we consider clear, cloud-free atmospheres. The envelope metal mass fraction,  $Z_{\rm env}$, is estimated using \texttt{easyCHEM} \citep{easychem} assuming chemical equilibrium for a given atmospheric metallicity in $\times$ solar units, log(Fe/H). The interior and the atmosphere are coupled at a pressure of 1000 bar, by using an iterative algorithm to ensure that the planet's interior radius (from the planet centre up to 1000 bar), and boundary temperature converge to constant values \citep{Acuna21}. The total planet radius is the sum of the converged interior radius and the thickness of the atmosphere, which is computed by integrating the gravity and pressure profiles from the bottom of the atmosphere (1000 bar) to the transit pressure \citep[20 mbar, see][]{Lopez14,Grimm18,Mousis20}. The total planet mass, envelope mass, CMF and $Z_{\rm planet}$ are recalculated taking into account the atmospheric mass (i.e $M_{\rm p} = M_{\rm interior} + M_{\rm atm}$), which has a significant effect in planets whose atmosphere (P > 1000 bar) is massive.

For the MCMC retrieval, we consider three observable parameters, which are the planetary mass and radius ($M_{\rm p}$~= $32.7\pm 5.9$, $R_{\rm p}$ = $6.25\pm 0.24$~$\rm R_{\rm \oplus}$), and age ($age \ = \ 2.20^{+0.61}_{-0.28}$ Gyr). We compute the log-likelihood as described in Eqs. 6 and 14 of \citet{Dorn15} and \citet{Acuna21}, respectively. We use the \texttt{emcee} \citep{emcee} MCMC package as a sampler. Our priors are uniform: $\mathcal{U}(0,0.99)$ for CMF; $\mathcal{U}(-2,2.4)$ for log(Fe/H); and $\mathcal{U}(50,250)$ K for the internal temperature. For the interior planet mass, $M_{\rm interior}$, we use a Gaussian prior with a mean and standard deviation equal to the observed mass. This choice eases the convergence of the MCMC retrieval as the difference between the interior mass and the total mass is negligible due to the low atmospheric mass. We assume a constant global equilibrium temperature of 1000 K, which is the maximum limit of our grid of atmospheric models. This is close enough to TOI-5005 b's equilibrium temperature (1040 K) to have a negligible effect on the metal content estimate.

\begin{figure}
    \centering
    \includegraphics[width=0.48\textwidth]{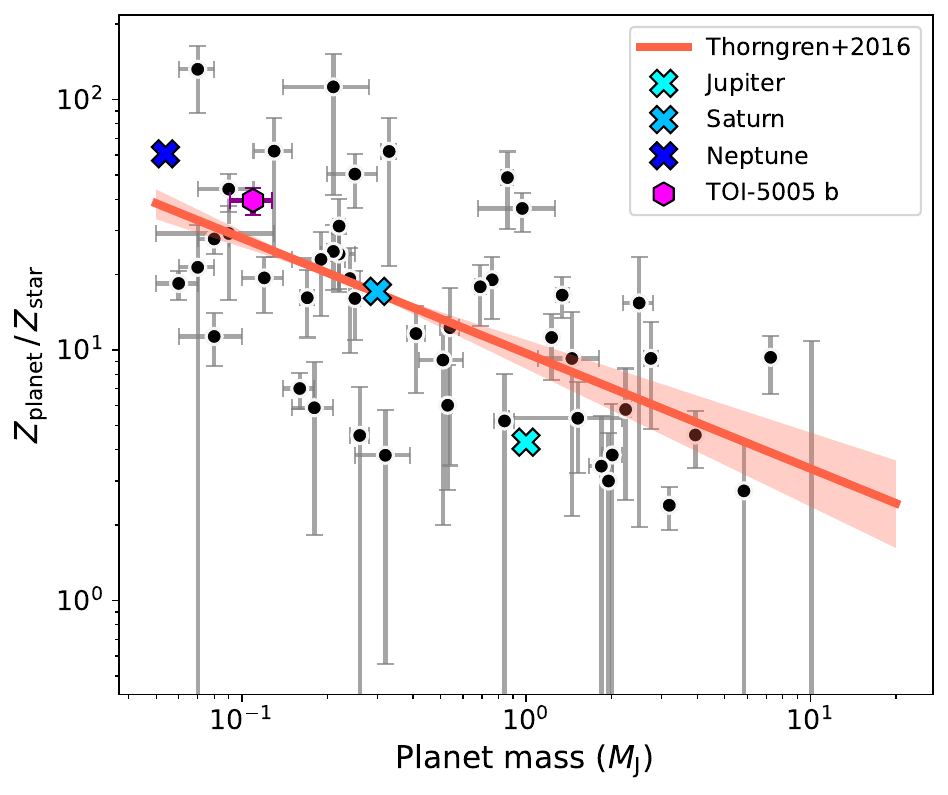}
    \caption{Heavy element enrichment of planets relative to their host stars as a function of mass. The black data points are from \citet{Thorngren16}, and the planet mass and enrichment of TOI-5005~b were inferred in Sects.~\ref{subsec:joint_analysis}  and \ref{sec:internal_structure}, respectively}
    \label{fig:enrichment_vs_mass}
\end{figure}

We summarize the results of the interior structure retrieval in Table \ref{tab:interior_retrieval}, and show the 1D and 2D posterior distributions in Fig.~E.6 (available on Zenodo). TOI-5005 b's mean metal mass fraction is $Z_{\rm planet} = $ 0.76, with 1$\sigma$ estimates ranging between 0.65 and 0.80. The metal content is distributed between the core and the envelope. In the latter, the envelope mass fraction is compatible with atmospheric metallicities ranging from subsolar (log(Fe/H) = - 1.17, 0.12 $\times$ solar) to 81 $\times$ solar within 1$\sigma$, although a metallicity of $\times$ 250 solar cannot be ruled out (2$\sigma$). For reference, Neptune and Uranus have metal mass fractions $Z_{\rm planet} > $ 0.80, and atmospheric metallicities between 60 and 100 $\times$ solar, whereas Saturn presents a $Z_{\rm planet} = 0.2 $ and an atmospheric metallicity of $\times$ 10 solar \citep[][and references therein]{MV23}. Thus, TOI-5005 b has an overall metal mass fraction slightly lower than that of the Solar System ice giants. To break the degeneracy between the CMF and $Z_{\rm env}$, atmospheric characterization data is necessary to constrain the atmospheric metallicity, log(Fe/H). This could also help unveil the extent of mixing in the interior: if the atmospheric metallicity is found to be high (> 10 $\times$ solar) it would suggest that metals are fairly well mixed in the interior \citep{Thorngren19}. In addition, we calculate the stellar metal mass fraction as $Z_{\star} = 0.014 \times 10^{\rm [Fe/H]_{\star}}$ \citep{Thorngren16}, and used it to infer the heavy element enrichment relative to the host star: $Z_{\rm planet}/Z_{\star}$ = 37.5 $\pm$ 5. In Fig.~\ref{fig:enrichment_vs_mass}, we plot such ratio versus the measured planet mass for the TOI-5005 system and include the sample and mass-metallicity relations obtained by \citet{Thorngren16}. TOI-5005 b is compatible with such a relation, which indicates that it could have formed via core accretion \citep[see][for further discussion]{Thorngren16}.

\subsection{Atmospheric mass-loss rate}
\label{subsec:mass-loss}
TOI-5005 b presents an intriguing case for the study of atmospheric evolution due to its unique combination of parameters. Despite its relatively high equilibrium temperature ($1040 \pm 20$~K), the planet's heavy mass contributes to a Jeans escape parameter $\Lambda_\mathrm{p} = R_\mathrm{p}/H$ of $38 \pm 7$, where $H$ is the scale height of the planet's atmosphere. Sub-Neptunes with Jeans escape parameters greater than 20--25 are not expected to experience significant atmospheric mass-loss \citep{Owen2016,Cubillos2017,Fossati2017,Vivien2022}. However, given its uniqueness, TOI-5005 b deserves a more detailed analysis.

The photoevaporation rate of super-Neptunes is expected to be in the energy-limited regime \citep{Owen2016-energy-limited}, and can be computed with \citep{Erkaev2007,Owen2013}
\begin{equation}
    \dot{M}=\epsilon \frac{\pi F_{\mathrm{XUV}} R_{\mathrm{p}}^3}{G M_{\mathrm{p}}}, \label{eq:mass-loss-rate}
\end{equation}
where $F_\mathrm{XUV}$ is the XUV flux received by the planet, $G$ is the gravitational constant, and $\epsilon$ is an efficiency parameter. The XUV flux from the star is not constant in time, and its value is approximated by the analytical fit of XUV luminosity as a function of age obtained by \cite{Sanz-Forcada2011}. The value of $\epsilon$, sometimes noted $\eta$ in the literature, is computed using Eq. (31) from \cite{Owen2017}. We perform a simple Monte Carlo analysis by assuming Gaussian priors on all stellar and planetary parameters from Tables \ref{tab:stellar_parameters} and \ref{tab:parameters_joint}, except for the stellar age, for which we use the posterior distribution from the \texttt{stardate} analysis ($\rm age \ = \ 2.20^{+0.61}_{-0.28}$ Gyr). We find a present-day photoevaporation rate of $(2.5 \pm 1.4) \times 10^{9}$ g\,s$^{-1}$ ($0.013\pm 0.008$~$\rm M_{\oplus}$~Gyr$^{-1}$). Since the XUV flux emission by the host star is decreasing with time, most of the photoevaporation happens within the first Gyr of the planet's evolution if the planet migrates early on and stays on its present-day orbit \citep{Bourrier2018-GJ436b,Owen2018,Attia2021}.

To quantify this, we estimate the total mass of H/He lost by TOI-5005 b following the approach of \cite{Aguichine2021}. We integrate the mass-loss rate $\dot{M}$ from $t=0$ to the planet's present-day age, assuming that $M_\mathrm{p}$, $R_\mathrm{p}$ and $T_\mathrm{eq}$ remained roughly constant and that variations in $\dot{M}$ are mainly caused by the changing stellar XUV luminosity as a function of star's age. In this case, we find that TOI-5005 b could have lost $0.36 \pm 0.18$ $\rm M_{\oplus}$~ of H/He, that is, $\sim 1.1\%$ of its current mass. This is not sufficient to substantially change the bulk composition of the planet but can alter the composition of its upper atmosphere.

The procedure described above is repeated by replacing Eq.~\ref{eq:mass-loss-rate} by mass-loss rates computed from the publicly available grid of models from \cite{Kubyshkina2018,Kubyshkina2021}, which is adapted to planets with masses 1--40 $\rm M_{\oplus}$. Mass-loss rates are computed by linear interpolation in this grid with respect to five parameters: stellar mass, incident XUV flux, planet equilibrium temperature, planet mass, and planet radius. This approach still uses the stellar XUV luminosity as a function of age from \cite{Sanz-Forcada2011}. The present-day mass loss rate computed from their model is $(3.3 \pm 2.3) \times 10^{10}$ g\,s$^{-1}$ ($0.17\pm 0.12$ ~$\rm M_{\oplus}$~Gyr$^{-1}$), and the total mass lost since formation, obtained by integrating $\dot{M}$, is $1.8 \pm 0.7$~$\rm M_{\oplus}$, which is comparable to the total H/He  in the envelope ($\sim 7.3$ $\rm M_{\oplus}$, see Sect. \ref{sec:internal_structure}).

These results are one order of magnitude greater than the estimate from the model of \cite{Owen2016}. We also computed mass-loss rates with the analytical fits to the model from \cite{Salz2016}, and found results compatible with those produced by the model from \cite{Kubyshkina2018}. However, the model of \cite{Salz2016} is only calibrated to moderate XUV fluxes, and is therefore outside of its validity range during the stellar saturation regime, so we do not present its results here. The discrepancy between the values from Eq. \ref{eq:mass-loss-rate} and the models of \cite{Salz2016} and \cite{Kubyshkina2018} is most likely due to the fact that Eq. \ref{eq:mass-loss-rate} computes the mass loss rate at the surface. In reality, XUV photons can be absorbed much higher in the atmosphere, up to nanobar levels, so that the surface to collect XUV is much bigger and the gravitational potential is weaker, resulting in much greater mass loss rates.

\subsection{Prospects for atmospheric characterization}
\label{subsec:prospects_atm}

We study the feasibility of atmospheric characterization of TOI-5005~b through the transmission and emission spectroscopy metrics, as well as modelling the spectrum of a primary and secondary atmosphere in both emission and transmission geometries. We determine the transmission spectroscopy metric \citep[TSM;][]{Kempton2018} using the planet's mass, radius, equilibrium temperature and J-band magnitude of the star. We find a TSM of 84 using the assumption of an H$_2$- and He-rich atmosphere. We also consider a secondary atmosphere with 500$\times$ metallicity, dominated by heavier species such as H$_2$O, CO and CO$_2$. For such a case, the higher mean molecular mass of the atmosphere reduces the TSM to 27. The emission spectroscopy metric (ESM) is 15 but is not strongly affected by the change in the mean molecular mass between the two cases.

\begin{figure*}
\centering
        \includegraphics[width=\textwidth,trim={0cm 0cm 0cm 0},clip]{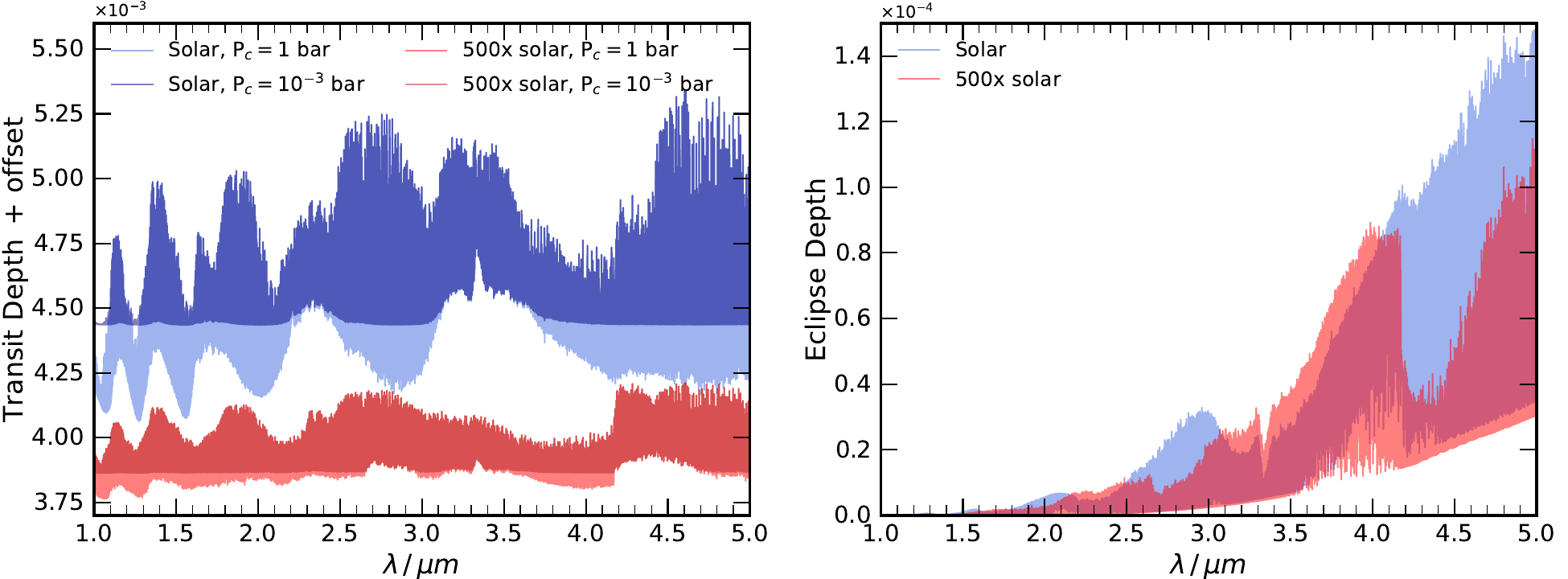}
    \caption{Modelled atmospheric spectra of TOI-5005~b assuming a solar and 500$\times$ solar metallicity atmosphere. Left panel: Transmission spectra, with cloud decks for each model at 1~bar and 1~mbar for each of the cases. Right panel: Eclipse depths of the dayside atmosphere.   Both show similar feature strengths and absorption from spectroscopically active species.}     
\label{fig:atm_model}
\end{figure*}

We generated model spectra in both transmission and emission geometries of TOI-5005~b using GENESIS \citep{Gandhi2017, Gandhi2020}. These are shown in Fig.~\ref{fig:atm_model} for both the primary atmosphere at solar metallicity as well as a secondary atmosphere at 500$\times$ solar metallicity. For all of the models, we assumed a temperature profile with 1.1 $\times$ T$_\mathrm{eq}$ (1140~K) in the deep atmosphere which monotonically cooled to 0.6 $\times$ T$_\mathrm{eq}$ (620~K) in the upper layers, in order to be as realistic as possible with potential spectroscopic signatures. It is likely that any clouds present in the atmosphere will reduce the signal from the planet's atmosphere further in transmission. Therefore, we also modelled the atmosphere with clouds at 1~mbar level for both the solar and high-metallicity cases. The model spectra show that despite the clouds, features do still appear due to absorption from H$_2$O, CO, CH$_4$ and CO$_2$ in the atmosphere. For the high-metallicity case, some features from strong lines originate at the $\sim \mu$bar level, indicating we may also be sensitive to photochemistry if the planet's atmosphere has a high metallicity. In emission in the infrared, we show the model spectra for the solar and 500$\times$ solar metallicity cases, which both offer similar feature strengths from the spectrally active species.

\section{Summary and conclusions}
\label{conclusions}

For this work we used TESS, HARPS, PEST, and TRAPPIST-South data to confirm the planetary nature of TOI-5005~b and characterize its orbital and physical properties. Our analysis shows that TOI-5005~b is a super-Neptune with a radius of $R_{\rm p}$ = $6.25\pm 0.24$~$\rm R_{\rm \oplus}$ ($R_{\rm p}$ = $0.558\pm 0.021$ $\rm R_{\rm J}$) and a mass of $M_{\rm p}$~= $32.7\pm 5.9$ $\rm M_{\oplus}$ ($M_{\rm p}$ = $0.103\pm 0.018$ $\rm M_{\rm J}$). These properties, together with its orbital period ($P_{\rm orb}$ = $6.3085044^{+0.0000092}_{-0.0000088}$ days), place TOI-5005~b in the Neptunian savanna near the ridge, a recently identified overdensity of Neptunes in the  period range of  $\simeq$3-5 days. Having a density of $0.74 \pm 0.16$ $\rm g \, cm^{-3}$, TOI-5005~b becomes a new member of a population of low-density Neptunes ($<1$ $\rm g \, cm^{-3}$) in the savanna, in contrast to the planets in the ridge, which typically show higher densities ($>1$ $\rm g \, cm^{-3}$). This trend, together with the different occurrence rates, supports the existence of different processes populating both regimes.

We also identified a periodic modulation in the TESS data that matches the orbital period of TOI-5005~b. Given the sub-Alfvénic star-planet distance, this signal can be interpreted as a reflection of magnetic star-planet interactions (MSPIs). We studied the possibility that the modulation has an instrumental origin by performing the strongest possible CBV-based correction. We found that the amplitude decreased, but the periodicity holds, which supports a stellar origin. Additionally, we studied how the modulation correlates with the planetary orbit, and found similar patterns to those observed in other systems with MSPIs. Overall, these results strongly support the MSPI explanation for the observed signal. However, we also consider that additional observations are needed to get a final confirmation. These can be performed in the optical with instrumentation less subjected (or subjected to different) systematics \citep[e.g. CHEOPS;][]{2021ExA....51..109B}, or in radio wavelengths to observe the electron cyclotron maser emission \citep[e.g. GMRT;][]{1991ASPC...19..376S}.

Our internal structure modelling of TOI-5005~b constrains the overall planetary metal mass fraction to a value slightly lower than that of the Solar System ice giants ($Z_{\rm planet}$  = $0.76^{+0.04}_{-0.11}$). This value, together with the measured $Z_{\rm star}$, makes this system consistent with the well-known mass-metallicity trend, which suggests that TOI-5005~b was formed via core accretion. However, the core mass fraction (CMF = $0.74^{+0.05}_{-0.45}$) and envelope mass fraction ($Z_{\rm env}$ = $0.08^{+0.41}_{-0.06}$) are  degenerate. To break this degeneracy, a direct measurement of the atmospheric metallicity is needed. This shows the relevance of atmospheric observations not only in probing the atmosphere itself, but also the planetary interior. Interestingly, having a transmission spectroscopy metric of 84, TOI-5005~b is amenable for atmospheric studies. We simulated model spectra in both transmission and emission geometries and found the appearance of features such as H$_2$O, CO, CH$_4$, and CO$_2$, which would be detectable even with clouds at 1 mbar level. We also estimated the present-day atmospheric mass-loss of TOI-5005~b, and found inconsistent values depending on the choice of photoevaporation model ($0.013\pm 0.008$~$\rm M_{\oplus}$~Gyr$^{-1}$ vs $0.17\pm 0.12$ ~$\rm M_{\oplus}$~Gyr$^{-1}$). Atmospheric escape mechanisms in super-Neptunes have received less attention compared to hot Jupiters and hot sub-Neptunes as modelling efforts have been primarily focused on these latter categories. This knowledge gap has led to discrepancies that hinder a comprehensive understanding of the evolutionary processes in such bodies. Hence, the detection of planets similar to TOI-5005 b holds the potential to stimulate research into these intermediate planets.

TOI-5005~b orbits the brightest host of its surrounding neighbours in the period-radius space, which makes it a key target for atmospheric and orbital architecture observations. TOI-5005 is thus an excellent option to be observed by the large-scale survey of orbital architectures ATREIDES (Bourrier et al., in prep.), and by the Near-Infrared Gatherer of Helium Transits (NIGHT) spectrograph \citep{2024MNRAS.527.4467F}. Combining our current knowledge with additional constraints on the spin-axis angle and mass-loss rate will provide a clearer picture of the overall evolution of this unusual planetary system. 

\section{Data availability}

Appendix E is available on Zenodo (\url{https://zenodo.org/records/13932107}).

\begin{acknowledgements}

We thank the anonymous referee for the constructive revision, which helped us to improve the clarity with which several results and discussions were presented. 
A.C.-G. is funded by the Spanish Ministry of Science through MCIN/AEI/10.13039/501100011033 grant PID2019-107061GB-C61. 
J.L.-B. is funded by the Spanish Ministry of Science and Universities (MICIU/AEI/10.13039/501100011033) and NextGenerationEU/PRTR grants PID2019-107061GB-C61 and CNS2023-144309. 
This material is based upon work supported by NASA’s Interdisciplinary Consortia for Astrobiology Research (NNH19ZDA001N-ICAR) under grant number 80NSSC21K0597.
This work has been carried out within the framework of the NCCR PlanetS supported by the Swiss National Science Foundation under grants 51NF40$\_$182901 and 51NF40$\_$205606. This project has received funding from the European Research Council (ERC) under the European Union's Horizon 2020 research and innovation programme (project {\sc Spice Dune}, grant agreement No 947634).
E.D.M. further acknowledges the support from FCT through the Stimulus FCT contract 2021.01294.CEECIND.
A.M. acknowledges funding support from Grant PID2019-107061GB-C65 funded by MCIN/AEI/10.13039/501100011033, and from Generalitat Valenciana in the frame of the GenT Project CIDEGENT/2020/036.
A.~C.~M.~C. acknowledges support from the FCT, Portugal, through the CFisUC projects UIDB/04564/2020 and UIDP/04564/2020, with DOI identifiers 10.54499/UIDB/04564/2020 and 10.54499/UIDP/04564/2020, respectively.
NCS acknowledges funding by the European Union (ERC, FIERCE, 101052347). Views and opinions expressed are however those of the author(s) only and do not necessarily reflect those of the European Union or the European Research Council. Neither the European Union nor the granting authority can be held responsible for them. This work was supported by FCT - Fundação para a Ciência e a Tecnologia through national funds and by FEDER through COMPETE2020 - Programa Operacional Competitividade e Internacionalização by these grants: UIDB/04434/2020; UIDP/04434/2020.
This paper made use of data collected by the TESS mission and are publicly available from the Mikulski Archive for Space Telescopes (MAST) operated by the Space Telescope Science Institute (STScI). Funding for the TESS mission is provided by NASA’s Science Mission Directorate. We acknowledge the use of public TESS data from pipelines at the TESS Science Office and at the TESS Science Processing Operations Center. Resources supporting this work were provided by the NASA High-End Computing (HEC) Program through the NASA Advanced Supercomputing (NAS) Division at Ames Research Center for the production of the SPOC data products. This publication makes use of data products from the Two Micron All Sky Survey, which is a joint project of the University of Massachusetts and the Infrared Processing and Analysis Center/California Institute of Technology, funded by the National Aeronautics and Space Administration and the National Science Foundation.
The material is based upon work supported by NASA under award number 80GSFC21M0002.
This publication made use of \texttt{TESS-cont} (\url{https://github.com/castro-gzlz/TESS-cont}), which also made use of \texttt{tpfplotter} \citep{2020A&A...635A.128A} and \texttt{TESS\_PRF} \citep{2022ascl.soft07008B}.
This work made use of \texttt{mr-plotter} (available in \url{https://github.com/castro-gzlz/mr-plotter}).
This work made use of \texttt{tpfplotter} by J. Lillo-Box (publicly available in www.github.com/jlillo/tpfplotter), which also made use of the python packages \texttt{astropy}, \texttt{lightkurve}, \texttt{matplotlib} and \texttt{numpy}.
Based on data collected by the TRAPPIST-South telescope at the ESO La Silla Observatory. TRAPPIST is funded by the Belgian Fund for Scientific Research (Fond National de la Recherche Scientifique, FNRS) under the grant PDR T.0120.21, with the participation of the Swiss National Science Fundation (SNF). The postdoctoral fellowship of KB is funded by F.R.S.-FNRS grant T.0109.20 and by the Francqui Foundation. This publication benefits from the support of the French Community of Belgium in the context of the FRIA Doctoral Grant awarded to M.T. M.G. F.R.S.-FBRS Research Director. F.J.P acknowledges financial support from the Agencia Estatal de Investigación (AEI/10.13039/501100011033) of the Ministerio de Ciencia e Innovación and the ERDF “A way of making Europe” through projects PID2022-137241NB-C43 and the Centre of Excellence “Severo Ochoa” award to the Instituto de Astrofísica de Andalucía (CEX2021-001131-S). E.J. is a Belgian FNRS Senior Research Associate.
This work has made use of data from the European Space Agency (ESA) mission {\it Gaia} (\url{https://www.cosmos.esa.int/gaia}), processed by the {\it Gaia} Data Processing and Analysis Consortium (DPAC, \url{https://www.cosmos.esa.int/web/gaia/dpac/consortium}).
We acknowledge the use of public TESS data from pipelines at the TESS Science Office and at the TESS Science Processing Operations Center. Resources supporting this work were provided by the NASA High-End Computing (HEC) Program through the NASA Advanced Supercomputing (NAS) Division at Ames Research Center for the production of the SPOC data products.
This research has made use of the Exoplanet Follow-up Observation Program (ExoFOP; DOI: 10.26134/ExoFOP5) website, which is operated by the California Institute of Technology, under contract with the National Aeronautics and Space Administration under the Exoplanet Exploration Program.
This research has made use of the NASA Exoplanet Archive, which is operated by the California Institute of Technology, under contract with the National Aeronautics and Space Administration under the Exoplanet Exploration Program. 
This research has made use of the SIMBAD database \citep{2000A&AS..143....9W}, operated at CDS, Strasbourg, France. This work also made use of \texttt{astropy} \citep{2022ApJ...935..167A}, \texttt{matplotlib} \citep{2007CSE.....9...90H}, \texttt{numpy} \citep{2020Natur.585..357H}, and \texttt{lightkurve} \citep{2018ascl.soft12013L}.

\end{acknowledgements}

%
%

\bibliographystyle{aa} 
\bibliography{references} 
\begin{appendix}

\section{\texttt{TLS} and \texttt{GLS} periodograms}

\begin{figure*}

\includegraphics[width=\textwidth]{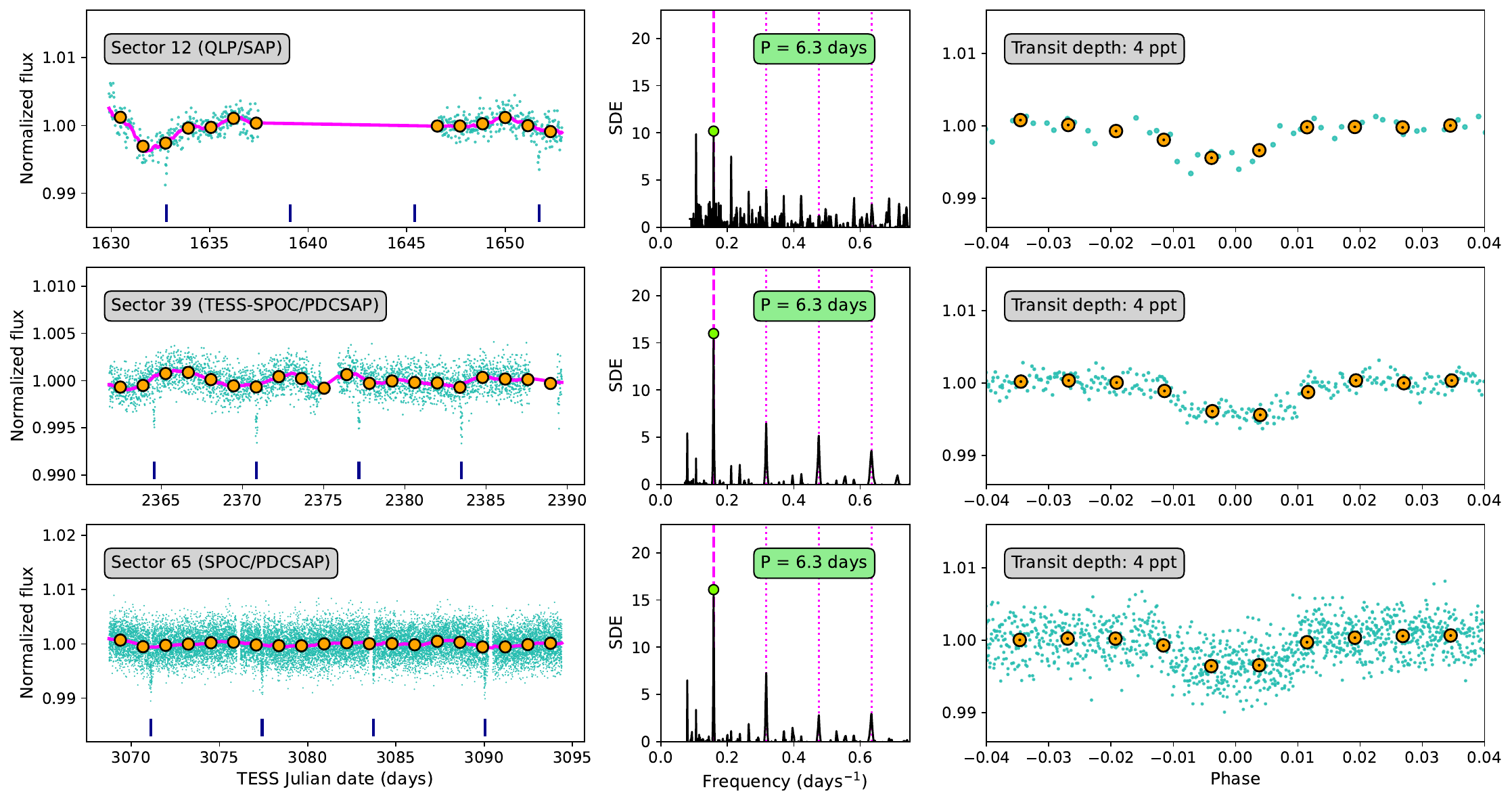}
         \caption{Transit-like signals within the TESS photometry of TOI-5005. Left panels: Photometric time series. The magenta line is the trend used to flatten the photometry before the periodogram computation, which was obtained through the time-windowed biweight method implemented in \texttt{wotan} \citep{2019AJ....158..143H} with a 1-day window length. The orange circles correspond to 1.3-day binned data. Centre panels: Transit  least squares periodograms of the corresponding time series. The vertical magenta dashed lines indicate the orbital period of TOI-5005.01, and the vertical magenta dotted lines indicate its second, third, and fourth harmonics. Right panels: Flattened time series folded to the maximum power periods. The orange circles correspond to 1 hr binned data.}
         \label{fig:tls_to_TESS}
\end{figure*}

\begin{figure*}
    \includegraphics[width=\textwidth]{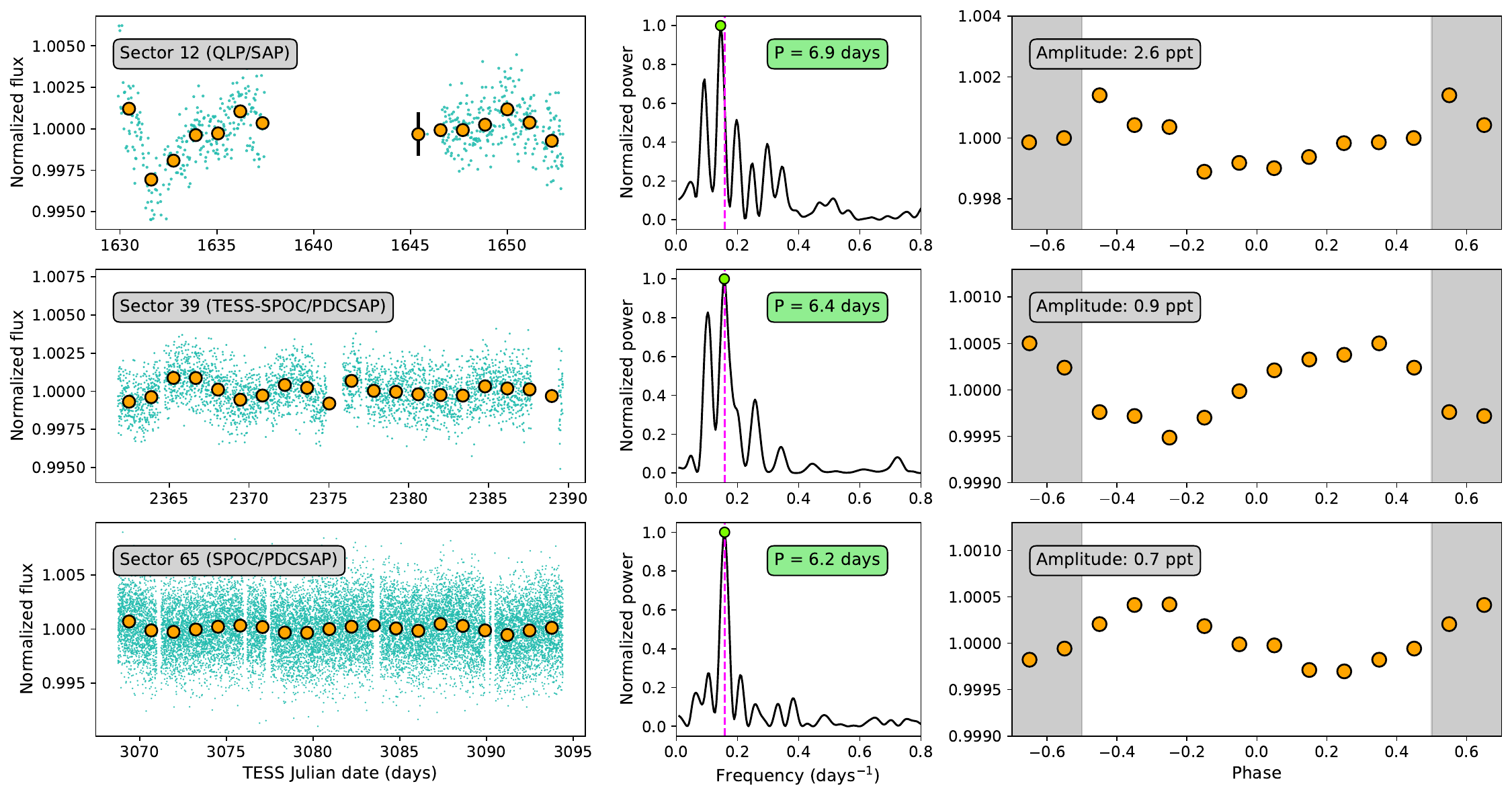}
         \caption{Sinusoidal signals within the TESS photometry of TOI-5005. Left panels: Photometric time series without the TOI-5005.01 transits. The orange circles correspond to 1.3-day binned data. Centre panels: Generalized Lomb-Scargle Periodograms of the time series. The vertical magenta dashed lines indicate the orbital period of TOI-5005.01. The green circles and boxes indicate the maximum power frequencies. Right panels: Photometric time series folded to the orbital period of TOI-5005.01. The phase is referred to the time of inferior conjunction. The orange circles correspond to data binned over phase windows of 0.1 in width.} 
         \label{fig:gls_to_TESS}
\end{figure*}

\begin{figure*}
    \includegraphics[width=\textwidth]{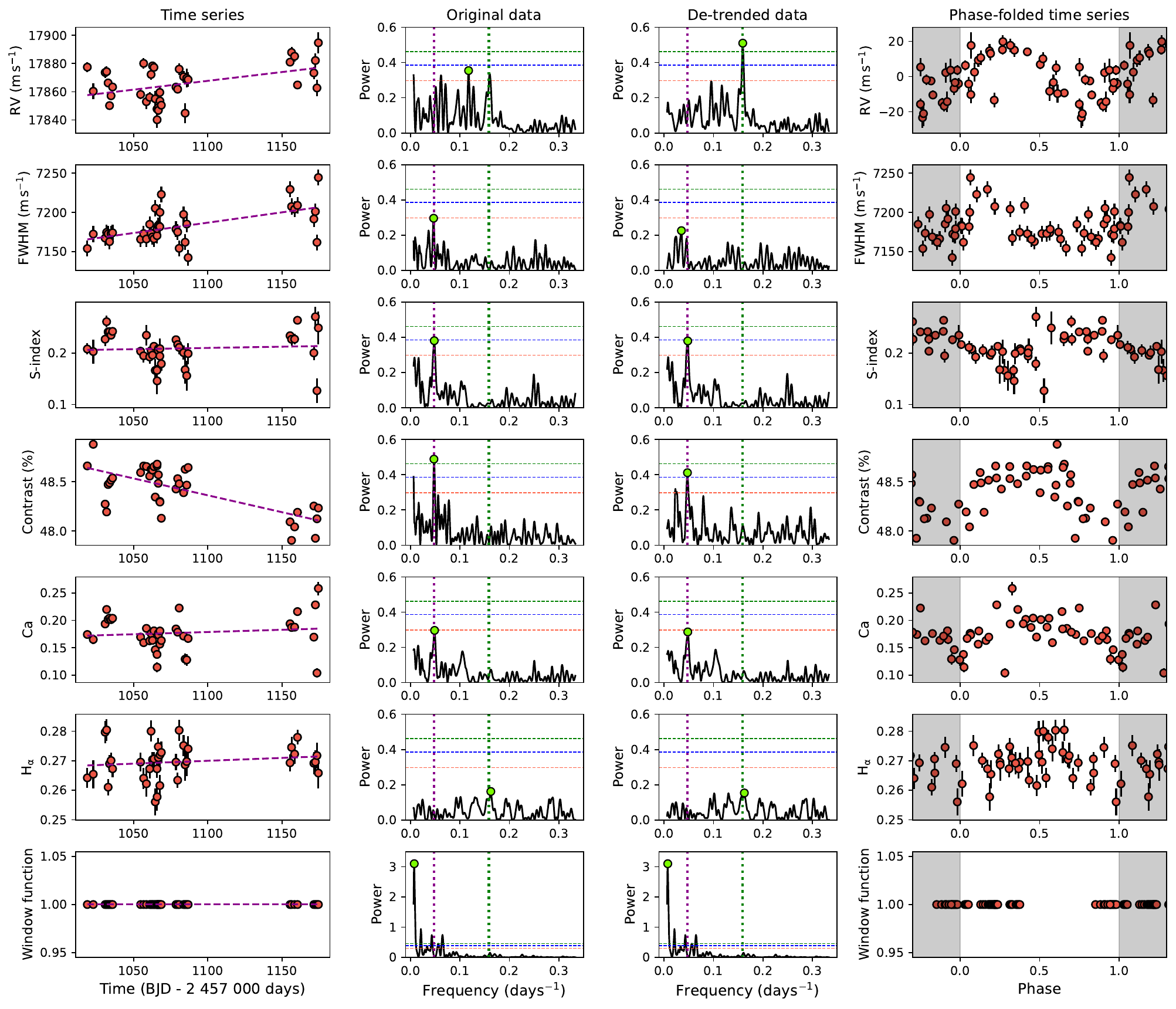}
    \caption{Sinusoidal signals within the HARPS RVs of TOI-5005. Left panels: Time series of the HARPS RVs, activity indicators, and the window function. The magenta dashed lines represent the linear trends fit to the data. Centre panels: GLS periodograms of the original time series and de-trended time series. The green dotted vertical lines indicate the location of the orbital period of TOI-5005.01 ($P_{\rm orb}$ = 6.3 days). The magenta dotted vertical lines indicate the $\simeq$21-day activity-related signal that most likely reflects the stellar rotation period. The horizontal dotted lines correspond to the 10 (orange), 1 (blue), and 0.1$\%$ (green) FAP levels. Right panels: HARPS data folded to the maximum power periods: 6.3 days (RVs), 21.3 days (FWHM), 20.8 days (S-index), 21.0 days (Contrast), 20.5 days (Ca), 6.2 days ($\rm H_{\alpha}$), and 129.8 days (window function). The RV panel shows the de-trended data and the indicators panels show the original time series. }
    \label{fig:gls_to_HARPS}
\end{figure*}

\section{Bayesian mathematical framework}
\label{bayesian}

We aim to estimate the probability that a physical model $M$ describes a dataset $D$. This probability is the conditional probability that $M$ is true given that $D$ is true, and it can be computed according to Bayes' theorem \citep{1763RSPT...53..370B} as

\begin{equation}
    P\left(M|D\right) = \frac{P\left(D|M\right) P\left(M\right)}{P\left(D\right)}.
    \label{eq:bayes_theorem}
\end{equation}

$P\left(D|M\right)$ is the probability of observing $D$ if the model $M$ is true, and it is commonly called likelihood. $P\left(M\right)$ is the probability that $M$ is true before $D$ was observed, and it is commonly called prior probability. $P\left(D\right)$ is the probability of observing $D$ for all possible values of the model parameters, and it is commonly called marginal likelihood or model evidence.

For an observed dataset $D$ with $N$ data points ($D_{1},...,D_{N}$), a synthetic set of $N$ values can be generated based on a physical parametric model $M$ ($M_{1},...,M_{N}$). The likelihood that a given data point $D_{i}$ is described by a synthetic point $M_{i}$ can be written as $P(D_{i}|M_{i})$, and the likelihood that the complete dataset is described by the model is the product of each probability

\begin{equation}
    P\left(D|M\right) = \prod_{i=1}^{N} P\left(D_{i}|M_{i}\right).
\end{equation}

The form of the likelihood function depends on the measurement distribution and the noise properties. Assuming normally distributed data with uncorrelated noise, it can be written as

\begin{equation}
    P(D|M) = \prod_{i=1}^{N} \left[ \frac{1}{\sqrt{2\pi \delta_{i}^{2}}}  \right] e^{-\frac{1}{2}\chi^{2}},
\end{equation}where 

\begin{equation}
    \chi^{2} = \sum_{i=1}^{N} \frac{\left(D_{i} - M_{i} \right)}{\delta_{i}^{2}}.
\end{equation}Here $\delta_{i}$ is the uncertainty of the data point $D_{i}$. Given that formal uncertainties can be underestimated, we consider $\delta_{i}^{2}$ = $\sigma_{i}^{2}+\sigma_{jit}^{2}$, being $\sigma_{i}$ the estimated uncertainties and $\sigma_{jit}$ a jitter term introduced to account for additional uncorrelated noise. 

Equation \ref{eq:bayes_theorem} can lead to very large or small numbers that cause computational numerical overflows, so it is more appropriate to consider it in logarithmic form; that is, ${\rm ln}\, P\left(M|D\right) = {\rm \ln}\,P\left(D|M\right) + P(M) - P(D)$. Therefore, in practice, we implemented the log-likelihood function as

\begin{equation}
    {\rm ln}\, P\left(D|M\right) = -\frac{1}{2} \left[  \sum_{i}^{N} {\rm ln} \, 2\pi \left(\sigma_{i}^{2}+\sigma_{jit}^{2}\right) + \chi^{2} \right].
    \label{eq:log-like_uncorrelated}
\end{equation}

In order to deal with correlated noise, we used models composed of a mean function $f(x_{i})$ and an autocorrelation function $k(x_{i}, x_{j})$, with $i,j = 1,...,N$. These models are known as GPs \citep[][]{2006gpml.book.....R}, and their log-likelihood functions can be written as

\begin{equation}
     {\rm ln}\, P\left(D|M\right) = -\frac{1}{2} r^{\rm T} K^{-1} r  - \frac{1}{2} {\rm ln} \, {\rm det} \, K - \frac{N}{2} {\rm ln}  (2\pi),
     \label{eq:log-like_correlated}
\end{equation}where 

\begin{equation}
    r = \left[D_{1} - f(x_{1}), D_{2} - f(x_{2}),..., D_{N} - f(x_{N})  \right],
\end{equation}and $K$ is the covariance matrix, whose elements are given by the autocorrelation function as $\left[K\right]_{ij}$ = $k(x_{i}, x_{j})$.

We implemented Eq. \ref{eq:log-like_uncorrelated} and computed it directly since its computational cost depends only on the evaluation of the models. However, in Eq. \ref{eq:log-like_correlated}, in addition to the evaluation of the mean function, it is necessary to compute the inverse and determinant of a $N\times N$ matrix, which scales the computational cost as $N^{3}$. Therefore, when dealing with correlated noise, we used the newly developed fast methods from \citet{2015ITPAM..38..252A} and \citet{2017AJ....154..220F}, which are implemented in the \texttt{george}\footnote{Available at \url{https://github.com/dfm/george}.} and \texttt{celerite}\footnote{Available at \url{https://github.com/dfm/celerite}.} Python packages.

We considered three types of prior distributions $P(D)$, which we defined for each particular analysis according to our previous knowledge of the phenomena to analyse. Those types are uniform, Gaussian, and truncated Gaussian priors. Uniform priors are also called weakly informative or uninformative priors since they only constrain that a given parameter $\theta_{i}$ lies inside a range $\left[a,b\right]$ with no preferred values (i.e. with equal probabilities). Uniform priors can be expressed as 

\begin{equation}
\centering
\mathcal{U}(a,b) = \left\{
        \begin{array}{ll}
            \left(b-a\right)^{-1}& \quad  a < \theta_{i} < b \\
            \quad 0 & \quad {\rm otherwise}
        \end{array}
    \right.
    \label{eq:unifor_priors}
\end{equation}

Gaussian priors are also called informative priors since they contain previous information on model parameters (i.e. their mean values and 1-$\sigma$ uncertainties). This information is based on previous measurements of an independent dataset not used within the analysis. We implemented Gaussian priors as 

\begin{equation}
    \mathcal{G}(a,b) = \frac{1}{\sqrt{2\pi b^{2}}} {\rm exp} \left[-\frac{\left(\theta_{i}-a\right)^{2}}{2b^{2}} \right],
    \label{eq:gaussian_priors}
\end{equation}where $a$ and $b$ are the median and standard deviation of the measured model parameter, respectively. Even if they are highly constrained, Gaussians can take any value from -inf to +inf. To avoid nonphysical values (i.e. negative stellar masses or radius, negative orbital periods of semi-major axes), we also used truncated Gaussian priors $\mathcal{TG}(a,b)$, which are Gaussian priors limited to a certain range of values. Given that nonphysical values are typically negative values, we mainly consider zero-truncated Gaussians, which we refer to as $\mathcal{ZTG}$(a,b).

Finally, to obtain the posterior distribution of every parameter $\theta_{i}$, we integrated the non-normalized posterior distribution $P(D|M)P(M)$ over the remaining parameters $\theta_{j\neq i}$, a process called marginalization over $\theta_{i}$.

\section{Uninformed CBV-based systematics correction}
\label{sec:cbv_correction}

The TESS PDC algorithm for systematics correction \citep{2012PASP..124.1000S,2012PASP..124..985S} consists of fitting SAP photometry to an ensemble of time series that represent the main instrumental trends in a given channel $\rm \left\lbrace Sector, Camera, CCD \right\rbrace$. Those time series are called co-trending basis vectors (CBVs) and are obtained through singular value decomposition (SVD) from the 50$\%$ most correlated SAP time series. CBVs contain information on a wide range of systematic effects detected within a given channel. However, not all those systematics necessarily affect the photometry of stars in different regions of the CCD in the same way. As detailed in \citet{2012PASP..124.1000S}, over-fitting is a major issue when dealing with CBVs. Coincidental correlations between instrumental effects and stellar variability can occur frequently, leading to stellar features being identified as systematics and consequently removed. To prevent the removal of astrophysical features, the PDC algorithm fits the CBVs to the observed data by using a Bayesian maximum a posteriori (MAP) approach. Bayesian inference allows for making conditioned fits that rely on a priori information on the expected phenomena to model. In the CBV correction context, the PDC algorithm assumes that the instrumental systematics affecting a given star are similar to the ones affecting its nearest stars. Hence, the algorithm imposes a priori constraints on the CBVs to be fitted to the raw data. This way, only photometric features that are also present in stars in the neighbourhood are removed.

The SPOC PDC procedure has been proven to be very successful in preserving stellar signals, allowing the determination of many rotation periods \citep[e.g.][]{2020ApJS..250...20C}. However, it could lead to under-fitting if a given star is not affected by instrumental systematics in the same way as an average nearby star. Therefore, we aim to perform the strongest possible CBV correction, which consists of making no assumptions about the expected systematics behaviour. As discussed above, this approach considerably increases the risk of over-fitting and removing real astrophysical features. However, if the 6.3-day signal prevails after performing such an uninformed CBV correction, it would strongly favour a stellar origin, since no combination of the major TESS instrumental features would explain it.

For each channel $\rm \left\lbrace Sector, Camera, CCD \right\rbrace$ we built a CBV-based instrumental systematics model as

\begin{equation}
    M_{\rm CBV} (t) = \sum_{i = 1}^{N_{m_{1}}} b_{i} m_{1,i} (t) + \sum_{j = 1}^{N_{m_{2}}} c_{j} m_{2,j} (t) + \sum_{k = 1}^{N_{m_{3}}} d_{k} m_{3,k} (t) + \sum_{l = 1}^{N_{s}} e_{l} s_{l} (t), 
    \label{eq:systematics_model}
\end{equation}where $m_{1,i}$, $m_{2,j}$, and $m_{3,k}$ are multiscale CBVs containing systematic trends in specific wavelet-based band passes \citep{Stumpe2014}, $s_{l}$ are CBVs containing short spike systematics, and $b_{i}$, $c_{j}$, $d_{k}$, and $e_{l}$ are the linear combination coefficients. All the CBVs were obtained from MAST via TESS bulk downloads.\footnote{ \url{https://archive.stsci.edu/tess/bulk_downloads/bulk_downloads_cbv.html}}

We modelled the SAP photometry of each sector through Bayesian inference as described in Sect.~\ref{sec:model_sel_param_det}. As discussed before, contrary to SPOC PDC, we did not impose any constraint on the linear combination coefficients, which we left to vary freely with uninformative priors $\mathcal{U} (-10^{4}, 10^{4})$. In Figs.~E.7, E.8, and E.9 (available on Zenodo), we show the posterior distributions obtained for each coefficient. In Fig.~E.10 (available on Zenodo), we show the  \texttt{GLS} periodograms of the SPOC PDCSAP photometry and our corrected photometry based on uninformative CBV coefficients (we refer to it as Free-CBVs PDCSAP). The S12 periodogram shows its highest peak (at $\simeq$4.6 days) surrounded by a forest of similar peaks, so no clear periodicity can be detected. The S39 and S65 periodograms, however, show maximum power peaks at 6.3 days, similar to the SPOC PDCSAP peaks. Although still present, the photometric modulation has decreased in amplitude, which is reflected within the smaller periodogram powers. Hence, the strongest possible CBV correction has absorbed part of the 6.3-day modulation detected within SPOC PDCSAP, but the absorption was not complete, still leaving clear evidence of a 6.3-day sinusoidal signal.

\section{Additional tables}


\onecolumn

\begin{table}
\renewcommand{\arraystretch}{1.23}
\setlength{\tabcolsep}{9.2pt}
\parbox{.47\linewidth}{
\caption{TESS Simple Aperture Photometry (SAP) of TOI-5005. The complete table is available at the CDS.}
\begin{tabular}{cccc}
\hline \hline
BJD (days)  & SAP & QF  & Sector \\ \hline
2458624.995 & 1.0026 $\pm$ 0.0016     & 18432   & TESS12 \\
...         & ...                     & ... & ...    \\ \hline
\end{tabular}
\label{tab:TESS_SAP}
}
\hfill
\parbox{0.47\linewidth}{
\caption{TESS QLP and PDCSAP photometry of TOI-5005. The complete table is available at the CDS.}
\begin{tabular}{cccc}
\hline \hline
BJD (days)  & QLP/PDCSAP & QF  & Sector \\ \hline
2458629.891 & 1.0047 $\pm$ 0.0013     & 0   & TESS12 \\
...         & ...                     & ... & ...    \\ \hline
\end{tabular}
\label{tab:TESS_QLP_PDCSAP}
}
\end{table}

\begin{table*}[]
\caption{PEST photometry and detrend parameters acquired on 5 May 2022. The detrend parameters correspond to the position (\textit{x},\textit{y}) and distance (\textit{dist}) to the detector centre of the target star, the full width at half maximum of the target point spread function (\textit{fwhm}), airmass ($\chi$), and background flux (\textit{sky}). The complete table is available at the CDS.}
\renewcommand{\arraystretch}{1.23}
\setlength{\tabcolsep}{9.88pt}
\begin{tabular}{ccccccccc}
\hline \hline
BJD (days)  & Flux   & Flux error & \textit{x} & \textit{y} & \textit{dist} & \textit{fwhm} & $\chi$ & \textit{sky} \\ \hline
2459705.010 & 1.0023 & 0.0024     & 1797.5795  & 769.5013   & 438.5767      & 6.1665        & 1.9437 & 1009.1895    \\
...         & ...    & ...        & ...        & ...        & ...           & ...           & ...    & ...          \\ \hline
\end{tabular}
\label{tab:pest_data}
\end{table*}

\begin{table*}[]
\caption{TRAPPIST-South photometry and detrend parameters acquired on 22 March 2024. The detrend parameters correspond to the displacement (\textit{dx},\textit{dy}) of the target star across the CCD, the full width at half maximum of the target point spread function (\textit{fwhm}), airmass ($\chi$), and background flux (\textit{sky}). The complete table is available at the CDS.}
\renewcommand{\arraystretch}{1.23}
\setlength{\tabcolsep}{14.85pt}
\begin{tabular}{cccccccc}
\hline \hline
BJD (days)   & Flux   & Flux error & \textit{dx} & \textit{dy} & \textit{fwhm} & $\chi$ & \textit{sky} \\ \hline
2460392.733 & 0.9982 & 0.0024     & -0.2685     & -0.5710     & 2.9760        & 1.2924 & 168.8333     \\
...          & ...    & ...        & ...         & ...         & ...           & ...    & ...          \\ \hline
\end{tabular}
\label{tab:trappist_south_data}
\end{table*}

\begin{table*}[]
\caption{HARPS radial velocities and activity indicators of TOI-5005 acquired between 15 March 2023 and 18 August 2023 under the programmes 108.21YY.001 and 108.21YY.002. The complete table is available at the CDS.}
\renewcommand{\arraystretch}{1.3}
\setlength{\tabcolsep}{8.4pt}
\begin{tabular}{ccccccc}
\hline \hline
RJD (days) & RV ($\rm m \, s^{-1}$) & FWHM ($\rm m \, s^{-1}$) & S-index           & Ca                  & $\rm H_{\alpha}$    & Contrast (\%) \\ \hline
60018.834  & 17877.4 $\pm$ 3.5      & 7153.9 $\pm$ 10.1         & 0.208 $\pm$ 0.011 & 0.1742 $\pm$ 0.0044 & 0.2642 $\pm$ 0.0035 & 48.662        \\
...  & ...     & ...       & ... & ... & ... & ...        \\ \hline
\end{tabular}
\label{tab:harps_rvs}
\end{table*}




\begin{table}[]
\centering
\small
\renewcommand{\arraystretch}{1.14}
\setlength{\tabcolsep}{25pt}
\caption{Inferred parameters of TOI-5005~b based on the joint analysis described in Sect.~\ref{subsec:joint_analysis}.}
\label{tab:parameters_joint}
\begin{tabular}{lll}
\hline
Parameter                                                   & Priors                              & Posteriors                            \\ \hline
\multicolumn{3}{l}{Orbital and physical parameters}                                                                                       \\ \hline
Orbital period, $P_{\rm orb}$ (days)                        & $\mathcal{U}(5.0, 7.0)$             & $6.3085044^{+0.0000092}_{-0.0000088}$ \\
Time of mid-transit, $T_{0}$ (JD)                           & $\mathcal{U}(2460089.0, 2460091.0)$ & $2460090.0356 \pm 0.0010$             \\
Orbital inclination, $i$ (degrees)                          & $\mathcal{U}(50.0, 90.0)$           & $89.53^{+0.33}_{-0.44}$               \\
Scaled planet radius, $R_{p}/ R_{\star}$                    & $\mathcal{U}(0.0, 0.1)$             & $0.0616 \pm 0.0012$                   \\
RV semi-amplitude, $K \, (\rm m\,s^{-1})$                   & $\mathcal{U}(0.0, 10^{5})$          & $11.6 \pm 2.1$                        \\
Ecc. parametrization, $\rm cos(\omega)\sqrt{e}$              & (fixed)                           & 0                                     \\
Ecc. parametrization, $\rm sin(\omega)\sqrt{e}$               & (fixed)                           & 0                                     \\
Planet radius, $R_{p} \, (\rm R_{\rm \oplus})$              & (derived)                           & $6.25 \pm 0.24$                       \\
Planet radius, $R_{p} \, (\rm R_{\rm J})$                   & (derived)                           & $0.558 \pm 0.021$                     \\
Planet mass, $M_{p} \, (\rm M_{\rm \oplus})$                & (derived)                           & $32.7 \pm 5.9$                        \\
Planet mass, $M_{p} \, (M_{\rm J})$                         & (derived)                           & $0.103 \pm 0.018$                     \\
Planet density, $\rho_{p}$ ($\rm g \, cm^{-3}$)             & (derived)                           & $0.74 \pm 0.16$                       \\
Transit depth, $\Delta$ (ppt)                               & (derived)                           & $3.80 \pm 0.15$                       \\
Transit duration, $T_{\rm 14}$ (hours)                      & (derived)                           & $3.144 \pm 0.046$                     \\
Relative orbital separation, $a / R_{\star}$                & (derived)                           & $15.29 \pm 0.50$                      \\
Orbit semi-major axis, $a$ (au)                              & (derived)                           & $0.06614 \pm 0.00045$                 \\
Planet surface gravity, $g$ $(\rm m\,s^{-2})$               & (derived)                           & $8.2 \pm 1.6$                         \\
Impact parameter, $b$                                       & (derived)                           & $0.13 \pm 0.10$                       \\
Incident flux, $F_{\rm inc} \, (\rm F_{\oplus})$            & (derived)                           & $226 \pm 12$                          \\
Equilibrium temperature [A=0], $T_{\rm eq} \, (\rm K)$      & (derived)                           & $1040 \pm 20$                         \\ \hline
\multicolumn{3}{l}{Limb-darkening coefficients}                                                                                           \\ \hline
Limb-darkening coefficient, $q_{\rm 1, TESS}$               & $\mathcal{ZTG}(0.32, 0.32)$         & $0.41^{+0.20}_{-0.16}$                \\
Limb-darkening coefficient, $q_{\rm 2, TESS}$               & $\mathcal{ZTG}(0.36, 0.14)$         & $0.31^{+0.13}_{-0.12}$                \\
Limb-darkening coefficient, $q_{\rm 1, PEST}$               & $\mathcal{ZTG}(0.43, 0.37)$         & $0.66^{+0.20}_{-0.22}$                \\
Limb-darkening coefficient, $q_{\rm 2, PEST}$               & $\mathcal{ZTG}(0.38, 0.12)$         & $0.43^{+0.12}_{-0.12}$                \\
Limb-darkening coefficient, $q_{\rm 1, TRAPPIST-South}$     & $\mathcal{ZTG}(0.27, 0.29)$         & $0.37^{+0.20}_{-0.17}$                \\
Limb-darkening coefficient, $q_{\rm 2, TRAPPIST-South}$     & $\mathcal{ZTG}(0.34, 0.15)$         & $0.30^{+0.14}_{-0.13}$                \\ \hline
\multicolumn{3}{l}{GP hyperparameters}                                                                                                    \\ \hline
$\eta_{\rm \sigma_{S12}}$                                   & $\mathcal{U}(0, 0.5)$               & $0.0060^{+0.0018}_{-0.0011}$          \\
$\eta_{\rm \sigma_{S39}}$                                   & $\mathcal{U}(0, 0.5)$               & $0.00423^{+0.00063}_{-0.00047}$       \\
$\eta_{\rm \sigma_{S65}}$                                   & $\mathcal{U}(0, 0.5)$               & $0.00353^{+0.00080}_{-0.00054}$       \\
$\eta_{\rm \rho_{S12}}$ (days)                              & $\mathcal{U}(0, 30)$                & $1.10^{+0.26}_{-0.18}$                \\
$\eta_{\rm \rho_{S39}}$ (days)                              & $\mathcal{U}(0, 30)$                & $0.633^{+0.125}_{-0.096}$             \\
$\eta_{\rm \rho_{S65}}$ (days)                              & $\mathcal{U}(0, 30)$                & $1.17^{+0.21}_{-0.16}$                \\ \hline
\multicolumn{3}{l}{Instrument-dependent parameters}                                                                                       \\ \hline
TESS LC level S12, $F_{\rm 0,S12}$                          & $\mathcal{U}(-0.1, 0.1)$            & $-0.00064^{+0.00175}_{-0.00195}$      \\
TESS LC level S39, $F_{\rm 0,S39}$                          & $\mathcal{U}(-0.1, 0.1)$            & $-0.00052^{+0.00098}_{-0.00096}$      \\
TESS LC level S65, $F_{\rm 0,S65}$                          & $\mathcal{U}(-0.1, 0.1)$            & $-0.00028^{+0.00112}_{-0.00111}$      \\
TESS LC jitter S12, $\sigma_{\rm TESS,S12}$                 & $\mathcal{U}(0, 0.005)$             & $0.000831 \pm 0.000030$               \\
TESS LC jitter S39, $\sigma_{\rm TESS,S39}$                 & $\mathcal{U}(0, 0.005)$             & $0.000750 \pm 0.000039$               \\
TESS LC jitter S65, $\sigma_{\rm TESS,S65}$                 & $\mathcal{U}(0, 0.005)$             & $0.000082^{+0.00084}_{-0.00057}$      \\
HARPS RV jitter, $\sigma_{jit,\rm HARPS}$ $(\rm m\,s^{-1})$ & $\mathcal{U}(0, 100)$               & $7.6^{+1.4}_{-1.2}$                   \\ \hline
\multicolumn{3}{l}{PEST and TRAPPIST-South detrend parameters}                                                                            \\ \hline
$c_{\rm PEST, x}$                                           & $\mathcal{U}(-10^{4}, 10^{4})$      & $0.001 \pm 0.020$                     \\
$c_{\rm PEST, y}$                                           & $\mathcal{U}(-10^{4}, 10^{4})$      & $-0.011 \pm 0.016$                    \\
$c_{\rm PEST, dist}$                                        & $\mathcal{U}(-10^{4}, 10^{4})$      & $0.0151 \pm 0.0055$                   \\
$c_{\rm PEST, fwhm}$                                        & $\mathcal{U}(-10^{4}, 10^{4})$      & $-0.0047 \pm 0.0021$                  \\
$c_{\rm PEST, \chi}$                                        & $\mathcal{U}(-10^{4}, 10^{4})$      & $0.0008 \pm 0.0013$                   \\
$c_{\rm PEST, sky}$                                         & $\mathcal{U}(-10^{4}, 10^{4})$      & $-0.00033 \pm 0.00032$                \\
$c_{\rm TRAPPIST-South, dx}$                                & $\mathcal{U}(-10^{4}, 10^{4})$      & $0.0000078 \pm 0.0000013$             \\
$c_{\rm TRAPPIST-South, dy}$                                & $\mathcal{U}(-10^{4}, 10^{4})$      & $0.0000052 \pm 0.0000015$             \\
$c_{\rm TRAPPIST-South, fwhm}$                              & $\mathcal{U}(-10^{4}, 10^{4})$      & $-0.0041 \pm 0.0010$                  \\
$c_{\rm TRAPPIST-South, sky}$                               & $\mathcal{U}(-10^{4}, 10^{4})$      & $0.0027 \pm 0.0017$                   \\
$c_{\rm TRAPPIST-South, \chi}$                              & $\mathcal{U}(-10^{4}, 10^{4})$      & $0.0017 \pm 0.0013$                   \\ \hline
RV linear drift                                             &                                     &                                       \\ \hline
Systemic velocity, $v_{\rm HARPS}$ $(\rm m\,s^{-1})$        & $\mathcal{U}(17800, 17900)$         & $17858.6 \pm 2.3$                     \\
Slope, $\gamma_{\rm HARPS}$ $(\rm m\,s^{-1}\,day^{-1})$     & $\mathcal{U}(0, 1)$                 & $0.143 \pm 0.031$                     \\ \hline
\end{tabular}
\end{table}

\section{Additional figures}

This Appendix is available in Zenodo.\footnote{\url{https://zenodo.org/records/13932107}}

\end{appendix}

\end{document}